\documentclass[fleqn,array]{aa}
\usepackage{natbib}
\bibpunct{(}{)}{;}{a}{}{,}
\usepackage{graphicx}	
\usepackage{amsmath}	
\usepackage{amssymb}	
\usepackage{multicol}        
\usepackage{bm}		
\usepackage[english]{babel}
\usepackage{array}
\usepackage[fleqn]{nccmath}
\newcolumntype{P}[1]{>{\centering\arraybackslash}p{#1}}
\usepackage[varg]{txfonts}
\title{Young stellar populations in early-type dwarf galaxies;}
\subtitle{occurrence, radial extent and scaling relations}

\author{E. Hamraz\fnmsep\thanks{E-mail: hamraz@astro.rug.nl} \inst{1,2},  R. F. Peletier\inst{1}, H. G. Khosroshahi\inst{2}, E. A. Valentijn\inst{1}, M. den Brok\inst{3}, A. Venhola\inst{1,4}}
\institute{Kapteyn Astronomical Institute, University of Groningen, Postbus 800, NL-9700 AV Groningen, the Netherlands\\
\and School of Astronomy, Institute for Research in Fundamental Sciences (IPM), Tehran, 19395-5746, Iran\\
\and Leibniz-Institut f\"{u}r Astrophysik Potsdam, An der Sternwarte 16, 14482 Potsdam, Germany\\
\and Astronomy Research Unit, University of Oulu, Oulu, FI-90014 Finland}

\date{Received 17/01/2019; accepted 28/03/2019}
\titlerunning{Young stellar populations in early-type dwarf galaxies}
\authorrunning{E. Hamraz et al.}

\abstract
{}
{To understand the stellar population content of dwarf early-type galaxies (dEs) and its environmental dependence, we compare the slopes and intrinsic scatter of color-magnitude relations (CMRs) for three nearby clusters, Fornax, Virgo and Coma. Additionally we present and compare internal color profiles of these galaxies to identify central blue regions with younger stars.} 
{We use the imaging of the HST/ACS Fornax cluster in the magnitude range of $-18.7 \leq M_{g^{\prime}} \leq -16.0$, to derive magnitudes, colors and color profiles, which we compare with literature measurements from the HST/ACS Virgo and Coma Cluster Survey. We benefit HST accuracy to investigate and parameterize the $(g^{\prime}-z^{\prime})$ color profiles of these dEs.}
{Based on analysis of the color profiles, we report a large number of dEs with young stellar populations in their center in all three clusters. While for Virgo and Coma the number of blue-cored dEs is found to be $85\% \pm 2\%$ and $53\% \pm 3\%$ respectively, for Fornax, we find that all galaxies have a blue core. We show that bluer cores reside in fainter dEs, similar to the trend seen in nucleated dEs. We find no correlation between the luminosity of the galaxy and the size of its blue core. Moreover, a comparison of the CMRs of the three clusters shows that the scatter in Virgo's CMR is considerably larger than in the Fornax and Coma clusters. Presenting adaptive smoothing we show that the galaxies on the blue side of the CMR often show evidence for dust extinction, which strengthens the interpretation that the bluer colors are due to young stellar populations. We also find that outliers on the red side of the CMR are more compact than expected for their luminosity. We find several of these red outliers in Virgo, often close to more massive galaxies. No red outlying compact early-types are found in Fornax and Coma in this magnitude range while we find three in the Virgo cluster. We show that the CMR of the Fornax and Virgo clusters are slightly bluer than that of Coma. We suggest that the large number of outliers and larger scatter found for the Virgo cluster CMR is a result of Virgo’s different assembly history.}
{}


\begin{document} 
\maketitle
%

\section{Introduction}
\label{section:Introduction}

\par Our understanding of galaxy formation and evolution is based on numerical models of the Universe. Parameters in these models are adjusted in order to reproduce a number of observations, such as the galaxy luminosity function, several scaling relations, appearance of galaxies, presence of AGNs, etc. \citep[for a review see][]{Silk2012}. One of the important scaling relations is the color-magnitude relation (CMR), connecting the magnitude of a galaxy, i.e. an indicator of its mass, to its color, an index of its stellar populations. This relation is fundamental, since it tightly connects a macroscopic property, galaxy mass, to small scale quantities such as the constituting stellar populations. The CMR of the early-type galaxies is often called the {\it 'red sequence'}, and is very pronounced in observations of galaxy clusters \citep[e.g.][]{Eisenhardt2007,Stott2009,Sanchez2014}, and in large surveys such as SDSS \citep [e.g.][]{Baldry2004}.

\par Galaxies were positioned either on the red sequence or the blue cloud, a densely populated area in color-magnitude space where galaxies are found to show star formation. Between the red sequence and the blue cloud, one finds a relatively underpopulated region, called the green valley. It is known that the red sequence is built up over a rather long period starting with the most massive galaxies. This process is called downsizing \citep{Thomas2005,Choi2014}. The red sequence has been identified in (proto) clusters up to z=2 \citep[e.g.][]{Muzzin2009,Taylor2009}. Recently, there has been a considerable amount of work going on to understand the faint end of the red sequence, and to find out when dwarf galaxies end up on the red sequence \citep[see][for an extended discussion about this]{BoselliGavazzi2014,Boselli2014,Roediger2017,Schombert2018}. 

\par To understand the CMR better, and therefore one of the influential aspects of galaxy formation, it is important to understand the origin of its scatter. Measuring the scatter is best done in galaxy clusters, since the distance errors are minimized, as all galaxies are roughly at the same distance. \cite{Bower1992b} measured a scatter in the relation between $M_V$ and both $U-V$ and $V-K$ of $\sim$0.05-0.06 mag for both the Virgo and Coma cluster, using a sample of giant ellipticals and S0 galaxies. For ellipticals only, the scatter reduces to $\sim$0.03-0.04 mag, by lowering the magnitude range to only 2 magnitudes. For all cases the scatter is larger than the uncertainties in the data. \cite{Eisenhardt2007} found similar result for the Coma cluster. There is thus a consensus that the CMR of galaxies is tight, with non-zero intrinsic scatter that cannot be explained by observational uncertainties. It is generally accepted that this scatter is caused by young stellar populations, making the color bluer than expected for a given magnitude \citep{Schweizer1992}. Hence the CMR and its scatter in a galaxy cluster is a very useful tool to measure the evolutionary status of its galaxies. 

\par It is generally known that dwarf galaxies have more extended star formation histories, and move to the red sequence only after their more massive counterparts \citep[Downsizing, e.g.][]{Thomas2005}. For nearby clusters, where the red sequence is well in place for massive galaxies, dwarf galaxies are ideal to study their evolution. However, not many studies discuss the CMR in the mass range of dwarf galaxies, since they have low surface brightness, and wide-field imaging surveys have only recently become available. \cite{JanzLisker} used SDSS data for early-types of the Virgo cluster and showed that the scatter around the CMR increases toward fainter magnitudes and that the scatter is intrinsic. \cite{Roediger2017} recently showed CMRs in optical colors for the galaxies in the center of the Virgo cluster. Their scatter is significantly lower than the one reported by \cite{JanzLisker}, and they also went considerably deeper in magnitude. The fact that the precision of the magnitudes in both papers is similar for galaxies brighter than M$_g$=-16 indicates that the decrease in scatter in \citet{Roediger2017} is due to two points: for dwarfs brighter than M$_g$=-16 the difference in samples seems to dominate (the fact that \cite{Roediger2017} only have galaxies in the center of Virgo), while for fainter objects the SDSS colors are not accurate enough.

\par In addition to the position of each galaxy on the CMR of its cluster, a detailed study of the color profile of each individual dE may reveal more hints to their formation and evolution. dEs are small, low-luminosity galaxies $(M_{B}\geq -18)$ \citep{Fergouson1994} with a shallow potential well that makes them sensitive to their environment. Early studies \citep[e.g.][]{Fergouson1994} assumed that dEs are galaxies without any gas. However, as our understanding of these galaxies improved, it was found that not all the dEs are quiescent galaxies with smooth surface brightness, elliptical isophotes and regular appearance any more, since some spiral structure, recent or ongoing star formation, gas and dust content are reported in some of them \citep{Looze2010,Lisker2006,Grebel2001,Jerjen2000,Hodge1973}. Some studies have even announced detection of H\,{\sc i} in some of these galaxies \citep{Conselice2003etal,Gavazzi2003}. In addition, \citet{Lisker2006} by using SDSS spectra showed clear signs of star formation in these galaxies. dE star formation histories are diverse and are dependent on environment \citep[e.g.][]{Tolstoy2009}.

\par The fact that early-type dwarfs show signs of star formation is not very new \citep{Vigroux1984}. Studying young population in early-type galaxies, \cite{Peletier1993} reported to find dust and recent star forming region in the center of the dE NGC205 (Messier 110). He presented radial color profiles in various colors and showed that it drops toward center of the galaxy \citep[Fig3,][]{Peletier1993}. \citet{Durrell1997} investigated two dwarf galaxies in Virgo with CFHT and found one of them to have a bluer nucleus than the galaxy. Using the advantage of higher resolution data, \citet{Lotz2004} for a sample of dE in the Leo Group, the Virgo and Fornax clusters, observed with HST WFPC2, showed that nuclei of dEs are bluer than their underlying galaxies. In the Virgo cluster and by using SDSS data for galaxies with $M_{B}\leq-13$, \citet{Lisker2006} noticed that about $5\%$ of dEs have a blue center. This number is more than $15\%$, among dEs with $M_{B}\leq-15$. \citet{Pak} reported about $70\%$ of dEs in their Ursa Major sample to have a young core with blue UV-optical color, an indication of recent or ongoing star formation in their centers \citep{Gu2006,DeRijcke}.

\par Studying the formation of stellar populations in the inner parts of dEs could lead us to find some clues about the origin of these galaxies and the role of the environment on their evolution. On the one hand, galaxy properties correlate with their environment. Ram pressure stripping, strangulation, merging, tidal interaction and harassment are among the important processes happening in the cluster and changing the properties of the galaxies. \cite{BoselliGavazzi2014} explained that the strong gravitational potential well in the core of these galaxies, might keep the gas needed for recent star formation in the stripped galaxies and make the center blue. Recently, \cite{Zabel} as part of ALMA Fornax Cluster Survey, studied the CO(1-0) line as the cold molecular gas tracer which shows star formation. They reported disturbed morphologies and kinematics for several of their CO detected galaxies. Ram pressure stripping is considered the more probable candidate for these disturbed morphologies which shows the role of the cluster environment of their evolution. Infalling late-type galaxies that interact with the cluster environment and transform to blue-cored dEs are one of the most popular formation scenarios \citep{Boselli2008,DeRijcke2010,Lisker2006,Mastropietro2005}. In some other scenarios, it could be gas-rich irregulars which have merged and formed dEs. \citet{Pak} suggested that the blue core dEs could be a possible object in transition of a late-type galaxy to a red early-type. On the other hand, considering similarities between the Virgo and Fornax nuclei versus their different environment, \citet{Turner2012} concluded that the formation of  blue-cored dEs is more dependent on local factors of the galaxy than its residing environment.

\par Benefiting from the high resolution imaging of HST, an object as small as a nucleus of a dwarf galaxy can be distinguished in the nearby clusters further than our local group. Since the size of these blue cores are comparable with the nuclei found by HST survey of the nearby clusters, reviewing their results will shed a light on this investigation. \cite{Lotz2004} used F555W and F814W bands and studied nuclei and globular clusters of 69 dwarf ellipticals in the Virgo and Fornax Clusters and the Leo Group. Defining the nuclei as a bright compact object in the region of $1.5^{\prime\prime}$ from isophotal center, they noted that most of the nuclei are bluer than their host galaxies. They did not find any correlation between the projected distance from the center of the cluster and the properties of the nuclei. They claimed that brighter dEs have deeper potentials and are able to attract gas in to their centers, which can explain the reason why redder and brighter dEs have redder and more luminous nuclei. Comparing the nuclei of dEs in Fornax and Virgo, they could not find any considerable differences. 

\par There are more studies which used HST images to go down to the size of the nuclei. Following the ACS Virgo cluster survey, \cite{Cote2006} investigated sizes, colors and some other properties of the nuclei of dEs in the Virgo cluster using F475W and F850LP bands. \citet{Cote2006} concluded that the number of galaxies with a bluer nucleus in their sample is between $66\%$ and $82\%$. They found many more nucleated dEs than previous studies. They came to the conclusion that since the color and the luminosity of the nuclei are strongly correlated and they rarely relate to their host galaxies, their formation and enrichment are more caused by internal factors.

\par The nuclei of the early-types in the Fornax cluster have also been studied using HST resolution. As part of the ACSFCS collaboration, \cite{Turner2012} found that most of low and intermediate luminosity early-type galaxies in their sample show an excess light in their central part. Similar to the previous studies, they also confirm that the nuclei are mostly bluer than their host galaxies, at least $72\%$ of them, and bluer hosts have bluer nuclei and vice-versa. Comparing their sample with \cite{Cote2006}, they concluded that the similarities of the nuclei of these two clusters, is an argument showing that the environment does not play a role in their formation and evolution. 

\par Our aim in this paper is to make a detailed study and comparison of the color of dEs, in the magnitude range of $-18.7 \leqslant M_{g^{\prime}} \leqslant -16.0$, i.e., the range of the dwarfs in ACSFCS, in three nearby clusters; Virgo, Fornax and Coma. To do this, we chose a sample of very accurate archival HST imaging data. Therefore, we can make a direct comparison between the HST-derived CMRs in these three clusters and study the origin of the scatter and outliers in CMR. Furthermore, we can relate the position of a galaxy on the CMR with features in its image, such as the presence of dust, young stars or galaxy truncations. We will also study the detailed color of each dE through its color profile and especially concentrate more on very inner part. Since the cores of the galaxies still contain information about the violent processes that have happened to their central regions, studying them will shed light on our understanding of the formation and evolution of dEs. Here, we aim to determine the radial gradient in the old population (seen in the outer part) as well as the central young population. 

\par Ground based telescopes in general cannot go further inwards than $1^{\prime\prime}$ which is about 100 pc at the distance of Fornax. Distinguishing the core of a dwarf early-type in nearby clusters (as small as a few 10 pc), is rarely possible with the sharpest ground based telescopes \citep{Grant2005}. The ACS camera on HST makes it possible to resolve a small nucleus with $R_{e}\sim0^{\prime\prime}.025$ \citep{Cote2006}. Using the advantage of HST resolution and sufficient separation in wavelength of the F475W and F850LP HST bands to study young stellar populations, now we can investigate in more detail the color of the very inner part of dwarf early-types in nearby clusters. To discuss the effect of the cluster environment, we choose three clusters, with distinct characteristic, that have been studied by HST: Virgo, Fornax and Coma.

\par The outline of this paper is as follows. In Section~\ref{section:Sample Selection and Data}, we present the selected sample of dwarf early-type galaxies from the Fornax, Virgo and Coma clusters. The analysis of photometric images of our sample is covered in Section~\ref{section:Analysis}. The results of the analysis contain various photometric parameters, CMRs, color profile of the galaxies in the samples and a comparison of the parameters in these three clusters is given in Section~\ref{section:Results}. Section~\ref{section:Discussion}, discusses and summarizes the results. More tables and outcome plots can be found in the Appendix.

\section{Sample Selection and Data}
\label{section:Sample Selection and Data}
\par We selected three clusters for which high quality HST Advanced Camera for Surveys (ACS) data is available. For Fornax we obtained archival data from HST Mikulski archive. The list of galaxies are chosen from the ACS Fornax Cluster Survey \citep[ACSFCS;][]{JordanF2007}. The ACSFCS contains a magnitude-limited sample of 43 galaxies in the Fornax cluster selected from the Fornax Cluster Catalog \citep[FCC;][]{Ferguson} and observed with the ACS on HST. The field of view is about $202^{\prime\prime} \times 202^{\prime\prime}$ and the pixel scale is $0.049^{\prime\prime}$ 
\par Here we included all the early-type dwarf galaxies in the sample of \cite{JordanF2007}. The magnitude limits are given by commonly accepted upper limit for dwarf galaxies \citep[$M_B=-18$;][]{Binggeli1985} and lower magnitude limit of the Fornax ACSFCS sample. Using the distance modulus of m-M=31.50 of the Fornax Cluster (Table~\ref{Tab:Clusters parameters}), these limits correspond to $13.5\leq m_B \leq15.5$. After calculating $g^{\prime}$-band magnitudes on the ACS images, and using the same distance modulus, we found that the galaxies have magnitudes between $M_{g^{\prime}}$ = -16.0 and -18.7.  This sample selection leads to 26 dwarf early-type galaxies for which we used their images in the F475W and F850LP bands. These two bands are similar to the SDSS $g^\prime$ and $z^\prime$ bands \citep{Sirianni2005}. We therefore used the magnitude limit $-18.7 \leqslant M_{g^{\prime}} \leqslant -16.0$ to select corresponding objects from Virgo and Coma.
\par For the Virgo cluster, we used HST Mikulski archival data. We selected our sample from the ACS Virgo Cluster Survey \citep[ACSVCS;][]{Cote2004}, which observed a hundred early-type galaxies in the Virgo cluster. They have chosen their sample from Virgo Cluster Catalog \citep[VCC;][]{Binggeli1987} and considered the early-type galaxy classification in \cite{Sandage1984} and some other conditions. We limited the selection to the same magnitude range as above for Fornax which provides us 55 early-type galaxies. 
\par For the Coma cluster, we used the Coma ACS Survey \citep{Carter2008}, for which the photometry is described in \citet{Hammer}. Since the sample of \citet{Hammer} consist of all types of galaxies in the fields of the Coma survey, even the non-members, we extracted the early-type members by cross-matching their sample with that of \citet{denBrok2011} whose sample of early-type galaxies are spectroscopically confirmed members as well as objects selected by eye as possible members. The same magnitude range similar to Fornax and Virgo limits this sample to 40 early-type galaxies. The two filters used for the Coma HST observations were F475W and F814W. The transformation between the colors, (F475W-F814W) to (F475W-F850LP) is discussed in Section~\ref{Subsection:The color-Magnitude Relation}. 
\par
\begin{figure}
\includegraphics[width=0.5\textwidth,height=5.9cm] {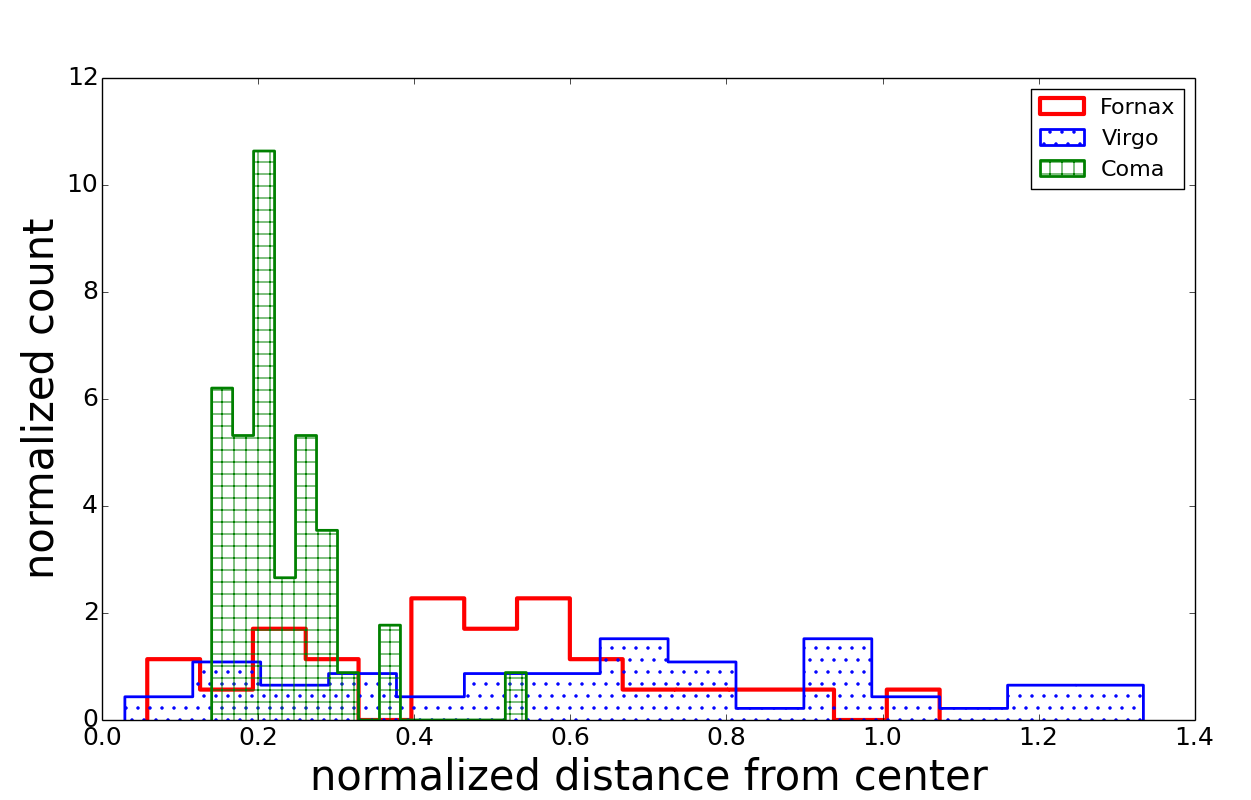}
\caption{Histogram of the projected distance from the center normalized to the Virial radius of the cluster. Different colors are used for each cluster.}
\label{fig:Distance from the center}
\end{figure}

 Our final sample consists of bright early-type dwarf galaxies in the magnitude range of $-18.7 \leq M_{g^{\prime}} \leq -16.0$ ($M_{\star}\sim 10^{9}-10^{10}~M_{\odot}$) observed in $g^{\prime}$ and $z^{\prime}$ by the ACS on HST. The total number of the selected galaxies are 26, 54 and 40 in the Fornax, Virgo and Coma clusters respectively. One should note that in the Fornax and Virgo clusters, HST was pointed at individually selected galaxies, while in Coma, only a limited part of the cluster was observed, mostly in the central regions. This difference, however, does not make the samples incompatible, since both samples are still complete in this magnitude range. Although one should consider the Coma sample not as a sample representing the entire cluster, but mostly the cluster center. This should not be a problem, since Coma is used as a comparison cluster containing mostly old galaxies, and the way it is included now the fraction of them is probably even larger. In Figure~\ref{fig:Distance from the center} we present the radial distances of each object normalized by the Virial radius of its cluster (Table~\ref{Tab:Clusters parameters}). This Figure represents how the selected dEs are distributed in each cluster. The spatial resolution of the observations (corresponding to 1 pixel) is ~5, 4 and 24 parsec in Fornax, Virgo and Coma respectively.

\begin{table}
\small
\caption{Parameters of the three clusters}
\begin{tabular}{lcccc}
\hline
Property & Fornax & Virgo & Coma & Reference \\
\hline
$R_{v}$(Mpc) & 1.4 & 1.55 & 1.99(/h)& 1,2,3 \\
Mass($M_{\odot}$) & $7\times10^{13}$ & $4.2\times10^{14}$ & $9.2\times10^{14}$ & 1,4,5\\
$\sigma_{v}(Km s^{-1})$ & 374 & 760  & 1200 & 1,6,7\\
Distance (Mpc) & 20 & 16.5 & 100 & 8,9\\
\hline
\end{tabular}
\tablebib{(1) \protect\cite{Drinkwater2001}, (2) \protect\cite{Ferrarese2012}, (3) \protect\cite{Kubo2007},(4) \protect\cite{McLaughlin1999}, (5) \protect\cite{Falco2014},(6) \protect\cite{Binggeli1987}, (7) \protect\cite{Colless1996}, (8) \protect\cite{Blakeslee2009},(10) \protect\cite{Carter2008}.} 

\label{Tab:Clusters parameters}
\end{table}

\section{Analysis}
\label{section:Analysis}
\subsection{Colors and magnitudes of bright early-type dwarfs} 
\label{subsection:color and magnitude of bright early-type dwarfs}
\par To calculate the color and magnitude of each galaxy in the Fornax cluster, we used Galphot, which is a surface photometry tool \citep{Franx1989,Jorgensen1992}. We modeled the galaxy in the F475W and F850LP filters with an initial run of Galphot on the inner isophotes with varying parameters such as center, ellipticity and position angle as a function of radius. Then we did the second run for each band by fixing these variables to the output values of the first run in F850LP band. The final output is the surface brightness profile of each galaxy in each band. Before running Galphot, the background of each image was calculated and subtracted using the mean of ten $4\times4$ (pixels square) boxes chosen randomly in the area around the masked galaxy, where we assumed that the contribution of the galaxy was minimal. We then slightly adjusted the background value, requiring the radial intensity growth curve to converge to a finite value. In this way, model-independent effective radii and total magnitudes were determined.

\par The $g^{\prime}-z^{\prime}$ (equal to F475W - F850LP) color is calculated in the AB photometric system. The zero points in $g^{\prime}$ and $z^{\prime}$ bands are from \citet{Sirianni2005}. The color is determined in the range of $1^{\prime\prime} < r < R_{e}$, to avoid background errors to become too dominant, and in most cases, to avoid the nuclear star clusters \cite[for a paper on nuclear clusters in Fornax from these data, see e.g.][]{Turner2012}. Table~\ref{tab:Fornax properties} lists our determined parameter values of the bright early-type dwarfs in Fornax: FCC numbers, $g^{\prime}-z^{\prime}$ colors and their uncertainty, total absolute magnitudes and effective radii in $g^{\prime}$ and $z^{\prime}$ bands. The average error in the color is 0.01 mag which combines background uncertainties in each band and zero point uncertainties.

\begin{table} 
\footnotesize
\caption{Determined parameters of the dEs in Fornax}
\begin{tabular}{p{0.15\columnwidth}p{0.11\columnwidth}p{0.11\columnwidth}p{0.2\columnwidth}p{0.11\columnwidth}p{0.11\columnwidth}}
FCC & $R^{\prime\prime}_{e}(g)$ &$R^{\prime\prime}_{e}(z)$ & $g^{\prime}-z^{\prime}$ & $M_{g^{\prime}}$ & $M_{z^{\prime}}$ \\
number &&&$(1^{\prime\prime}<r<R_{e})$&&\\
(1) & (2) & (3) & (4) & (5) & (6) \\
\hline
		FCC19 & 11.3 & 11.6 & 1.08 $\pm$ 0.01 & -16.46 & -17.57 \\
		FCC26 & 7.7 & 9.1 & 0.72 $\pm$ 0.02 & -16.60 & -17.47 \\
		FCC55 & 12.4 & 12.6 & 1.27 $\pm$ 0.01 & -18.00 & -19.29 \\
		FCC90 & 6.9 & 8.1 & 0.86 $\pm$ 0.01 & -16.94 & -17.89 \\
		FCC95 & 12.2 & 11.9 & 1.27 $\pm$ 0.02 & -16.99 & -18.25 \\
		FCC100 & 16.3 & 15.4 & 1.14 $\pm$ 0.02 & -16.24 & -17.33 \\
		FCC106 & 8.1 & 7.7 & 1.21 $\pm$ 0.02 & -16.72 & -17.89 \\
		FCC119 & 12.6 & 12.4 & 1.17 $\pm$ 0.02 & -16.64 & -17.80 \\
		FCC136 & 16.6 & 16.7 & 1.26 $\pm$ 0.02 & -17.02 & -18.29 \\
		FCC143 & 9.5 & 7.6 & 1.35 $\pm$ 0.01 & -17.84 & -19.09 \\
		FCC148 & 12.7 & 12.7 & 1.24 $\pm$ 0.01 & -18.69 & -19.92 \\
		FCC152 & 12.6 & 12.3 & 1.15 $\pm$ 0.01 & -17.85 & -18.99 \\
		FCC182 & 9.8 & 8.6 & 1.35 $\pm$ 0.01 & -17.02 & -18.29 \\
		FCC190 & 16.3 & 15.0 & 1.38 $\pm$ 0.02 & -18.29 & -19.61 \\
		FCC202 & 9.5 & 9.4 & 1.24 $\pm$ 0.02 & -16.49 & -17.74 \\
		FCC203 & 12.1 & 11.8 & 1.18 $\pm$ 0.03 & -16.20 & -17.36 \\
		FCC204 & 10.2 & 10.0 & 1.25 $\pm$ 0.03 & -16.73 & -17.97 \\
		FCC249 & 8.4 & 7.5 & 1.36 $\pm$ 0.01 & -18.44 & -19.75 \\
		FCC255 & 11.7 & 11.5 & 1.22 $\pm$ 0.01 & -17.97 & -19.19 \\
		FCC277 & 9.9 & 8.6 & 1.33 $\pm$ 0.01 & -18.18 & -19.40 \\
		FCC288 & 8.3 & 8.0 & 1.13 $\pm$ 0.02 & -16.53 & -17.58 \\
		FCC301 & 9.3 & 8.9 & 1.26 $\pm$ 0.01 & -17.80 & -19.04 \\
		FCC303 & 12.9 & 13.0 & 1.12 $\pm$ 0.02 & -16.23 & -17.36 \\
		FCC310 & 18.0 & 17.8 & 1.32 $\pm$ 0.01 & -18.28 & -19.60 \\
		FCC324 & 19.7 & 17.8 & 1.15 $\pm$ 0.03 & -16.50 & -17.60 \\
		FCC335 & 11.6 & 11.1 & 1.18 $\pm$ 0.01 & -17.34 & -18.49\\
		\hline
\end{tabular}
\tablefoot{Parameters of the dEs in Fornax determined here:
(1) Name of the galaxy from the Fornax Cluster Survey,
(2) and (3) Effective radius in $g^{\prime}$ and $z^{\prime}$ bands with an error of  about 6\%,
(4) $g^{\prime}-z^{\prime}$ color in the range of $1^{\prime\prime} < r < R_{e}$ and its error,
(5) and (6) Absolute magnitude in $g^{\prime}$ and $z^{\prime}$ bands. The magnitudes and colors have been corrected for galactic reddening.}
\label{tab:Fornax properties}
\end{table}

\par
\begin{figure}
\includegraphics[width=0.5\textwidth,height=5.9cm]{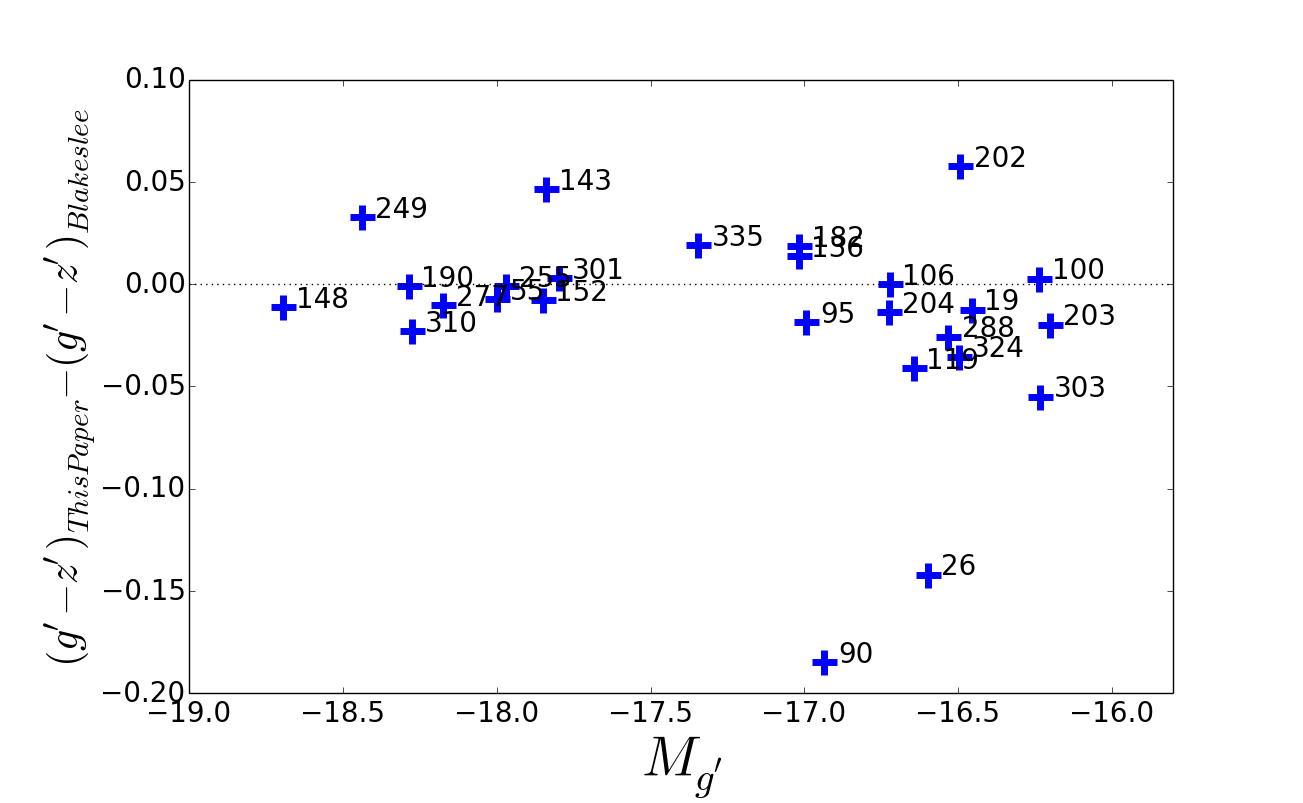}
\caption{Difference between tabulated $(g^{\prime}-z^{\prime})$ colors in Table~\ref{tab:Fornax properties} and calculated colors in \protect\cite{Blakeslee2009} for Fornax versus $g^{\prime}$-band magnitude. The two outliers, FCC90 and FCC26, have very blue extended cores in Figure~\ref{Fig:color profile of Fornax}.}
\label{fig:my color vs Blakeslee}
\end{figure}

 \cite{Blakeslee2009} calculated g-z colors of Fornax early-type galaxies from the ACS HST data as well. They measured colors using annuli up to where the surface brightness of the galaxy is at least $60\%$ of the sky level, whereas we consider the color up to one effective radius. They also excluded the innermost annuli $r<1^{\prime\prime}$. Figure~\ref{fig:my color vs Blakeslee} shows the comparison of our calculated color versus \cite{Blakeslee2009}. FCC26 and FCC90 are two very blue galaxies in our measurements. These galaxies have very extended blue cores which will be discussed later, in Figure~\ref{Fig:color profile of Fornax}. The fact that the apertures used by \cite{Blakeslee2009} are larger can explain this difference.

\par
\begin{figure}
	\includegraphics[width=0.5\textwidth,height=5.9cm]{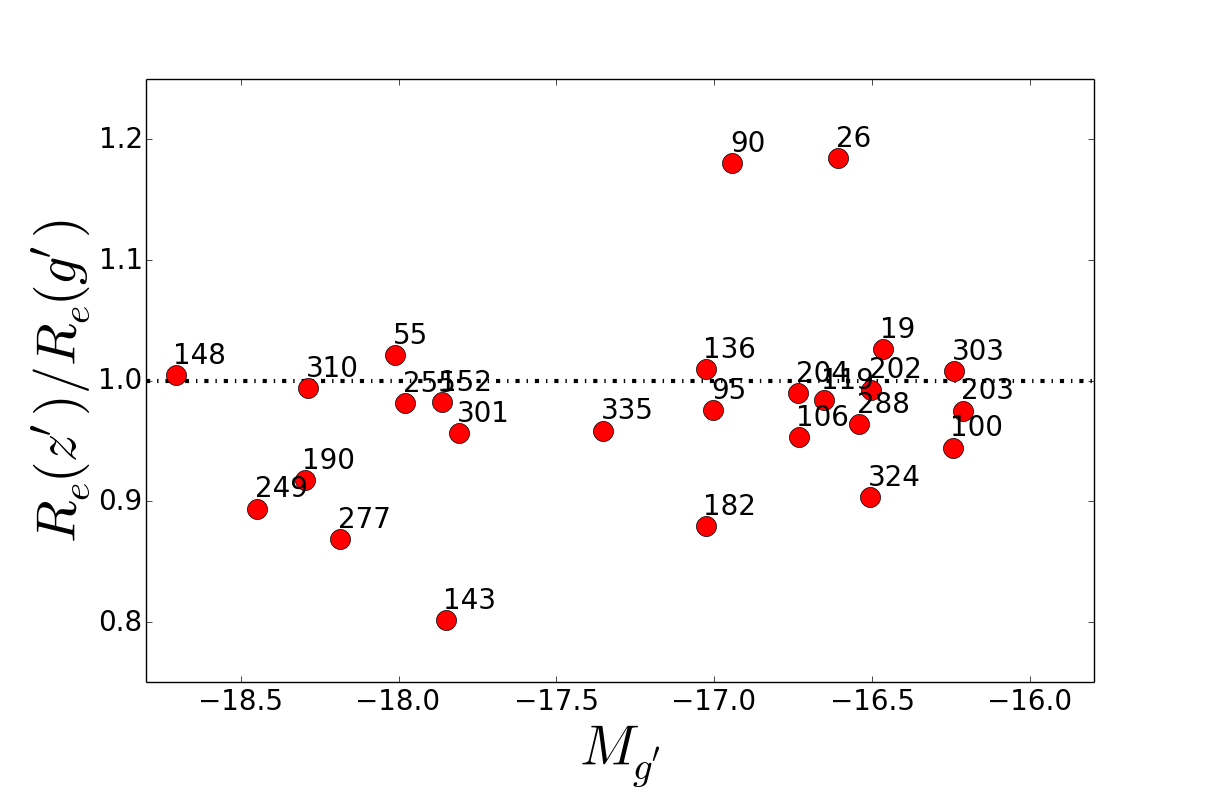}
	\caption{Ratio of effective radii in $g^{\prime}$- and $z^{\prime}$-bands versus absolute magnitude in $g^{\prime}$-band. The dashed line is $\frac{R_{e,z}^{\prime}}{R_{e,g}^{\prime}}=1$. The numbers on the data points are the FCC numbers.}
	\label{fig:Re vs mag}
\end{figure}

\begin{figure}
	\includegraphics[width=0.5\textwidth,height=5.9cm]{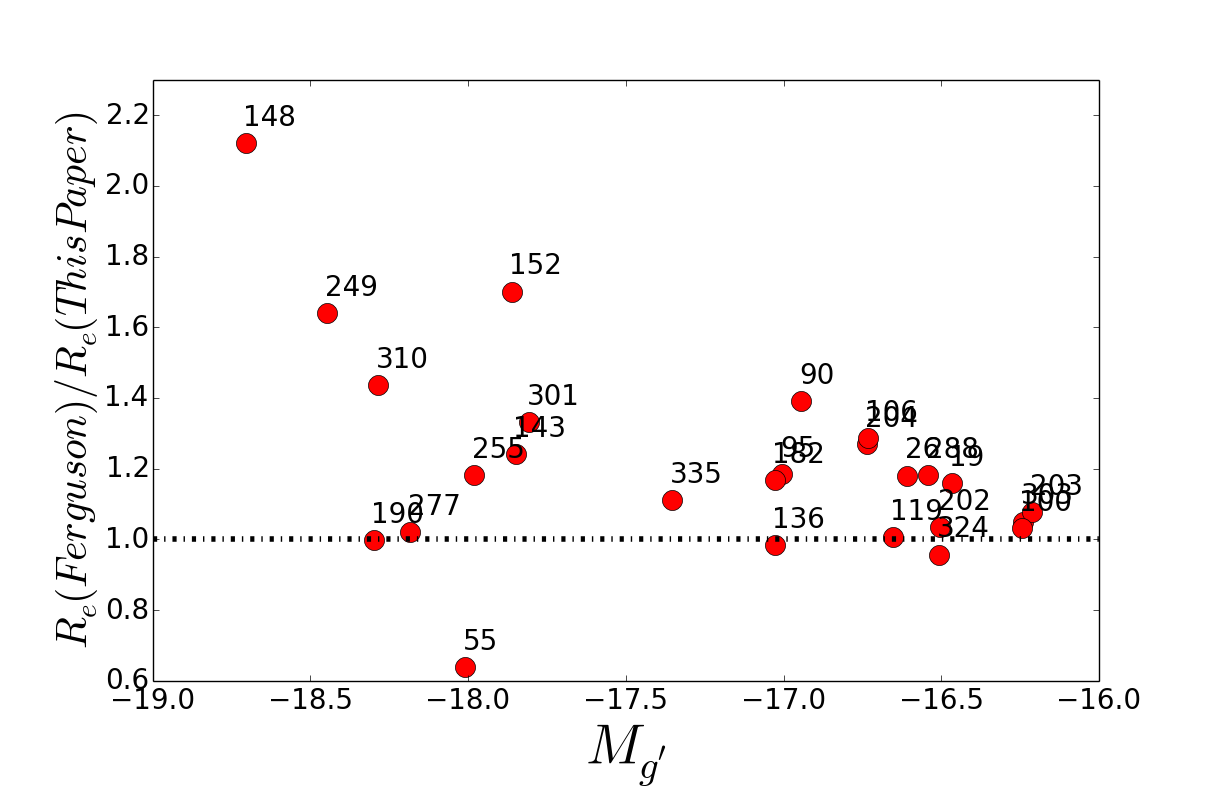}
	\caption{Ratio of effective radii of \citet{Ferguson} to the effective radii in $g^\prime$-band in this paper versus total magnitude of the galaxy in $g^\prime$-band. The dashed line is $R_{e}(Ferguson)=R_{e}(new)$.}
	\label{fig:Re}
\end{figure}

 To check the accuracy of the effective radii, we show the ratio of the radii in $g^{\prime}$- and $z^{\prime}$-bands as a function of galaxy magnitude in Figure~\ref{fig:Re vs mag}. Since FCC26 and FCC90 have a very blue centers, which will be discussed in Section~\ref{subsection:Radial color profiles}, $\frac{R_{e,z}^{\prime}}{R_{e,g}^{\prime}}$ is larger than 1. Most of the galaxies, however, show a ratio in the range of 0.8 to 1, implying a larger effective radius in the $g^{\prime}$-band. These galaxies generally become bluer going outwards, i.e, a negative color gradient which will be discussed in detail in Section~\ref{subsection:Comparison of the Clusters}. This range of ratios is not unexpected \citep[see e.g.][Table 3 for ratios between effective radii in B and R in more massive galaxies]{Peletier1994}. We also compared our determined effective radii in the $g^\prime$ band with the ones from \citet{Ferguson} in the B-band in Figure~\ref{fig:Re}. 

\par For the Virgo cluster we adopt the colors from \citet{Ferrarese2006}. They determined them in the same radial range as we described for Fornax above ($1^{\prime\prime} < r < R_{e}$). The photometric errors in Virgo given by \citet{Ferrarese2006} are much larger than what we derived here, since they gave errors on the total colors. However, inside R$_{e}$ the surface brightness of the galaxies is generally so high, that the errors in the magnitudes inside $R_e$ are almost exclusively caused by errors in $R_e$. When one determines the errors in the color inside $R_e$, they almost completely vanish, since the dependence on $R_e$ vanishes. By concentrating on colors we avoided the large (often 0.1 mag or larger) magnitude errors quoted in \citet{Ferrarese2006}. Instead, for errors of the colors in Virgo, we used the average error of the early-type dwarfs in Fornax. 

\par For the Coma cluster, the colors are tabulated by \citet{Hammer} from the HST/ACS data. As the closest magnitude system in that paper to ours are the Kron magnitudes, we used colors calculated in Kron aperture (1 Kron radius corresponds to 1.19 effective radius for an exponential surface brightness profile). For all the three clusters, the photometry was corrected for Galactic extinction using the dust maps of \citet{Schlegel1998} with the extinction law adopted by \citet{Jordan2004}(ACSVCS-II) from \citet{Sirianni2005}:

\begin{equation}
A_{g} = 3.634 E(B-V),
\label{eq:extinction rule1}
\end{equation}

\begin{equation}
A_{z} = 1.485 E(B-V).
\label{eq:extinction rule2}
\end{equation}

For Fornax and Coma we used the same foreground extinction for the galaxies in each cluster, $<E(B -V )> = 0.013$ and $<E(B -V )> =0.009$, as it is reported in \cite{JordanF2007} and \citet{Hammer}. For Virgo we used different correction values for each galaxy tabulated in \citet{Ferrarese2006} with a mean E(B-V) value of 0.028.

\begin{figure*}
\includegraphics[width=\textwidth,height=11.5cm]{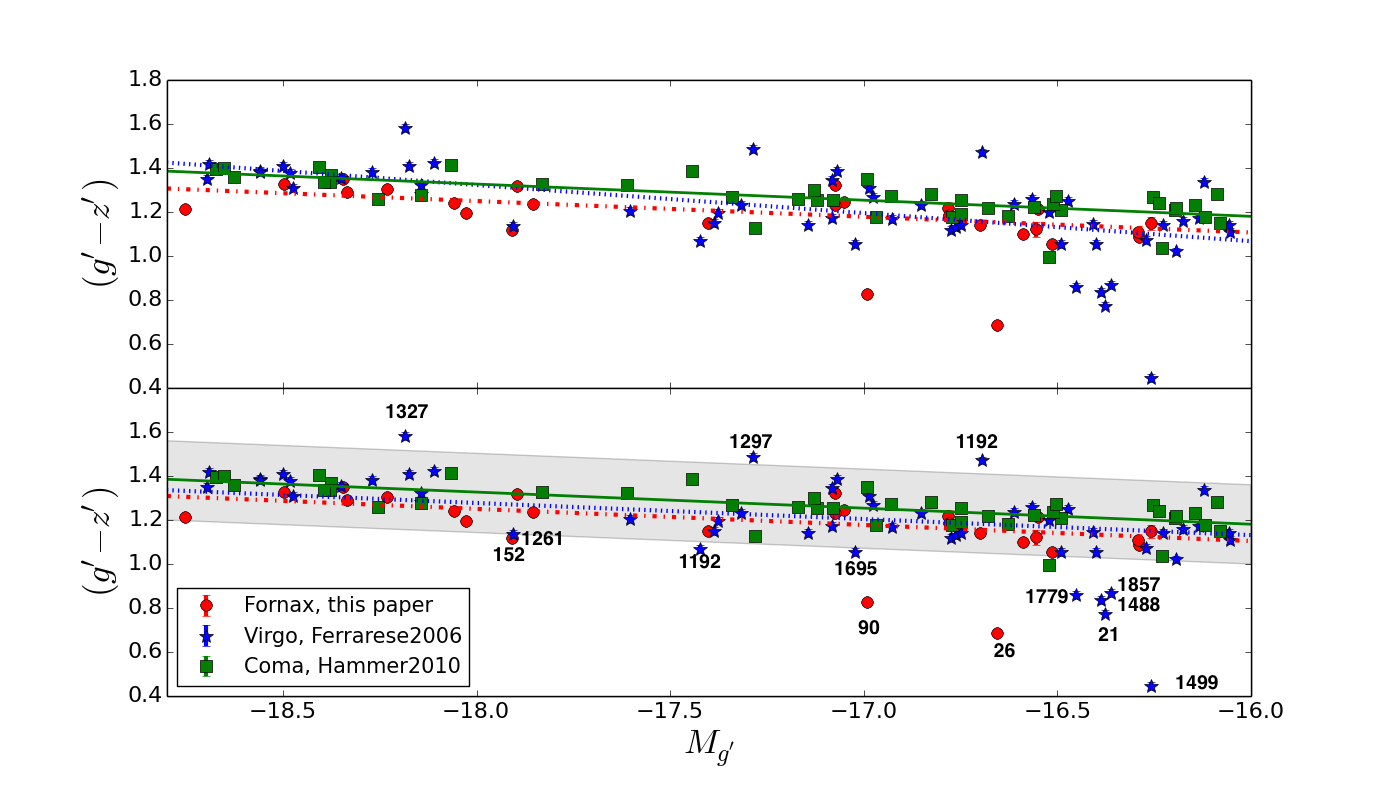}
\caption{Color-magnitude diagram of bright early-type dwarfs from Fornax (this study), from Virgo \citep{Ferrarese2006} and Coma \citep{Hammer}. The color codes of least squares fitted lines are the same as for the points; red for Fornax, blue for Virgo and green for the Coma cluster. The fitted lines in the upper panel are least squares fits for each cluster. In the lower panel we fixed the slope of the fitted lines to the slope of the CMR in Coma as determined in upper panel. All the parameters of the lines are given in Table~\ref{Tab:CMR}. VCC and FCC numbers of the red and blue outliers in the Virgo and Fornax clusters are indicated. The gray area shows the area of $\pm2\sigma$ away from the CMR of Coma. The colors and magnitudes have been corrected for galactic extinction}.
\label{fig:CMD}
\end{figure*}

\subsection{Color profile, unsharp masking and color maps}
\label{subsection:Color profile, unsharp masking, color maps}
\par To accurately investigate the substructure in the stellar populations of each galaxy, radial color profiles were determined. This process was done for the Fornax and Virgo galaxies, while for the Coma cluster we just adopted the output of Galphot from \cite{denBrok2011} who did the same procedure as above to calculate the color profiles. All the color profiles are sky subtracted. It is important to mention that the color obtained in the centers of the galaxies, which this paper is interested in, is very sensitive to the point spread function (PSF) of the data. We convolved the image of the galaxy in each band with the PSF of the other band before running Galphot on them, in the same way as was done in e.g. \citet{Peletier2012}. The PSFs used here are generated by the Tiny Tim HST PSF software \citep{Krist2011}. 

\par The unsharp masking technique is a well known method to study the substructure of a galaxy. An unsharp mask is the original image divided by the smoothed one. This method amplifies high-frequency components of the image which provides a more reliable and clear detection of gas, dust features and asymmetric star formation regions compared to the residual image \citep{Lisker2006}. We produce unsharp masked images of the Fornax galaxies by smoothing them with a two-dimensional circular Gaussian of a kernel size of $\sigma=20$ pixels (1 arcsec). In most cases the unsharp masked images show central irregularities which are caused by gas and dust features or young stellar populations regions.

\par Another method of extracting substructures is by identifying them on the color map of the galaxy which is made using the images of the galaxy in two different filters. In Figure~\ref{Fig:color profile of Fornax}, we present the unsharp masked and color maps of our Fornax sample.

\section{Results}
\label{section:Results}
\subsection{The color-magnitude relation}
\label{Subsection:The color-Magnitude Relation}

\par To compare the CMRs of these three clusters, we converted the Coma F814W data to the F850LP system using the following transformation, which is derived from Formula 1 in the Appendix of \cite{Roediger2017} :
\begin{equation}
(F475W - F814W) = 0.924(F475W - F850LP) - 0.027.
\label{eq:transfor color}
\end{equation}
The resulting (F475W - F850LP) color corresponds to the SDSS $g^{\prime}-z^{\prime}$ color, with possible minor differences due to small differences in filter transmission curves.

\par Figure~\ref{fig:CMD} shows the color-magnitude diagram for the Virgo, Fornax and Coma clusters. In the upper panel, we show the data points as well as three different least squares fits to dEs of each cluster, taking into account the uncertainties in the photometry. In the lower panel we fit the CMR of the two other clusters with the slope fixed to the slope of Coma's CMR. The parameters of these fits are given in Table~\ref{Tab:CMR fixed slope}. We applied an iterative sigma rejection algorithm ($\sigma=3$) to all three clusters before fitting the lines in both upper and lower panels, removing the bluest outliers.
\begin{table}
\caption{Parameters of the CMRs in Figure~\ref{fig:CMD} }
\label{Tab:CMR}
\label{Tab:CMR fixed slope}
\begin{tabular}{lccc}
		\hline
		\multicolumn{4}{l}{Upper panel: slope and intercept fitted}\\
		\hline
		Cluster & Slope(a) & y-intercept(b)& $\sigma(mag)$\\
		\hline
		Fornax & $-0.071\pm0.017$ & $1.250\pm0.016$ & 0.060 \\
		Virgo & $-0.128\pm0.022$ & $1.323\pm0.025$ & 0.129\\
		Coma & $-0.073\pm0.012$ & $1.327\pm0.011$ & 0.065 \\
		\hline
		\multicolumn{4}{l}{Lower panel: only intercept fitted and fixed slope}\\
		\hline
		Fornax & -0.073 & $1.250\pm0.013$ & 0.060 \\
		Virgo & -0.073 & $1.277\pm0.019$ & 0.144 \\
		Coma & $-0.073\pm0.012$ & $1.350\pm0.011$ & 0.065 \\
		\hline
	\end{tabular}
\tablefoot{Parameters of the fitted CMRs in Fig.~\ref{fig:CMD} (upper and lower panel); $(g^{\prime}-z^{\prime})=a(M_{g^{\prime}}+18)+b$ and $\sigma$ is the standard devition.}
\end{table}

\par The bottom panel of Fig.~\ref{fig:CMD} shows that Fornax and Virgo are bluer than Coma. The second result is that the scatter with respect to the CMR is higher for the Virgo cluster. The standard deviation $\sigma$ of the fitted lines to the Fornax, Virgo and Coma are $\sigma= 0.060, 0.144, 0.065$ mag respectively, after $3\sigma$ clipping. We discuss the interpretation of these results in next section.  
\begin{figure}
	\includegraphics[width=0.5\textwidth,height=5.9cm]{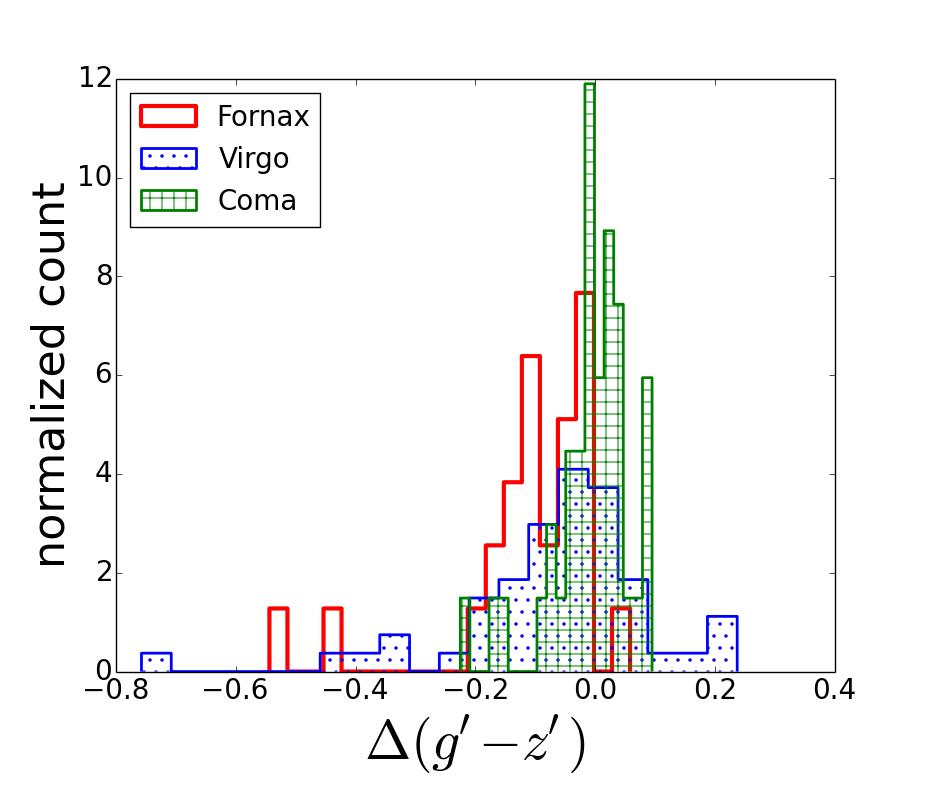}
	\caption{Distribution of the difference between the color of a galaxy in our sample and its expected color from the CMR of the Coma in Figure~\ref{fig:CMD}.}
	\label{fig:color residual histo}
\end{figure}

\subsection{Reasons for the Scatter in CMR}
\label{Subection:Reasons for the Scatter in the CMR}

\par Considering the cluster masses (Table~\ref{Tab:Clusters parameters}) and the morphology-density relation, one might expect that the scatter in Coma should be the smallest, given the fact that the Coma galaxies were mostly selected from the center of the cluster, followed by Virgo and then by Fornax. Since galaxy formation would go faster in the most massive clusters (downsizing), the scatter caused by residual star formation should become smaller, as clusters get more massive. However, in the previous Section it was shown that the scatter in Virgo is much higher than in Coma and Fornax.

\par Figure~\ref{fig:color residual histo} shows a histogram of color residuals with respect to Coma's CMR. For this we assumed that the CMR is a relation connecting the oldest galaxies for all magnitudes. A galaxy may lie blueward of the CMR when it has a younger effective age. In principle, a galaxy cannot lie on the red side of the CMR of early-types. We assumed that the CMR of Coma, as it is the reddest and the one with the least scatter, indicates this relation with the oldest galaxies. In this case the color residual is a measure of the age difference between the oldest galaxy at a certain magnitude and the galaxy itself. To find outliers from the CMR we fitted a normal distribution on the color residuals of all the three clusters together. Before fitting that, we clipped all the galaxies further than $3\sigma$. After clipping, the Gaussian distribution has $\sigma=0.09$. Here we defined red and blue outliers as galaxies that are located further than $2\sigma$ on both sides of the CMR, showed by a gray area in the lower panel of Figure~\ref{fig:CMD}. Fornax has only blue outliers: FCC 26, FCC 90 and FCC 152. Virgo has some outliers on both blue and red sides: VCC21, VCC1192, VCC1261, VCC1488, VCC1499, VCC1695, VCC1779, VCC1857 are on the blue side and VCC1327, VCC1297, VCC1192 are on the red side. Coma has one which is bluer: COMAi13005.684p275535.20. The names of these outliers are shown in the lower panel of Fig.~\ref{fig:CMD}.
\begin{figure*}
\centering
\includegraphics[width=15.5cm,height=8.3cm]{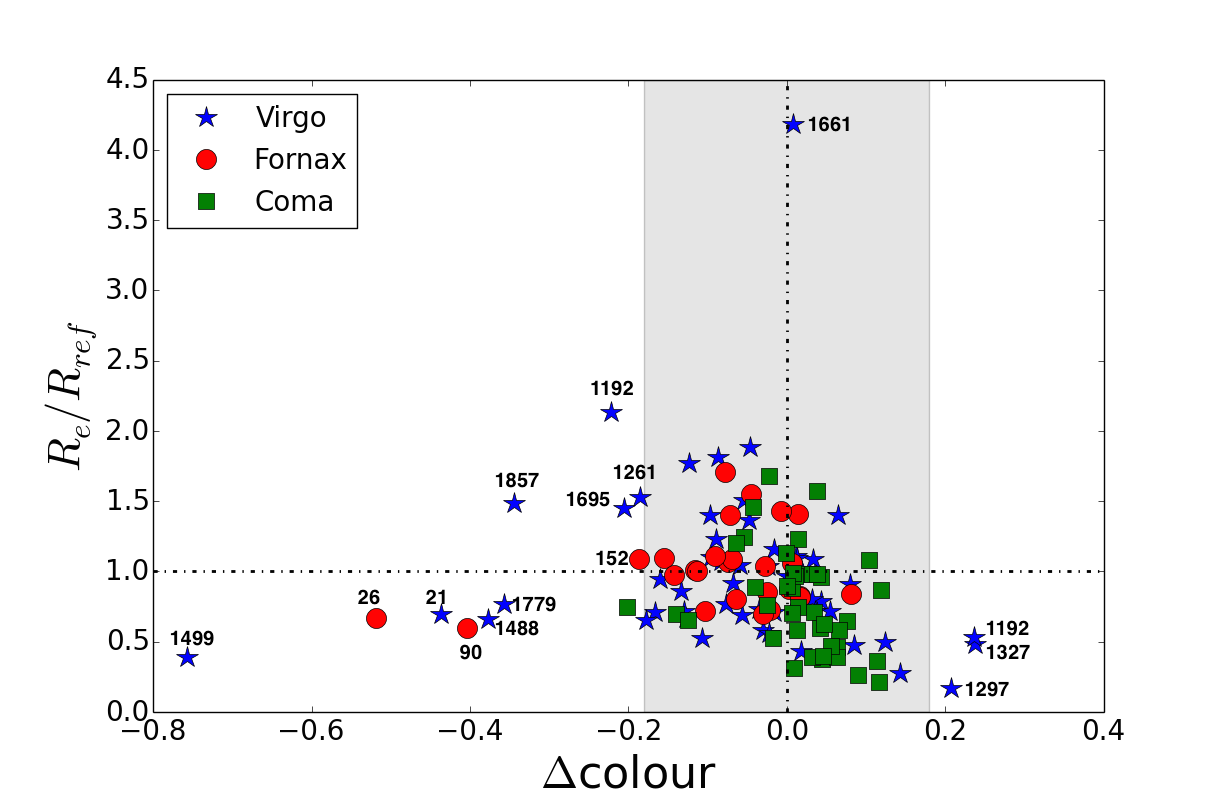}
\caption{$\Delta$color vs. normalized effective radius in $g^\prime$ for the bright early-type dwarfs in Fornax, Virgo and Coma. $\Delta$color is the difference between the color of a galaxy in our sample and its expected color from the CMR of the Coma cluster (color residual). Here the effective radius is divided by the expected radius from $R_{e}$ vs magnitude relations in each cluster, equation~\ref{eq:g mag versus radius in Fornax}, \ref{eq:g mag versus radius in Virgo} and \ref{eq:g mag versus radius inComa}. The vertical dashed line corresponds to the CMR of Coma and the horizontal dashed line is $R_{e}= R_{ref}$. The gray area corresponds to the gray area of Figure~\ref{fig:CMD}.}
\label{fig:Re vs delta}
\end{figure*}

\par Investigating the behavior of the outliers in Fig.~\ref{fig:Re vs delta}, we found that the effective radii and the colors of the outliers are strongly correlated. This figure was made by determining the color and the radius that an average galaxy of a certain magnitude should have, using the color-magnitude and magnitude-effective radius relation of the sample. A least squares fit gave us the following relation between $g^{\prime}$-band magnitude and effective radius for Fornax, Virgo and Coma clusters respectively: 
\begin{equation}
	r_{e}(kpc) = -0.020\times(M_{g^{\prime}}+18)+ 1.16
	\label{eq:g mag versus radius in Fornax}
\end{equation}
\begin{equation}
	r_{e}(kpc) = 0.010\times(M_{g^{\prime}}+18)+ 1.10
	\label{eq:g mag versus radius in Virgo}
\end{equation}
\begin{equation}
	r_{e}(kpc) = -0.682\times(M_{g^{\prime}}+18)+ 4.28
	\label{eq:g mag versus radius inComa}
\end{equation}
The effective radii of the galaxies in Coma are extracted from the HST Coma survey paper III \citep{Hoyos2011}. $\Delta color$ is the difference between the color of a galaxy from the CMR of the Coma cluster. We defined the normalized effective radius as the ratio of the effective radius in $g^{\prime}$ and the expected radius from the radius-magnitude relation. Plotting $\Delta color$ versus normalized effective radius, we found that there are three regions in Fig.~\ref{fig:Re vs delta}: the galaxies with small $\Delta$colors and normal radii, the ones that are redder than the CMR, which tend to have smaller radii than expected, and the ones that are bluer, which mostly tend to be smaller in size. The second group can be identified with the class of compact (red) dwarf ellipticals, while the third are dwarf ellipticals with young stellar populations, which also are more compact, as e.g. BCD galaxies. Note that the second group (of compact dwarf ellipticals) does not contain any galaxies in Fornax and Coma, but several objects in Virgo. Apart from some exceptions, there is a visible trend that as galaxies become redder they also become smaller. VCC1661 is the galaxy with the largest effective radius \citep[For a detailed study of this extended galaxy see][]{Koch2017}. 

\par We first discuss the red outliers in detail. They are generally compacts elliptical galaxies. These are objects like M32, with radii smaller than expected, and redder colors. They tend to be red, without any strong sign of young stellar populations \citep{Guerou2015,Chilingarian2009}. It is thought that these objects lost a significant amount of matter and light in an interaction, causing a truncated surface brightness profile. The resulting color, after the interaction, is redder than expected for its radius, since the original galaxy was larger and therefore had a redder color. Furthermore, the central regions are generally redder, due to the color gradients in the original galaxies surviving the interaction. 

\par To understand the outliers on the red side better, we take a closer look at them. All of these outliers are Virgo early-type dwarfs: VCC 1192, VCC 1297, VCC 1327. There are two additional red galaxies very close to the $2\sigma$ treshhold from Virgo: VCC 1627 and VCC1871. From \citet{Ferrarese2006}, VCC 1192 is a close companions of the massive elliptical, M49, with a projected distance of $4^{\prime}.2$ $(\sim 20.2~kpc)$. Similarly, VCC 1297 and VCC 1327 are at the projected distances of $7^{\prime}.3$ and $7^\prime.5$ $(\sim35.1,~36.1~kpc)$ from M87. All the three galaxies are near giant companions and could be tidally truncated by interactions with their massive neighbors. Interestingly, \cite{Guerou2015} classified VCC 1192 and VCC1297 as compact early-type galaxies. It looks as if most, maybe all, of the red outliers are compact dwarfs. In \cite{Guerou2015} there are more compacts from Virgo that are slightly redder or on the CMR in our plot: VCC1178, VCC1440, VCC147 and VCC1627, although they are not counted as red outliers here as they are in the $2\sigma$ shaded area. To conclude, we found 3 outlying red galaxies in the Virgo Cluster more than $2\sigma$ away from the CMR of Coma.

\par
\begin{figure*}
	\centering
	\includegraphics[width=4.3cm,height=3.8cm]{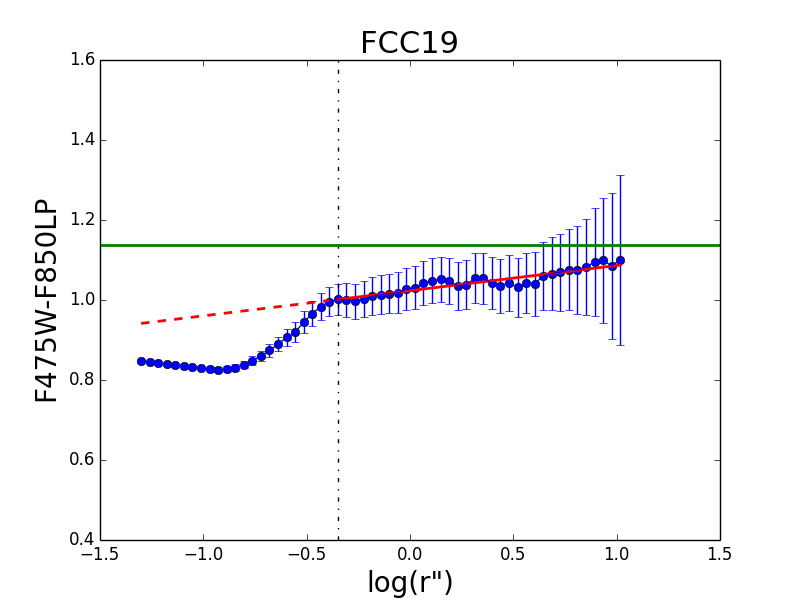}
	\includegraphics[width=4.3cm,height=3.8cm]{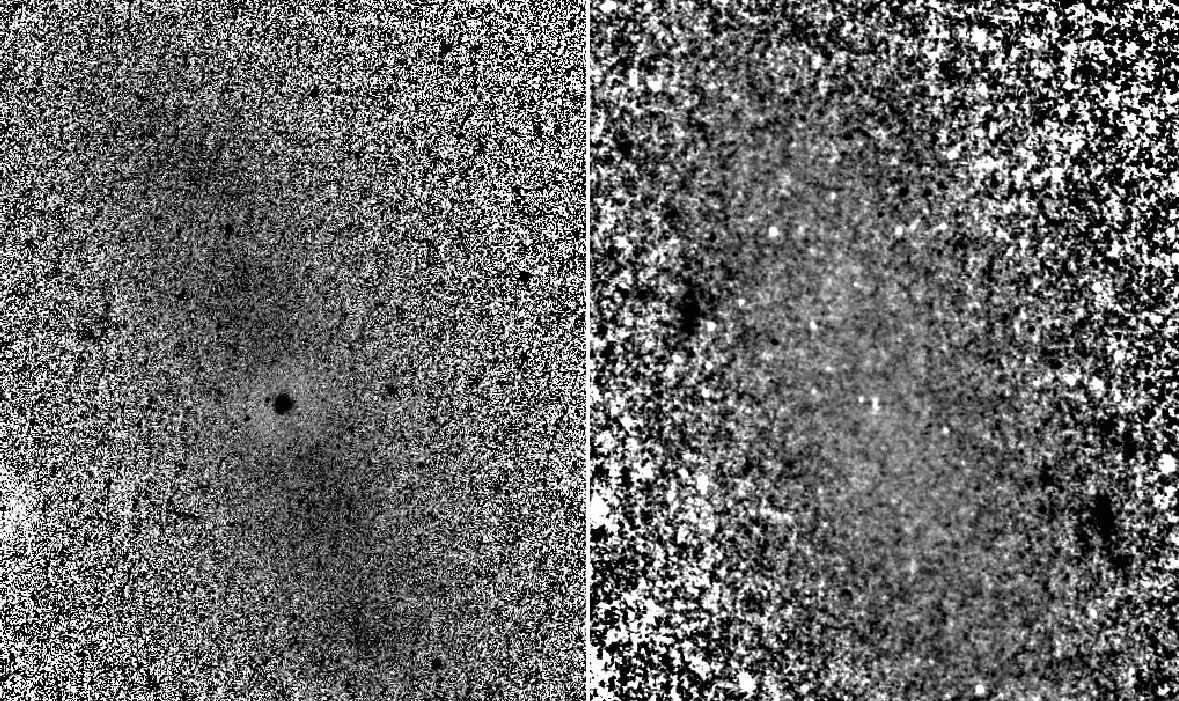}
	\includegraphics[width=4.3cm,height=3.8cm]{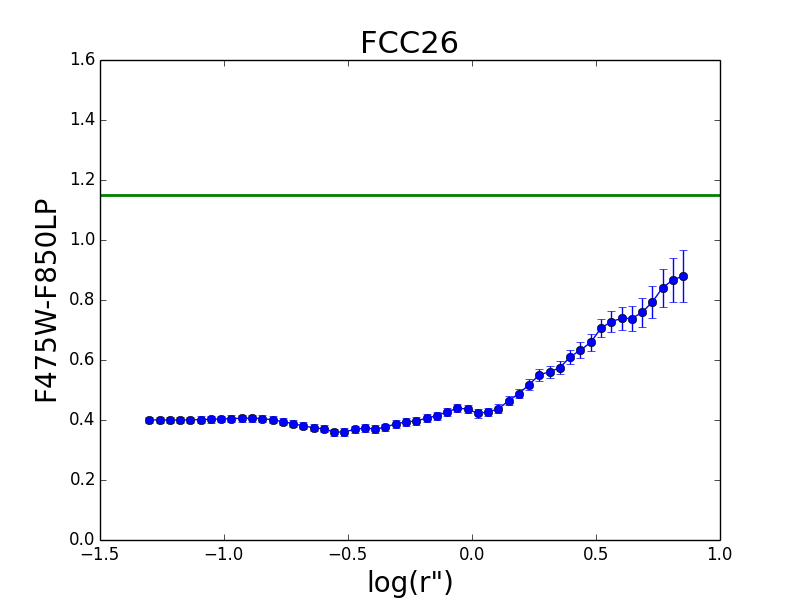}
	\includegraphics[width=4.3cm,height=3.8cm]{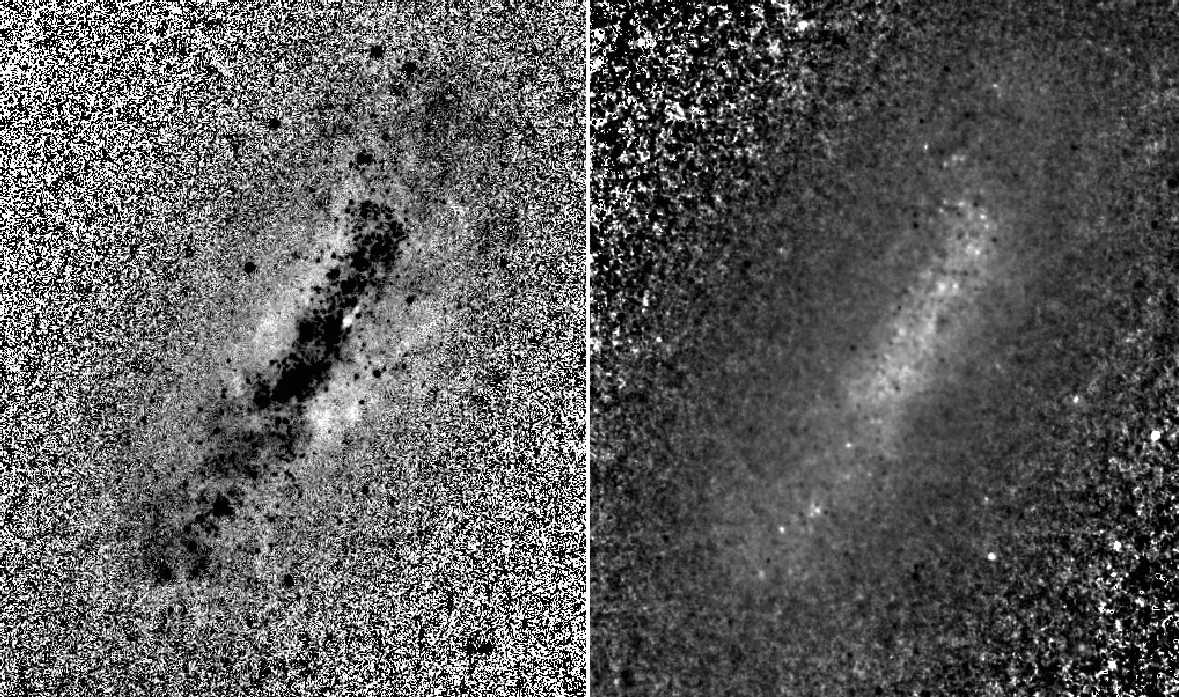}
	\includegraphics[width=4.3cm,height=3.8cm]{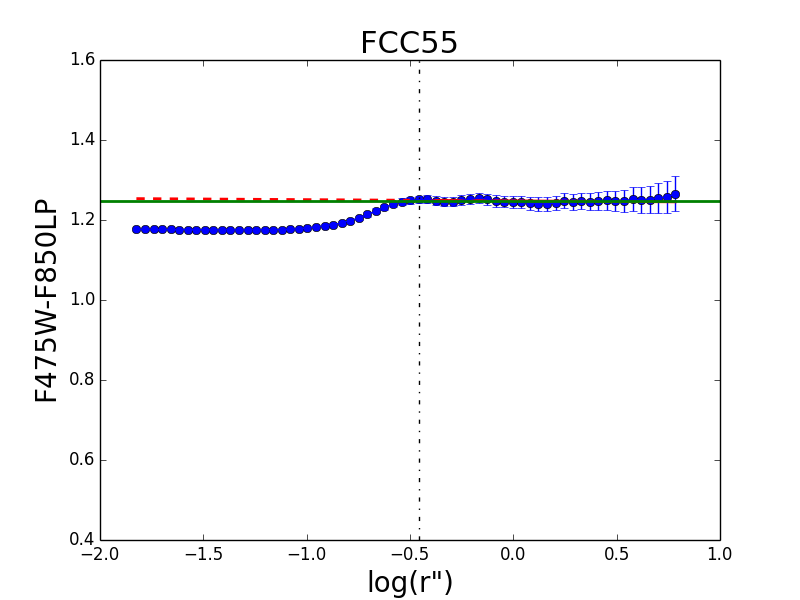}
	\includegraphics[width=4.3cm,height=3.8cm]{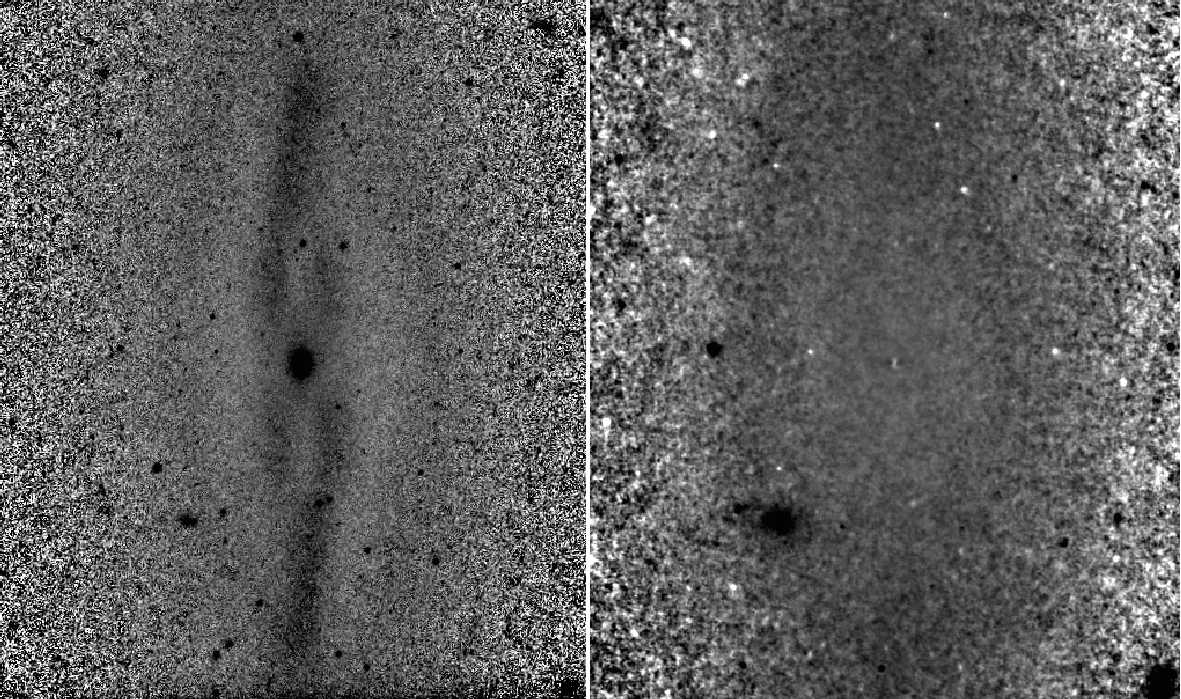}
	\includegraphics[width=4.3cm,height=3.8cm]{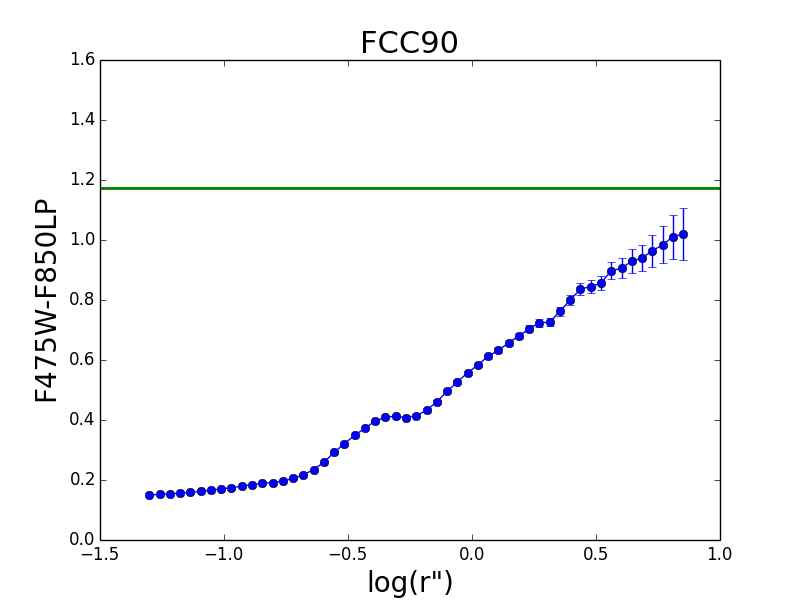}
	\includegraphics[width=4.3cm,height=3.8cm]{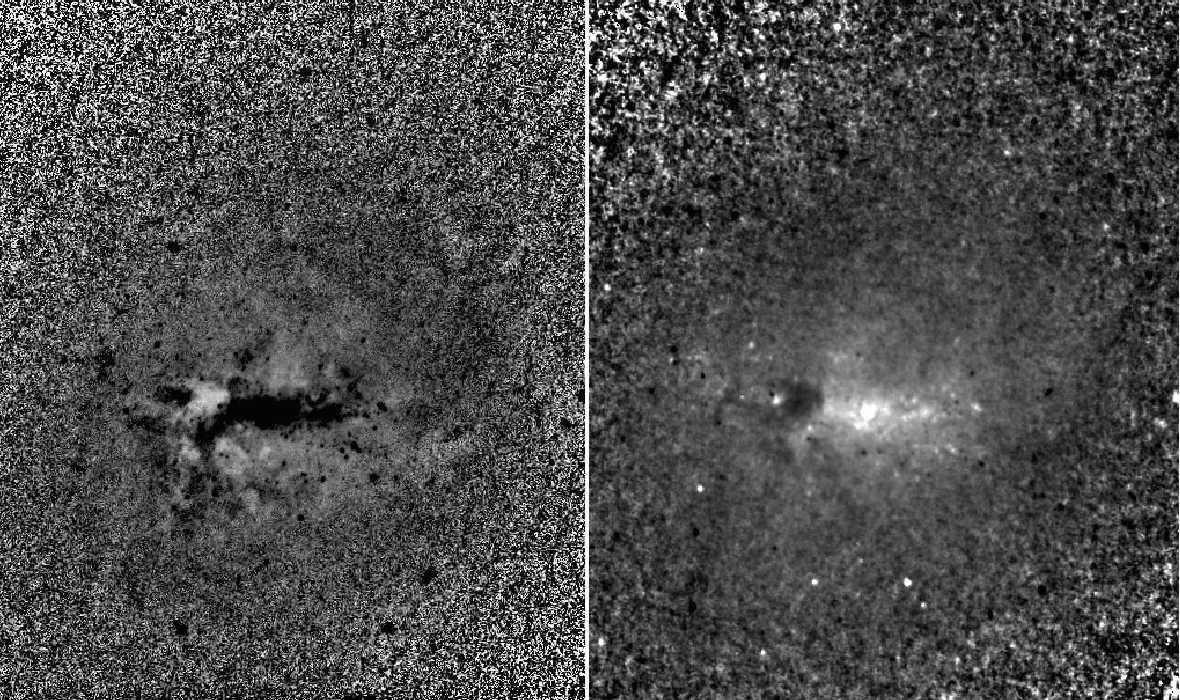}
	\includegraphics[width=4.3cm,height=3.8cm]{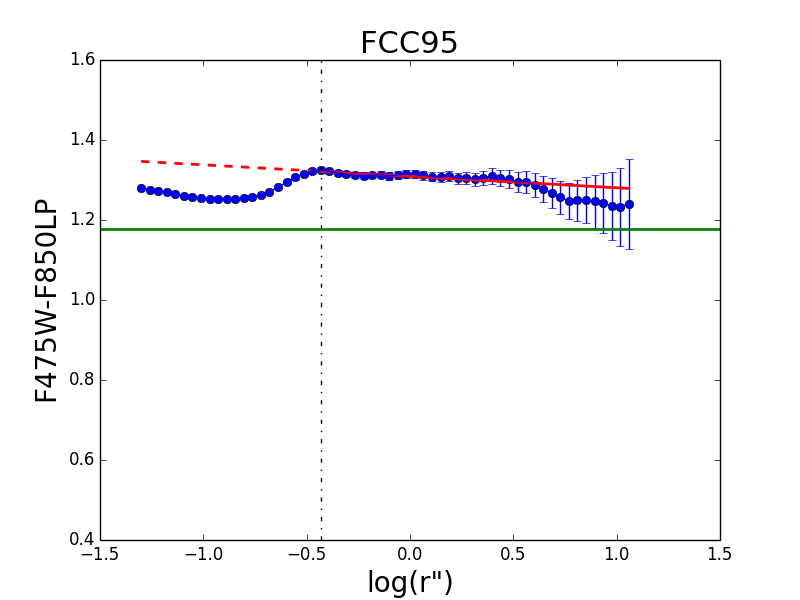}
	\includegraphics[width=4.3cm,height=3.8cm]{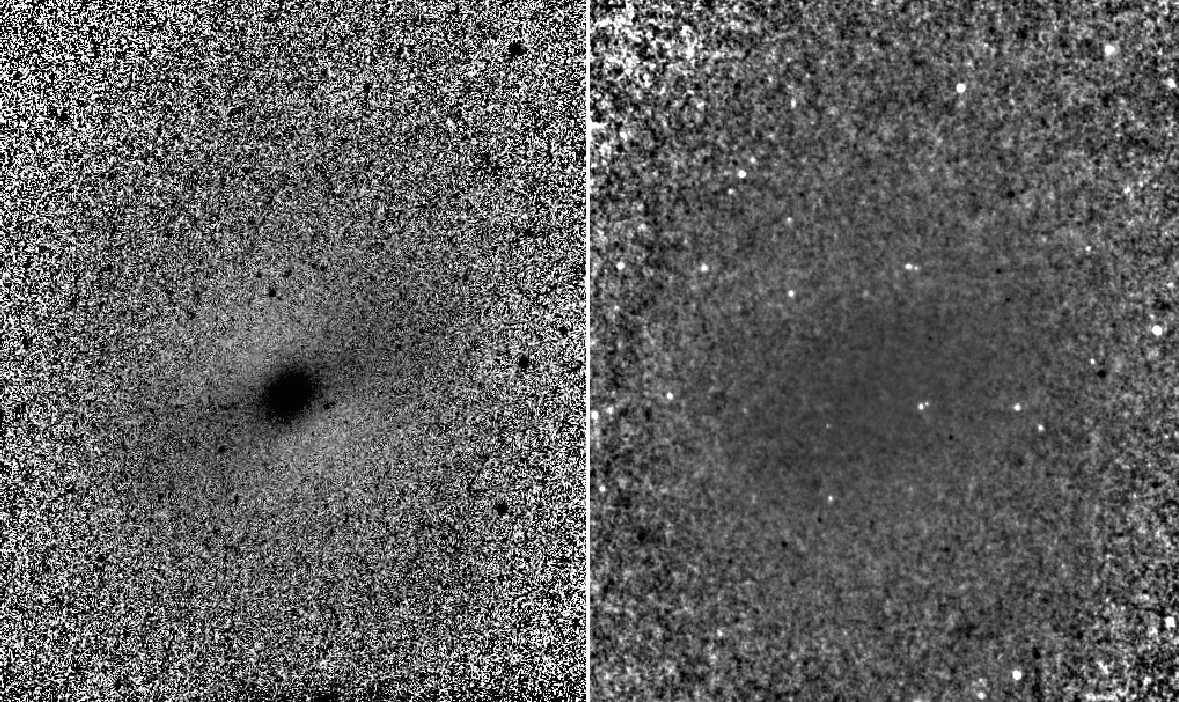}
	\includegraphics[width=4.3cm,height=3.8cm]{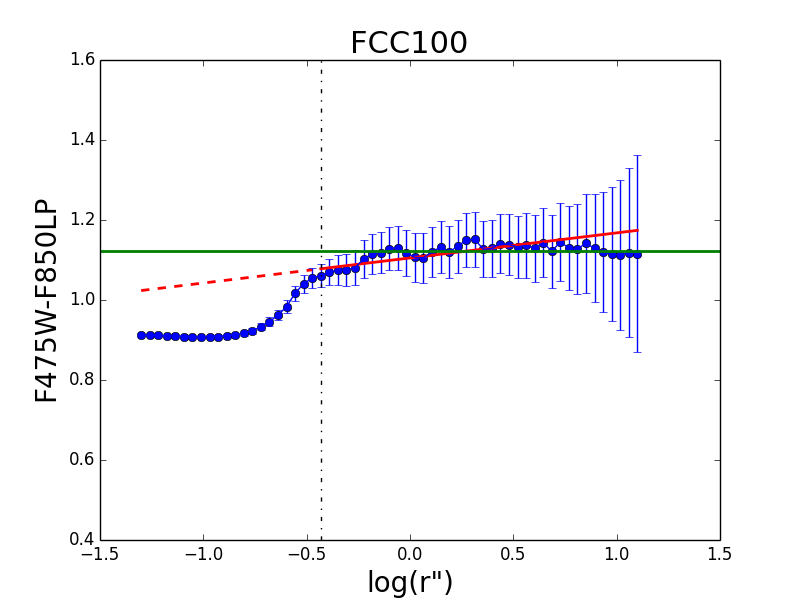}
	\includegraphics[width=4.3cm,height=3.8cm]{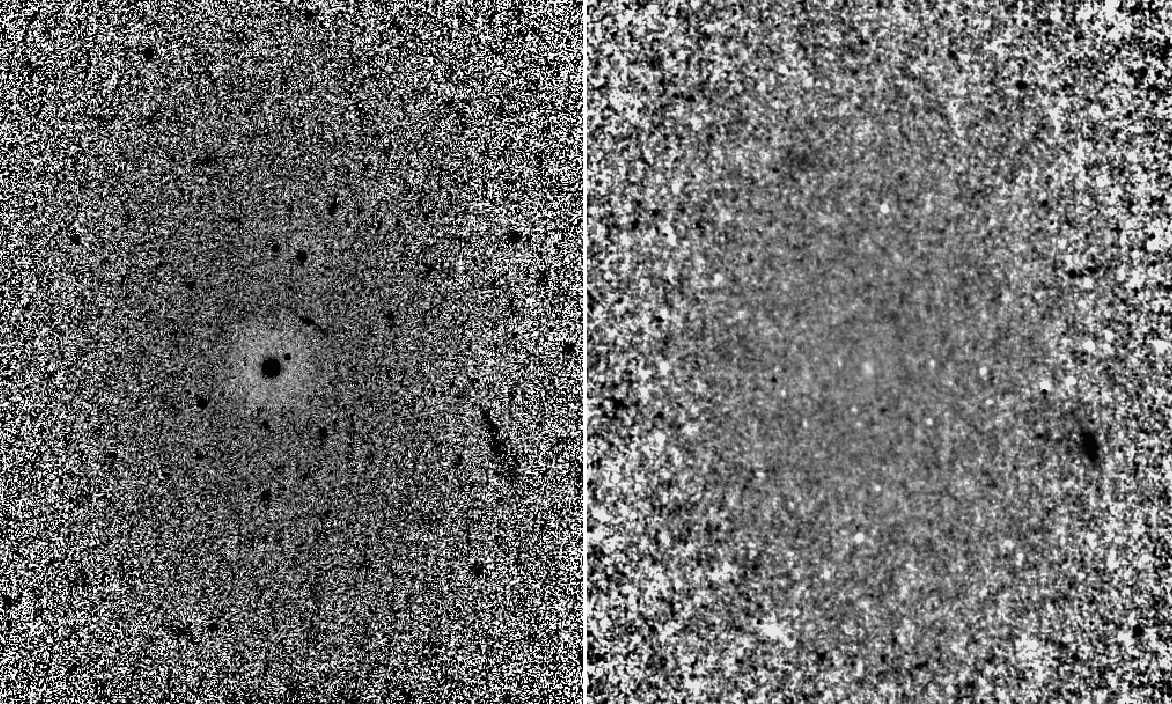}
	\includegraphics[width=4.3cm,height=3.8cm]{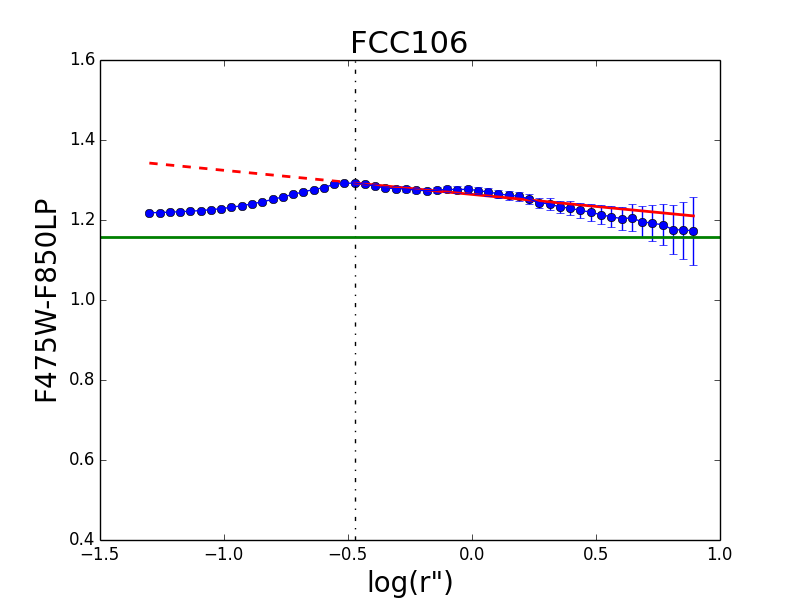}
	\includegraphics[width=4.3cm,height=3.8cm]{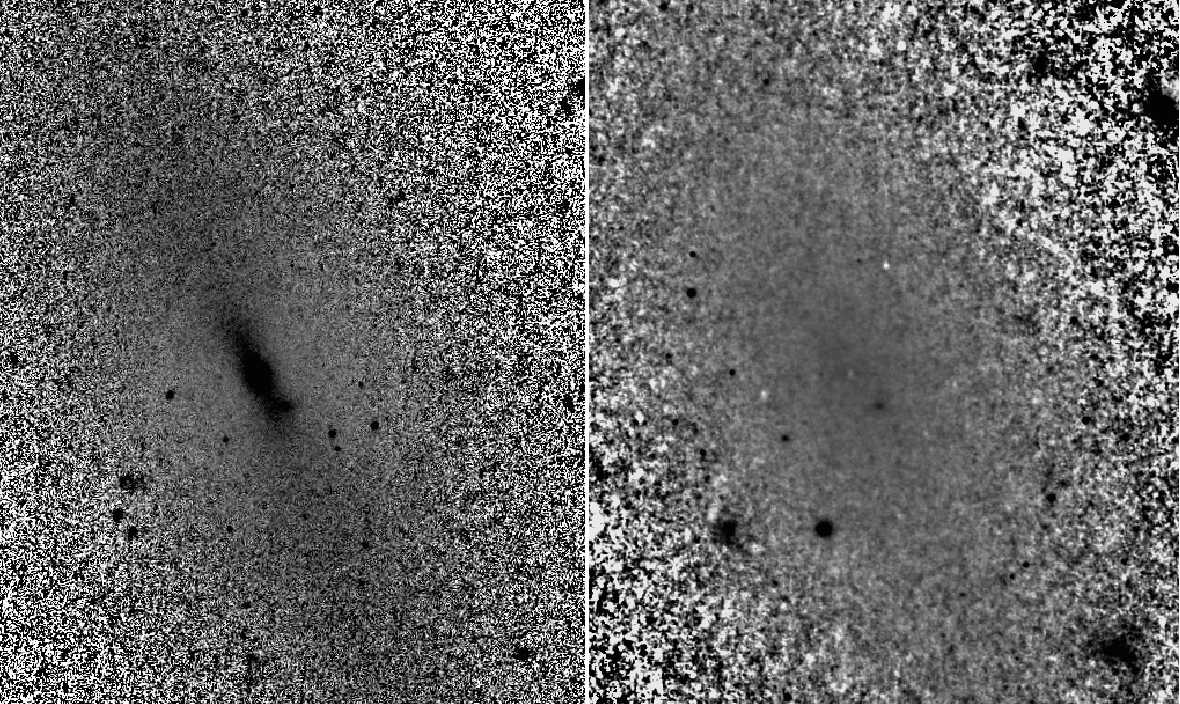}
	\includegraphics[width=4.3cm,height=3.8cm]{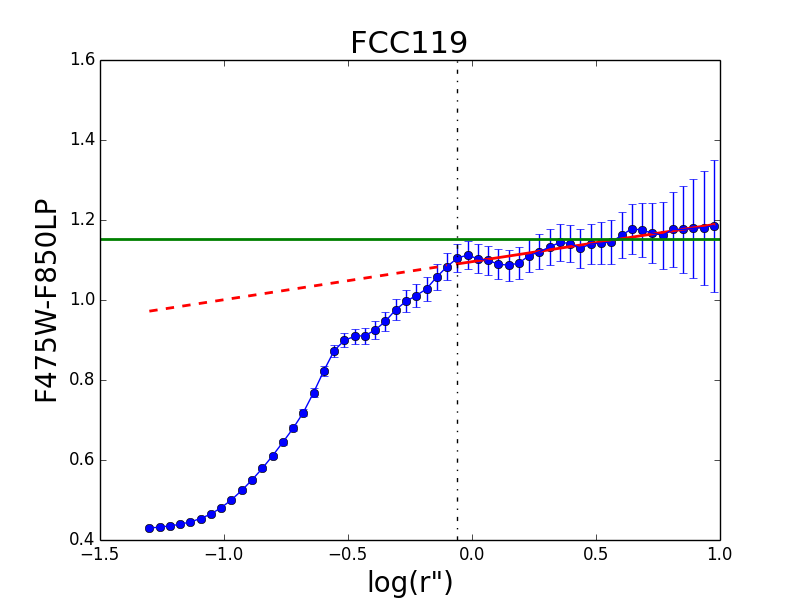}
	\includegraphics[width=4.3cm,height=3.8cm]{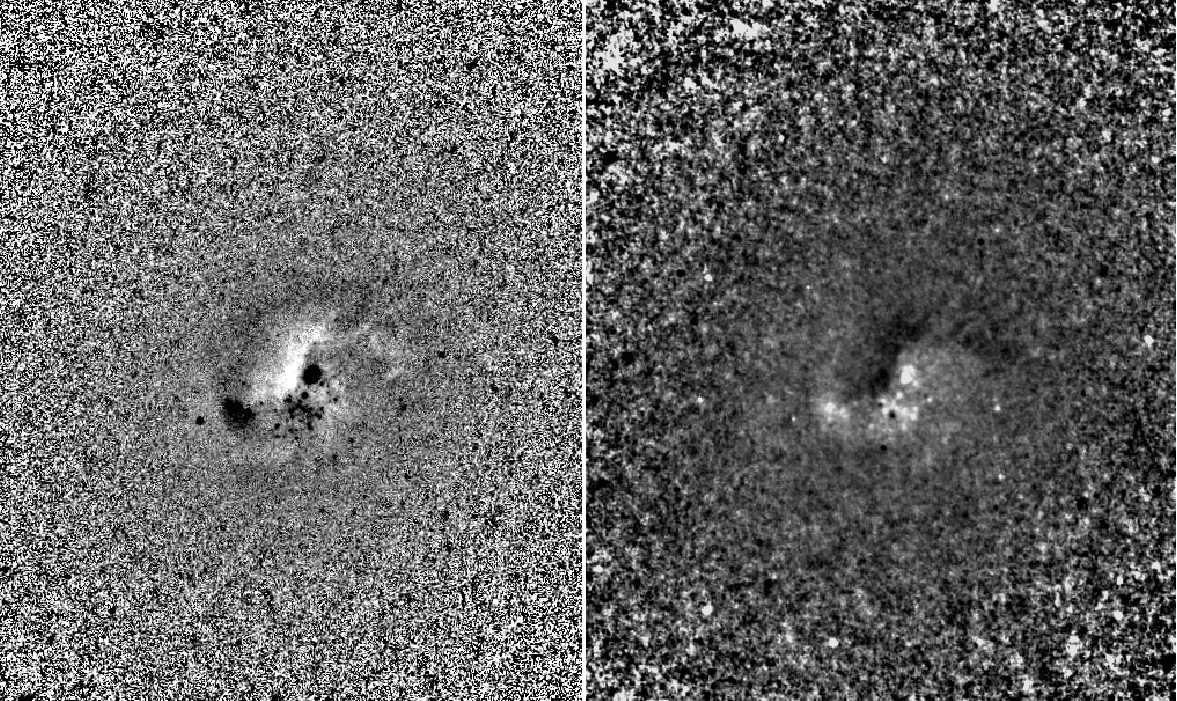}
	\includegraphics[width=4.3cm,height=3.8cm]{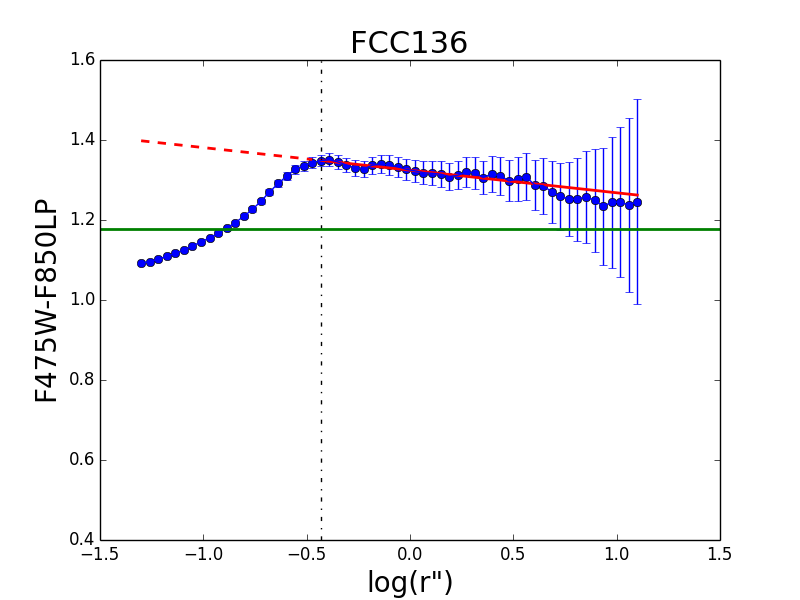}
	\includegraphics[width=4.3cm,height=3.8cm]{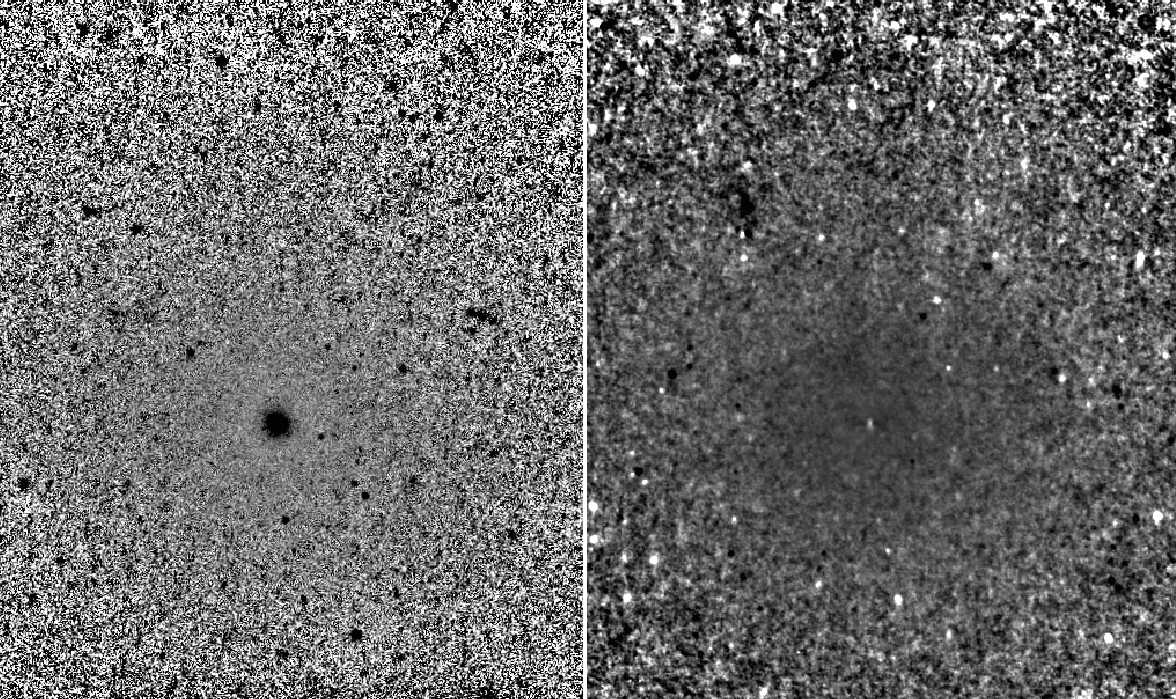}
	\includegraphics[width=4.3cm,height=3.8cm]{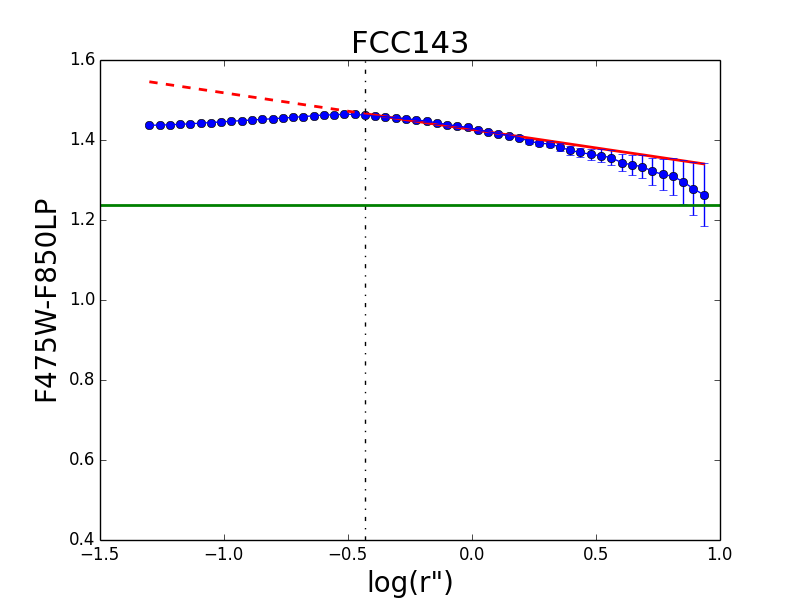}
	\includegraphics[width=4.3cm,height=3.8cm]{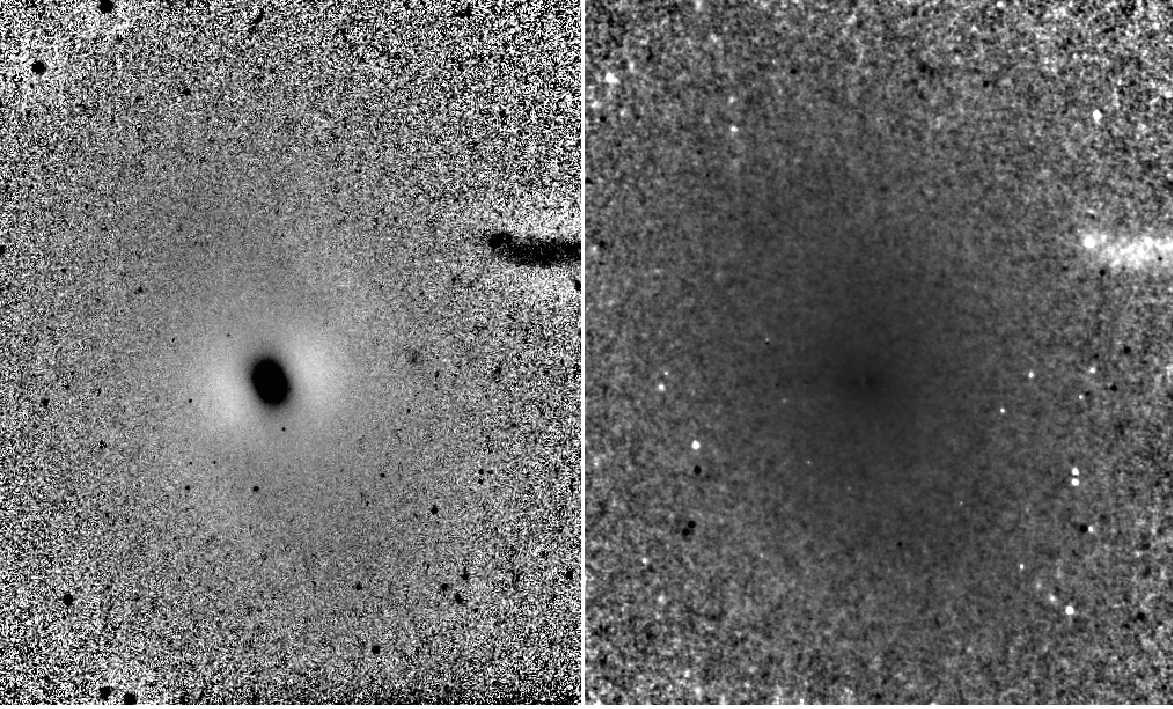}
	\includegraphics[width=4.3cm,height=3.8cm]{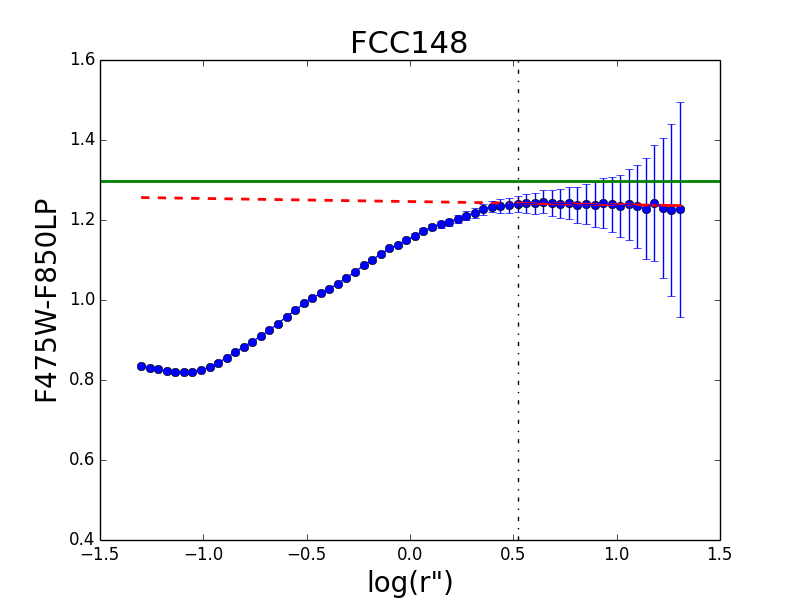}
	\includegraphics[width=4.3cm,height=3.8cm]{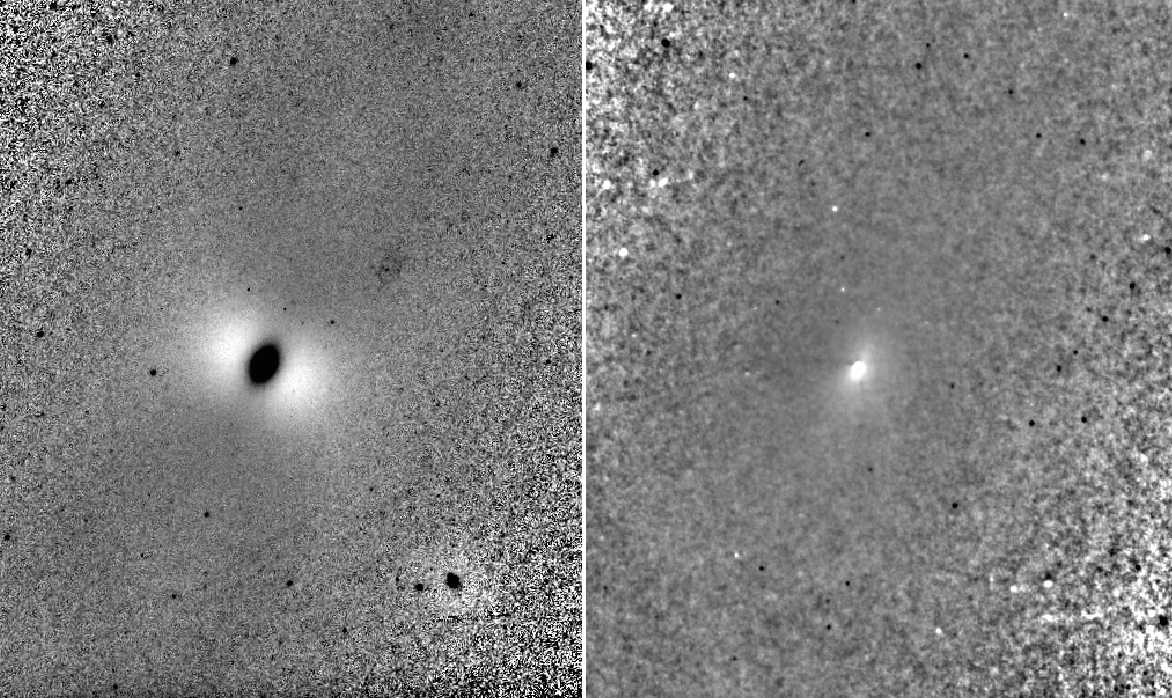}
	\includegraphics[width=4.3cm,height=3.8cm]{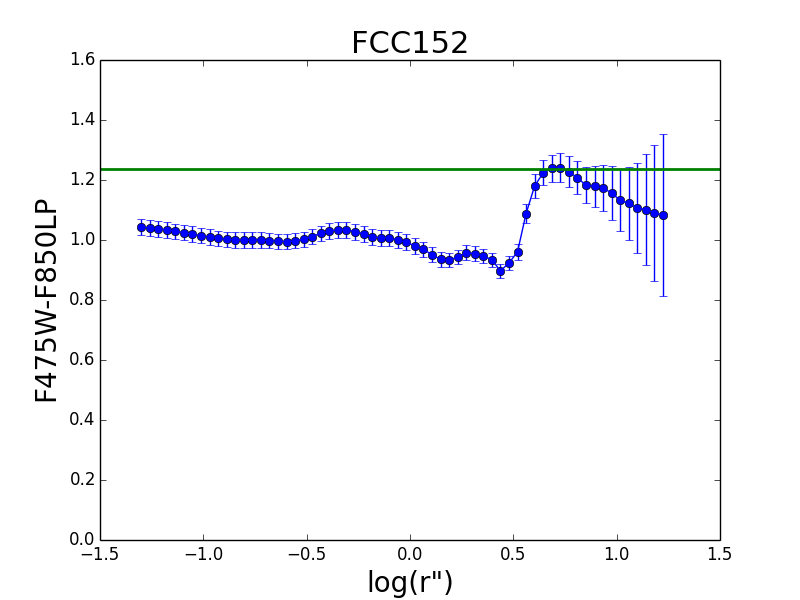}
	\includegraphics[width=4.3cm,height=3.8cm]{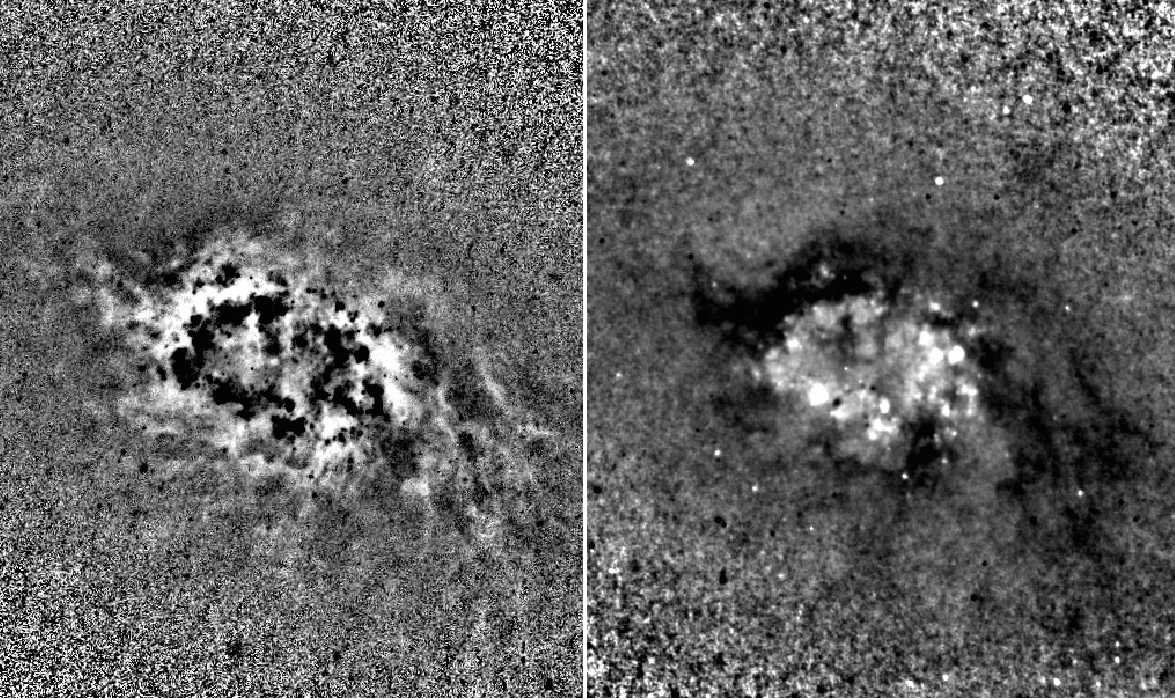} 
	\caption{Color profiles (color-log(r)), unsharp masked and color images of the Fornax early-type dwarf galaxies. In the color profiles, the green line shows the color of the CMR in that magnitude. The red line is fitted to the outer part of the profile and the black dashed line separates the inner part from the outer part of the galaxy. The width of the unsharp masked and the color images are $26^{\prime\prime}$ and the color of the unsharp masked is inverted.}
	\label{Fig:color profile of Fornax}
\end{figure*}

 On the blue side of the CMR, there are also several outliers which have been discussed extensively in the literature, by e.g. \cite{Lisker2007}. These are objects which still contain a considerable fraction of young stellar populations, even though they have been classified as dE. In Figure~\ref{Fig:color profile of Fornax}, we showed the unsharp masked images of these galaxies from the Fornax cluster. It is clear from their unsharp masked images that they have extended star forming regions. FCC26 and FCC90 both are also bright in GALEX UV indicating young stars \citep{Paz2007}. Dwarfs with younger stellar populations lie on the blue side of the CMR in most cases. In addition, VCC1499 is highlighted by \cite{BoselliGavazzi2014} as an example of a galaxy with a "post-star-burst" spectrum.

\par Since the colors give us crude information about the galaxy ages, we can estimate the age difference between galaxies in the clusters by assuming that galaxies consist of a single stellar population. For a typical metallicity of dEs, Z=0.004 \citep{Rys2015}, the models of \citet{BruzualCharlot} predict $\Delta(g^{\prime}-z^{\prime})/\Delta(log Age) = 0.575$. An offset of $\Delta(g^{\prime}-z^{\prime})=0.102$ (Table~\ref{Tab:CMR fixed slope}) for Fornax regarding the Coma CMR, corresponds to a difference in age of about $1.5$ Gyr. On average the early-type dwarfs in the center of Coma are thus 1.5 Gyr older than the ones in the Fornax and Virgo clusters.

\par We applied a Kolmogorov-Smirnov (K-S) test to investigate whether age distributions of the three clusters are similar within the errors. Each time we compare 2 clusters. The outcome probabilities, p-value, indicate whether these ages are drawn from the same distribution: p= 0.032, 0.0001 and 0.008 respectively for (Fornax, Virgo), (Fornax, Coma) and (Virgo, Coma) pairs. These tests indicate that the age distribution of dEs in Coma is statistically different from the age distribution of dEs in Fornax and Virgo, the latter two have a more similar age distributions.

\begin{figure*} 
	\centering
	\includegraphics[width=4.3cm,height=3.8cm]{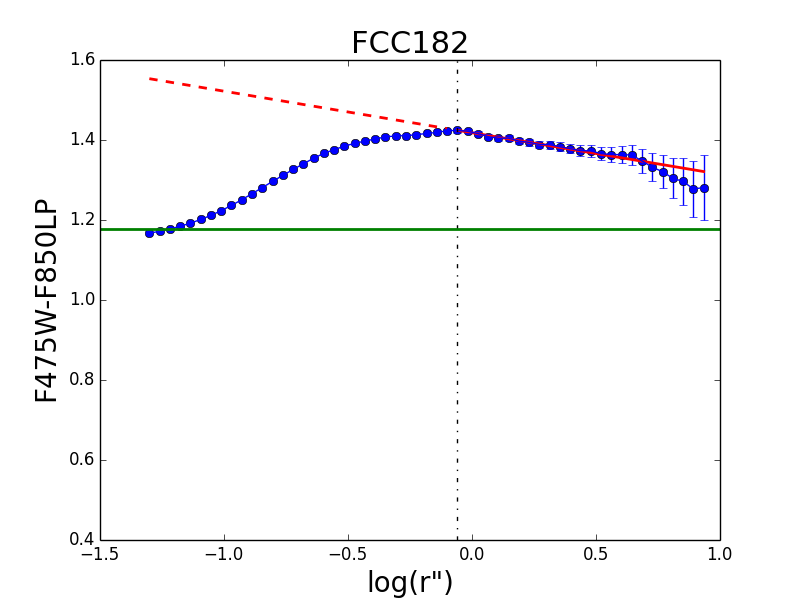}
	\includegraphics[width=4.3cm,height=3.8cm]{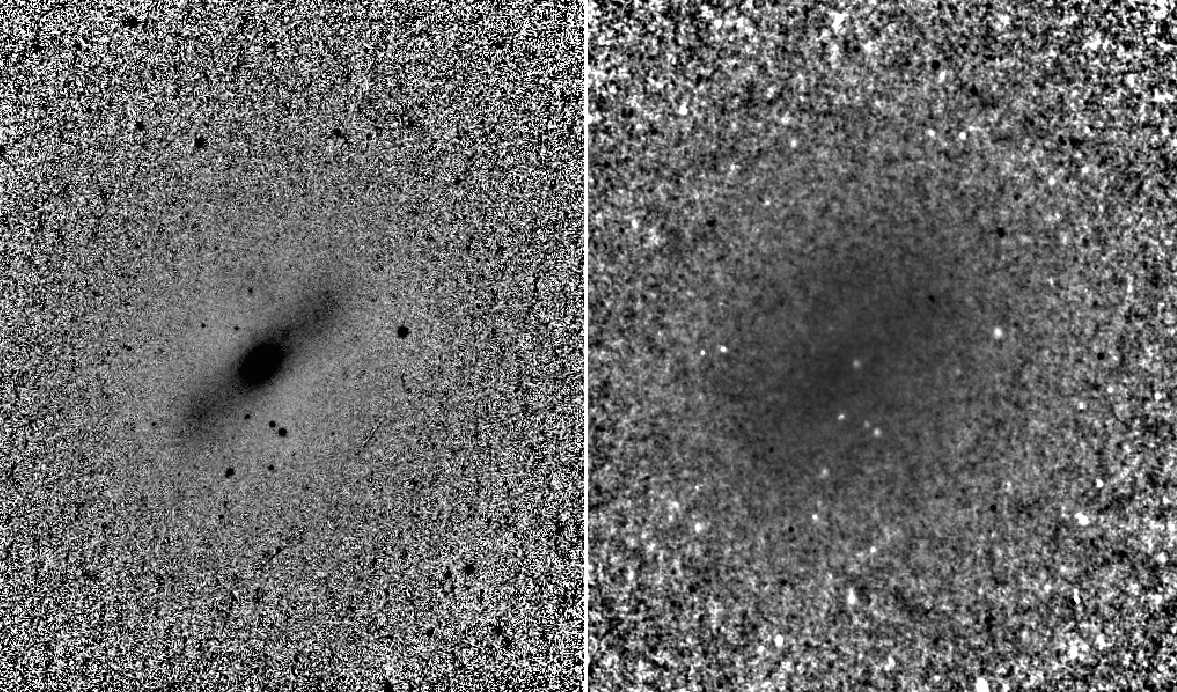}
	\includegraphics[width=4.3cm,height=3.8cm]{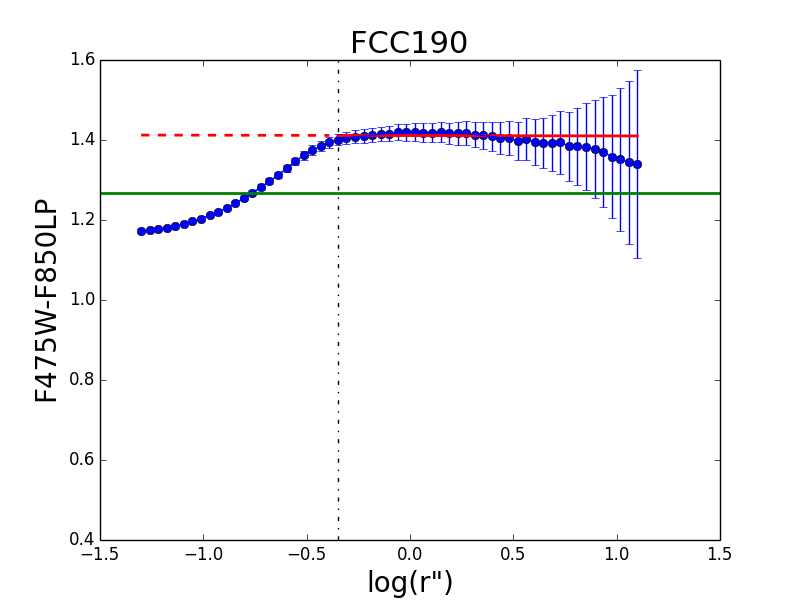}
	\includegraphics[width=4.3cm,height=3.8cm]{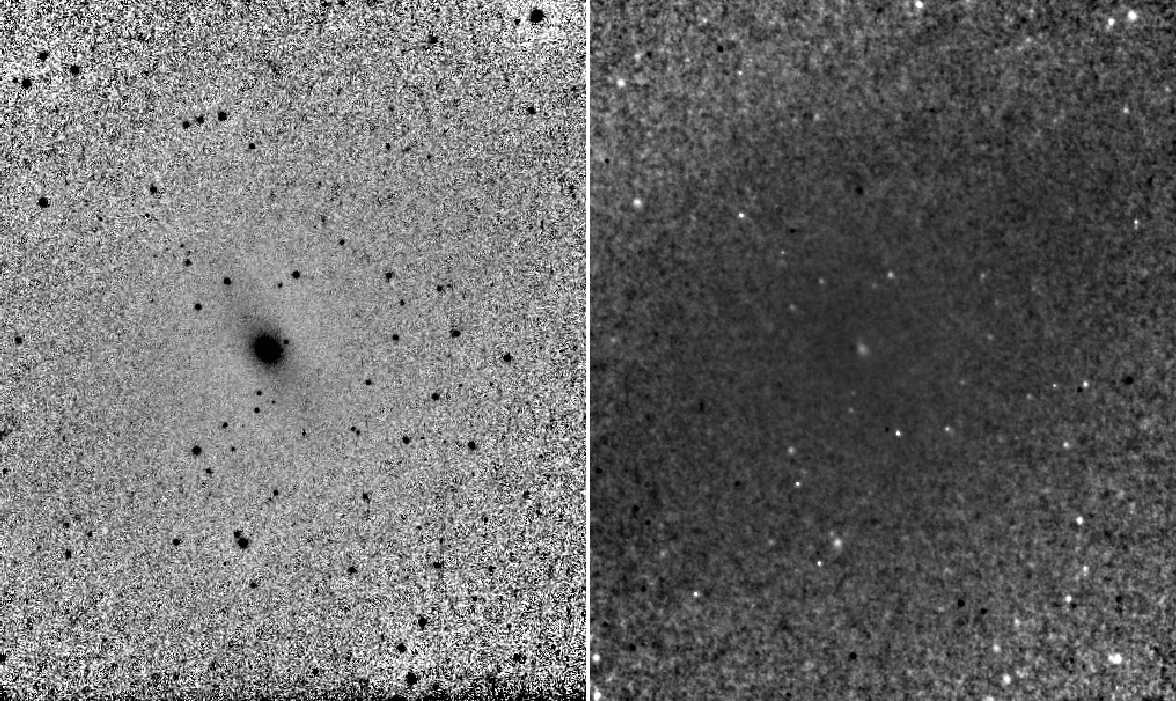}
	\includegraphics[width=4.3cm,height=3.8cm]{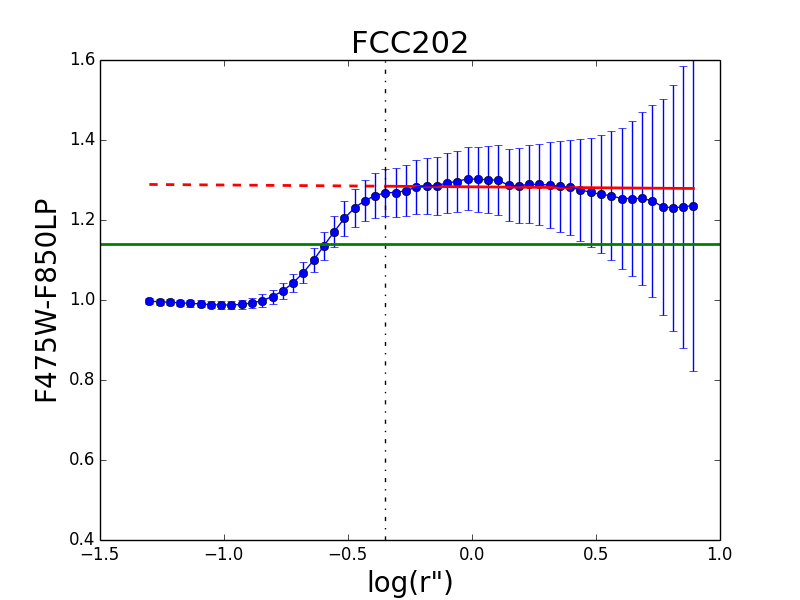}
	\includegraphics[width=4.3cm,height=3.8cm]{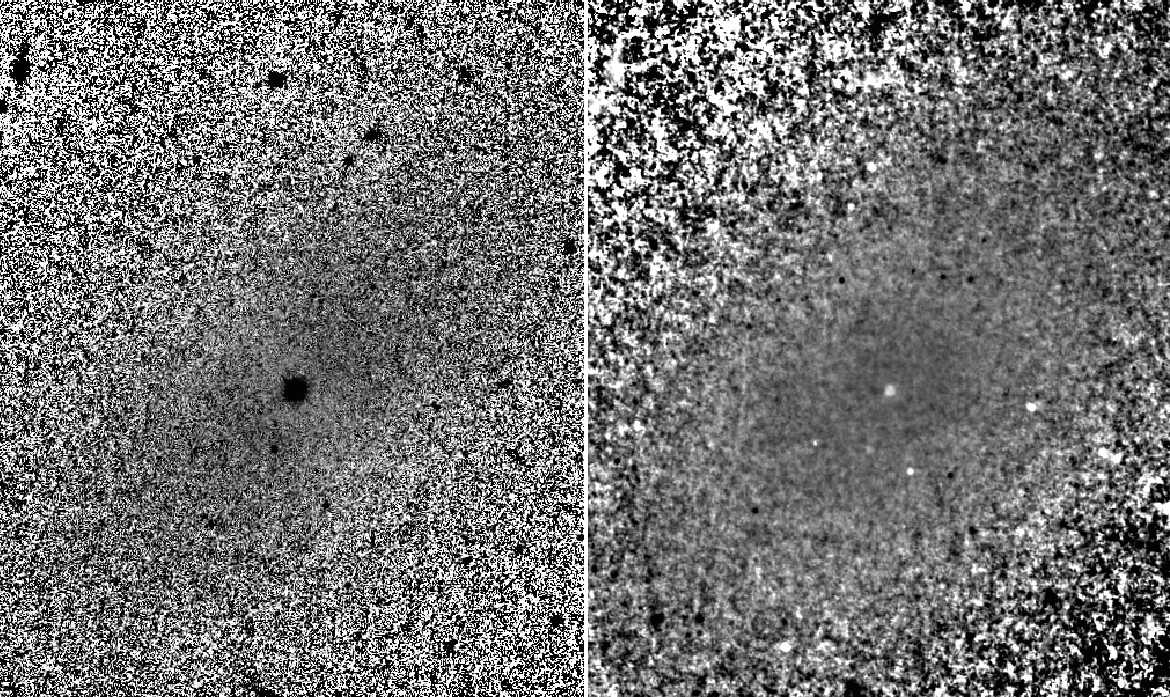}
	\includegraphics[width=4.3cm,height=3.8cm]{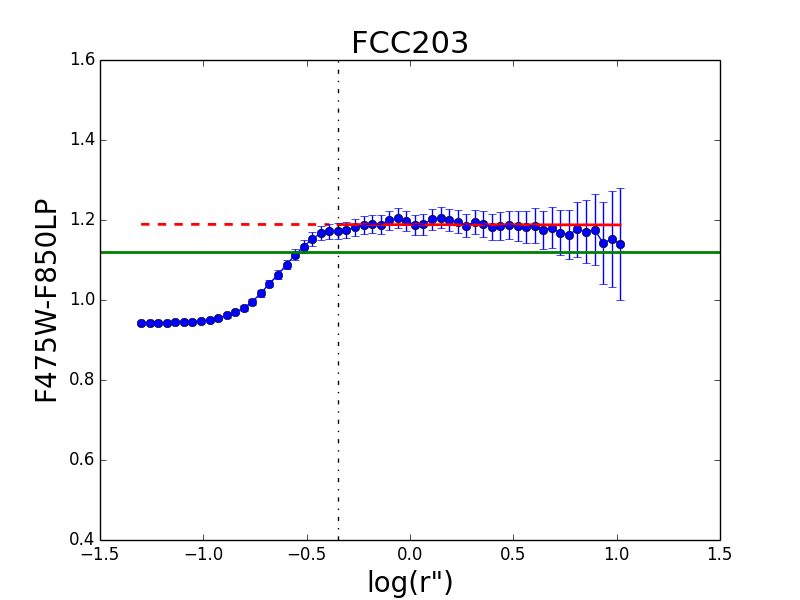}
	\includegraphics[width=4.3cm,height=3.8cm]{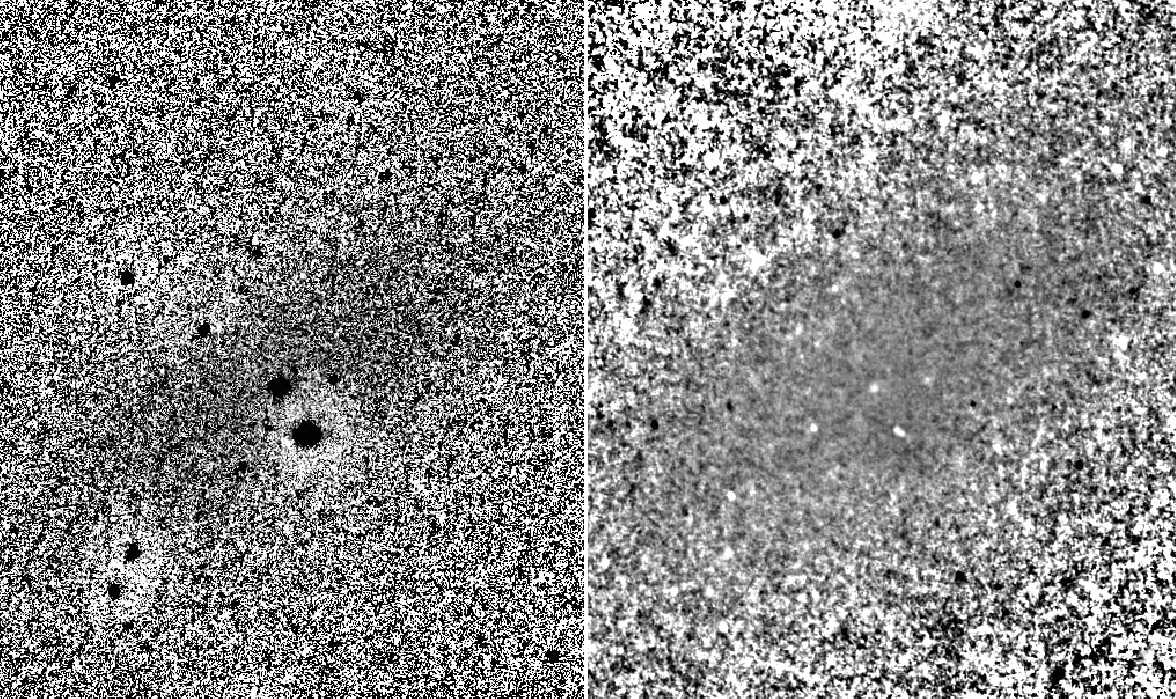}
	\includegraphics[width=4.3cm,height=3.8cm]{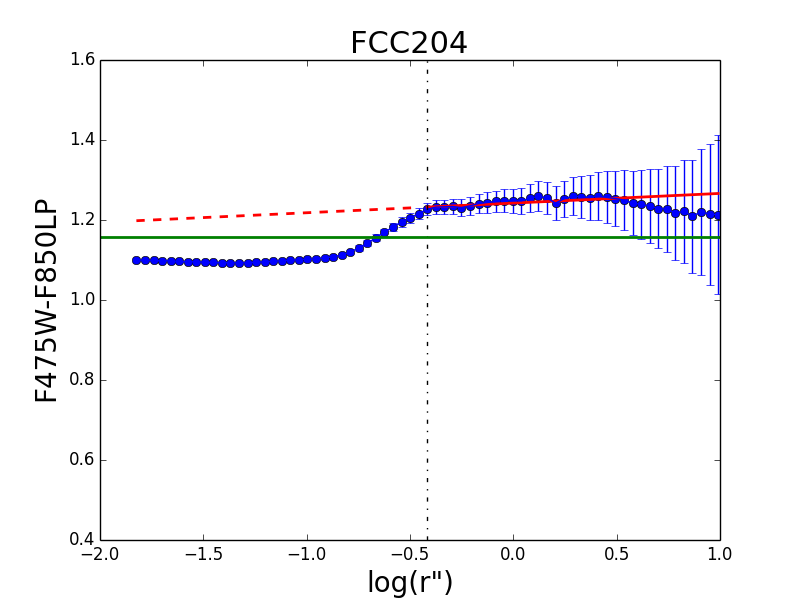}
	\includegraphics[width=4.3cm,height=3.8cm]{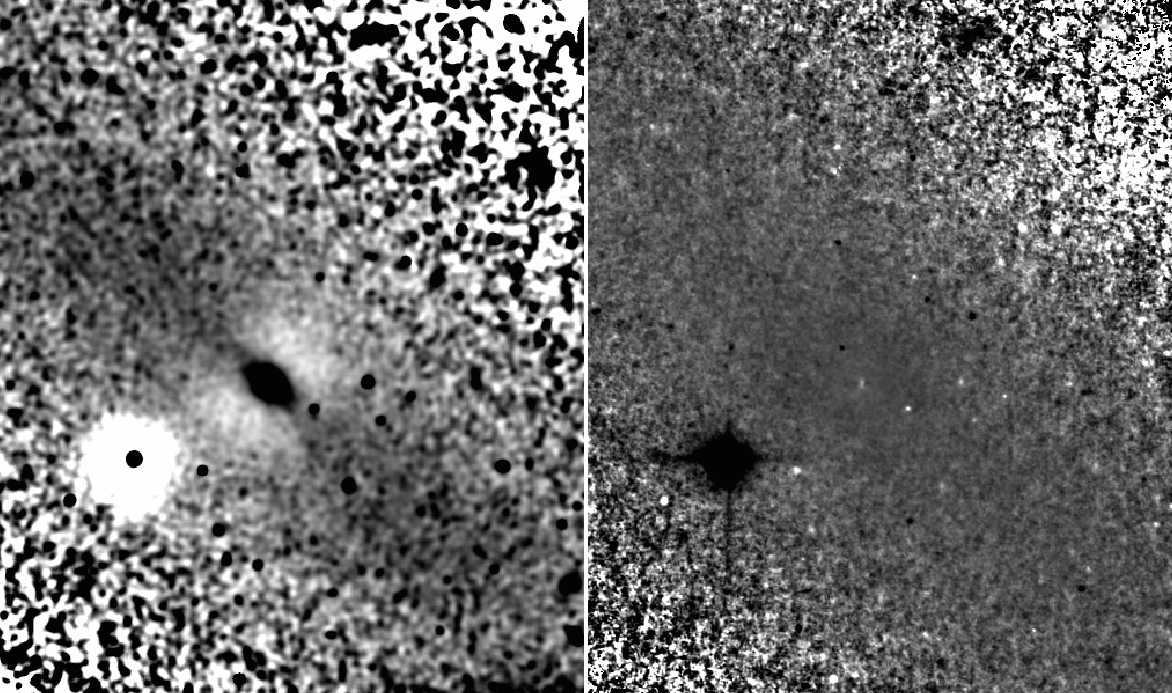}
	\includegraphics[width=4.3cm,height=3.8cm]{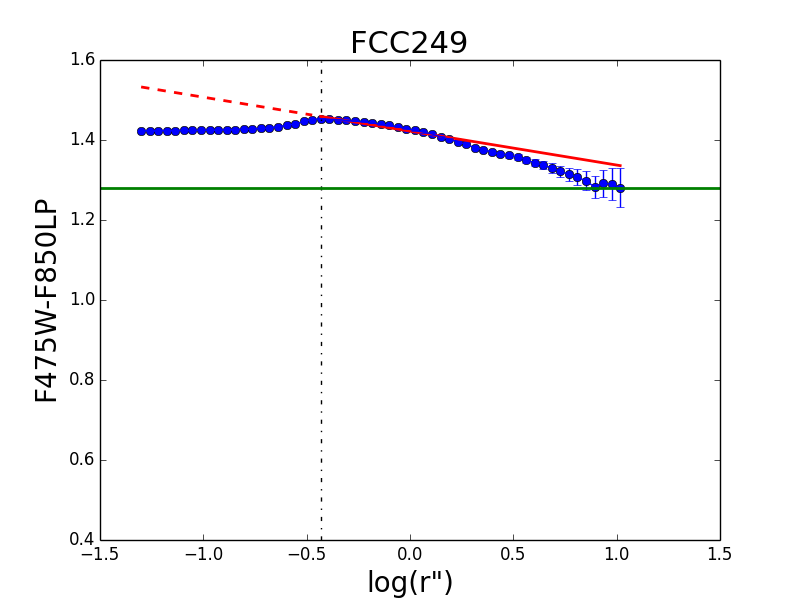}
	\includegraphics[width=4.3cm,height=3.8cm]{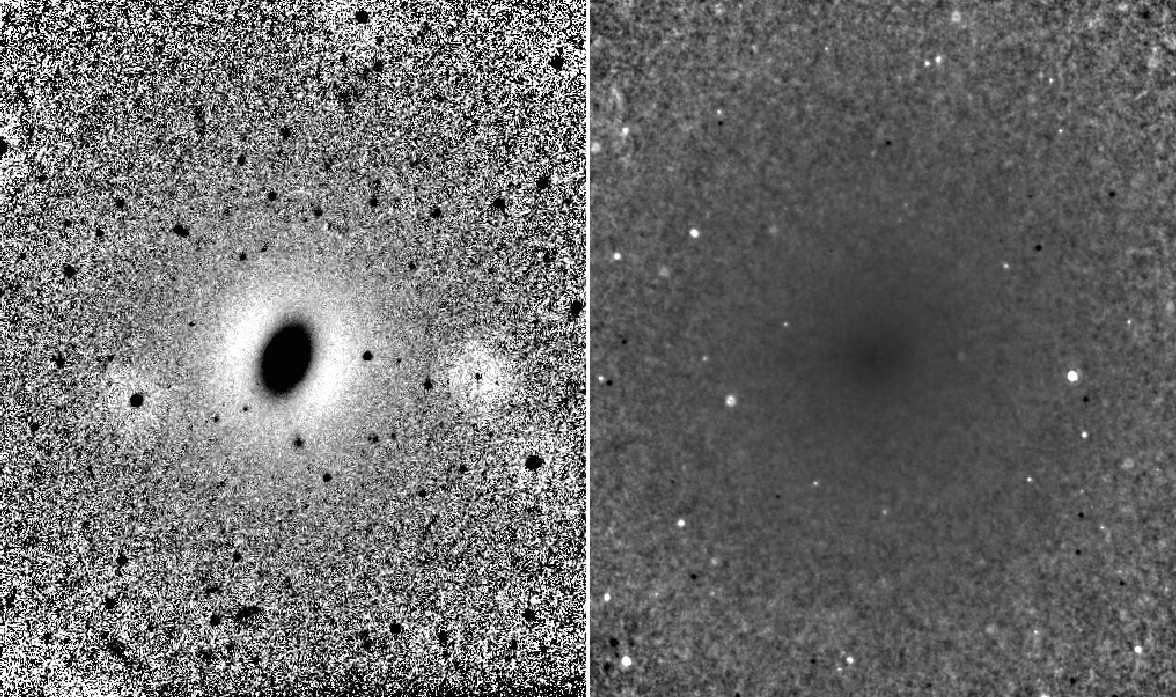}
	\includegraphics[width=4.3cm,height=3.8cm]{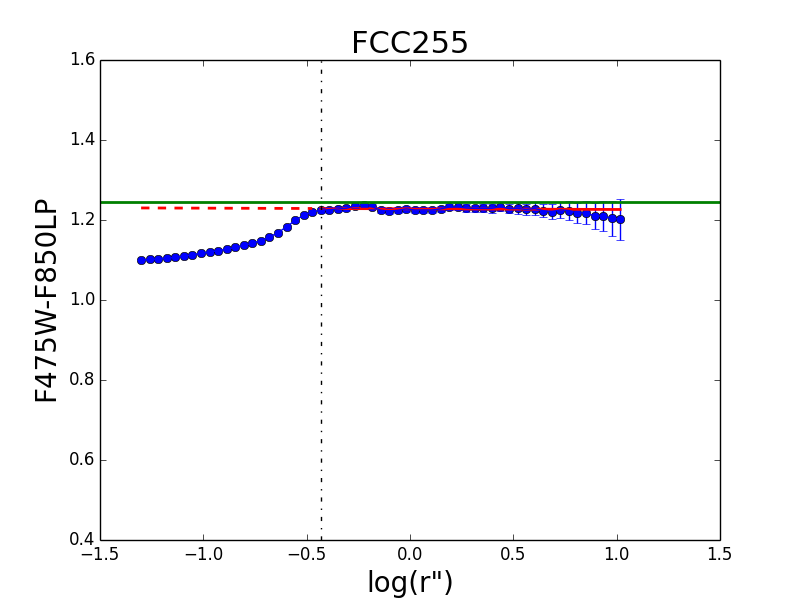}
	\includegraphics[width=4.3cm,height=3.8cm]{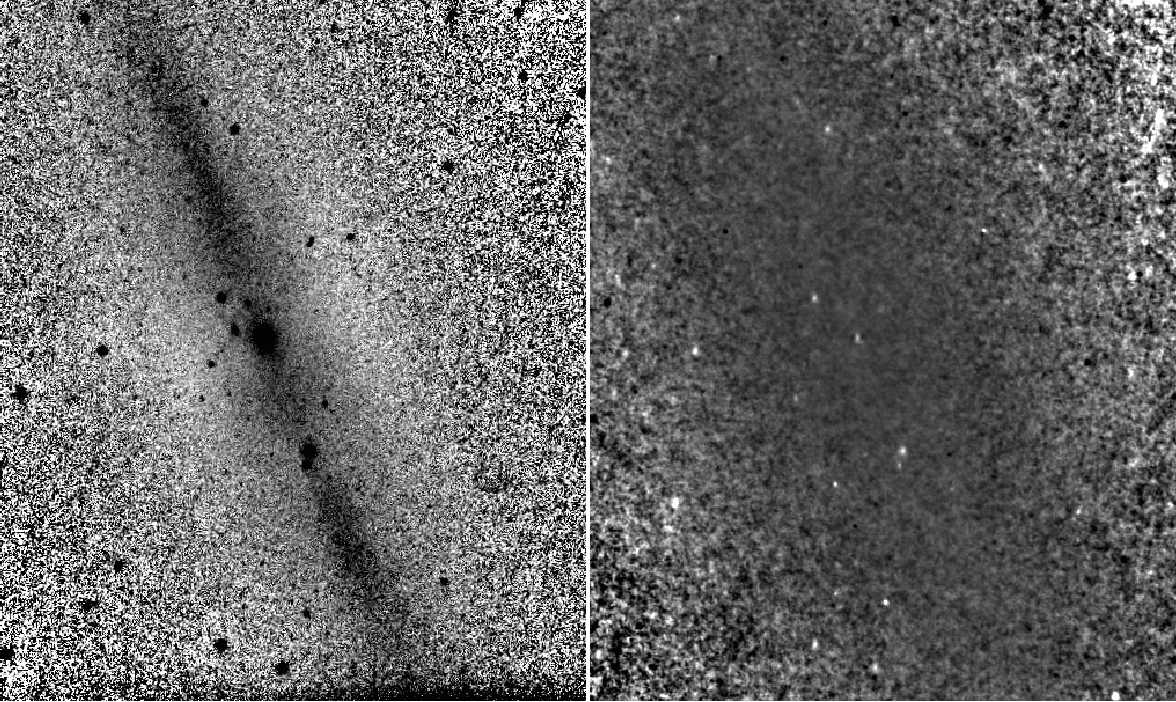}
	\includegraphics[width=4.3cm,height=3.8cm]{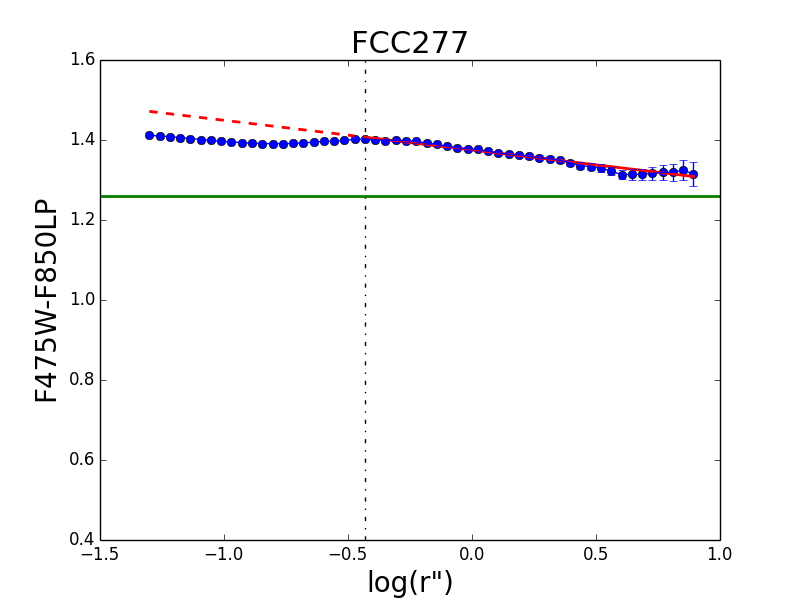}
	\includegraphics[width=4.3cm,height=3.8cm]{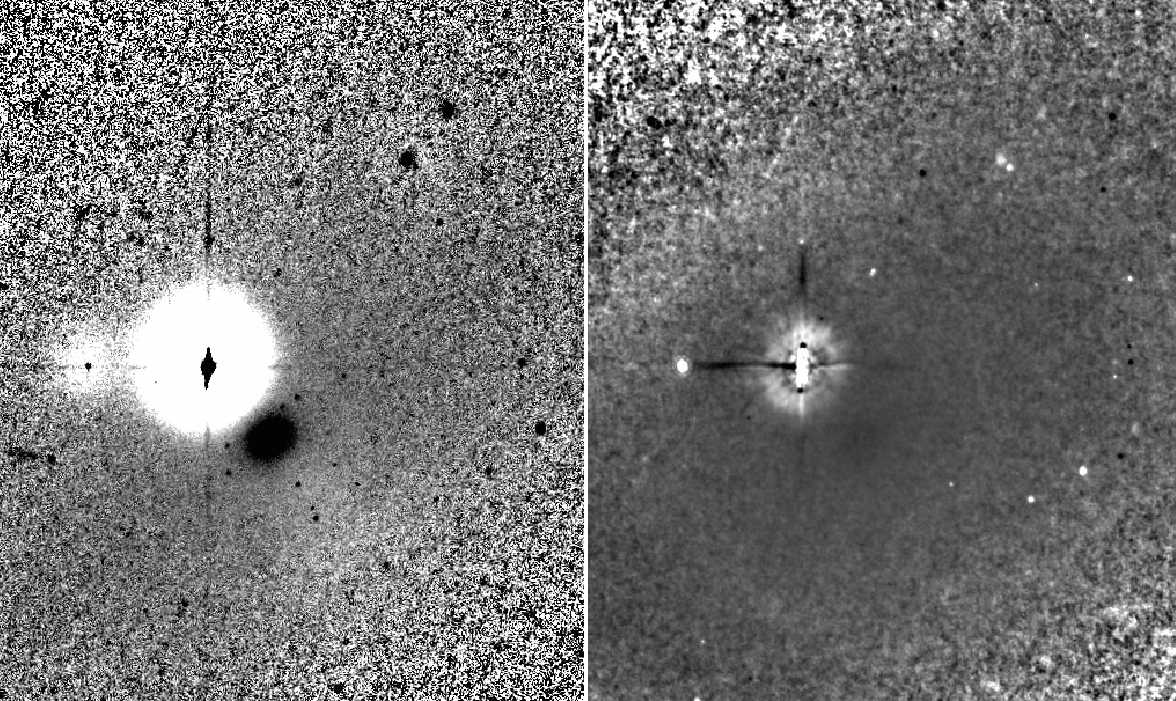}
	\includegraphics[width=4.3cm,height=3.8cm]{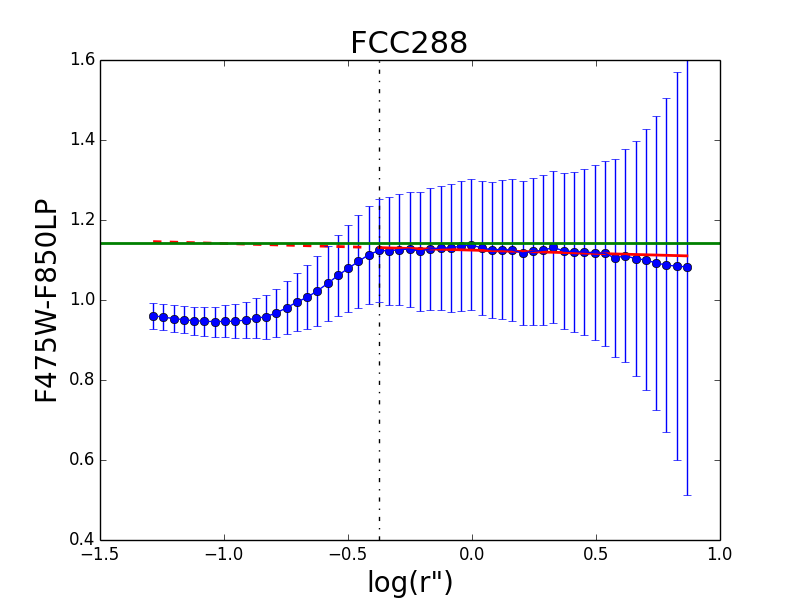}
	\includegraphics[width=4.3cm,height=3.8cm]{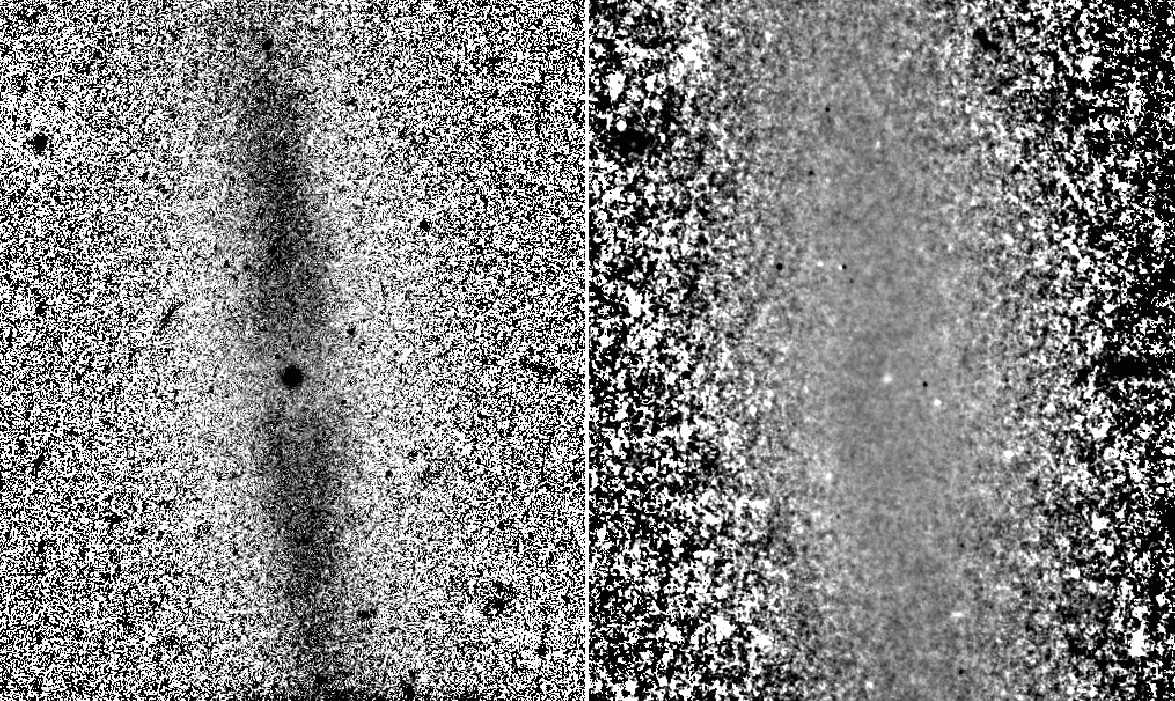}
	\includegraphics[width=4.3cm,height=3.8cm]{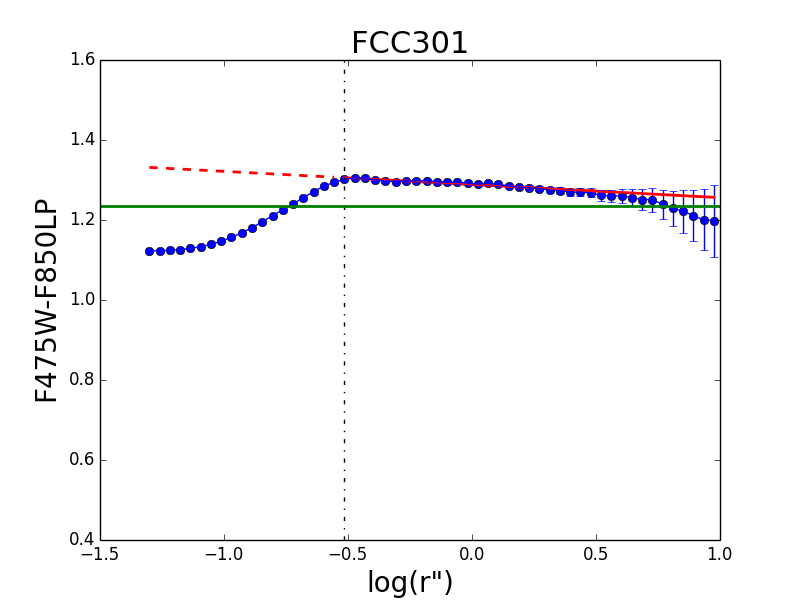}
	\includegraphics[width=4.3cm,height=3.8cm]{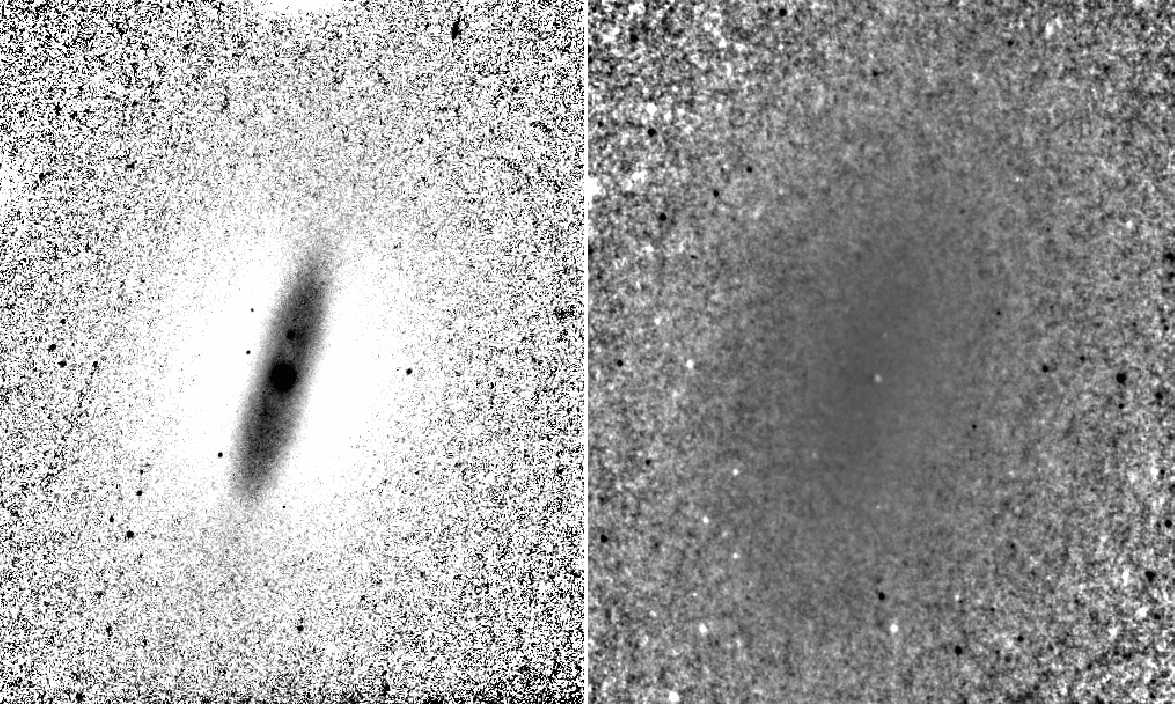}
	\includegraphics[width=4.3cm,height=3.8cm]{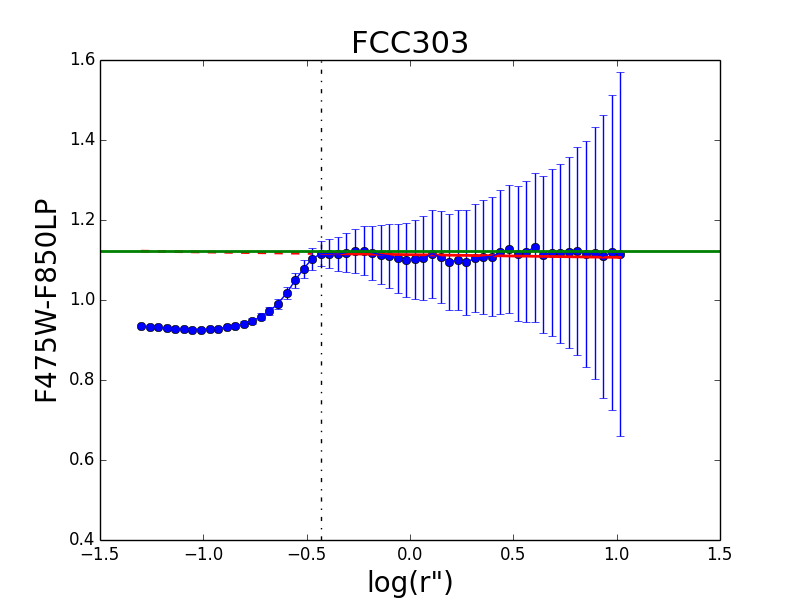}
	\includegraphics[width=4.3cm,height=3.8cm]{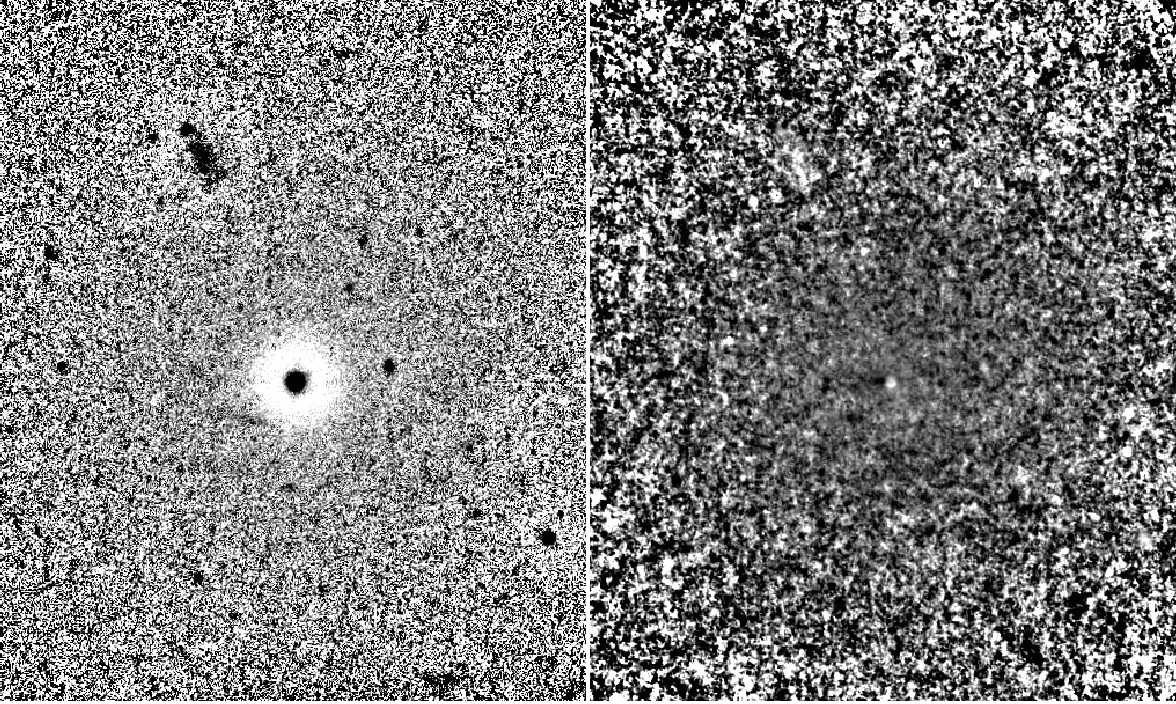}	\includegraphics[width=4.3cm,height=3.8cm]{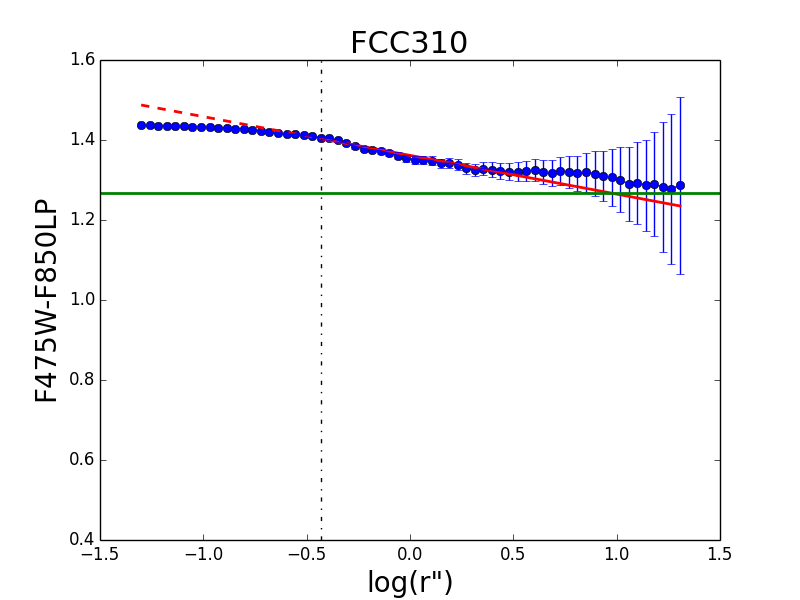}
	\includegraphics[width=4.3cm,height=3.8cm]{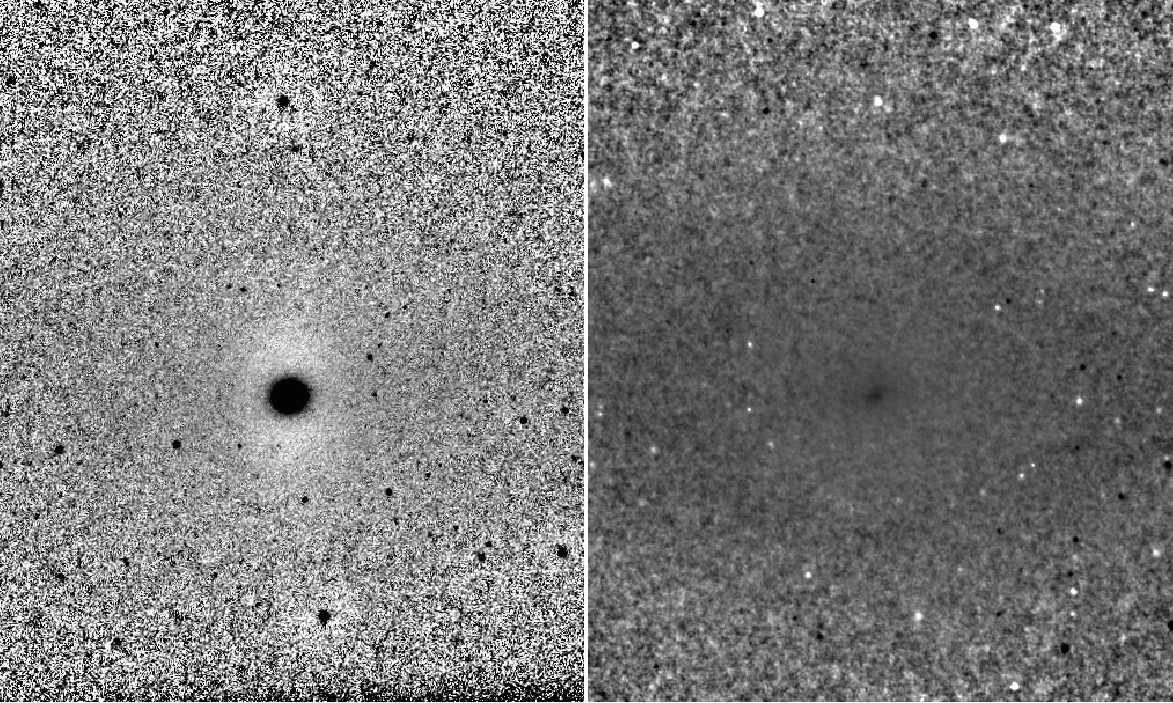}
	\caption{Fig. 8 continued}
	\label{Fig:color profile of Fornax_continue 1}
\end{figure*}
\begin{figure*} 
	\centering
	\includegraphics[width=4.3cm,height=3.8cm]{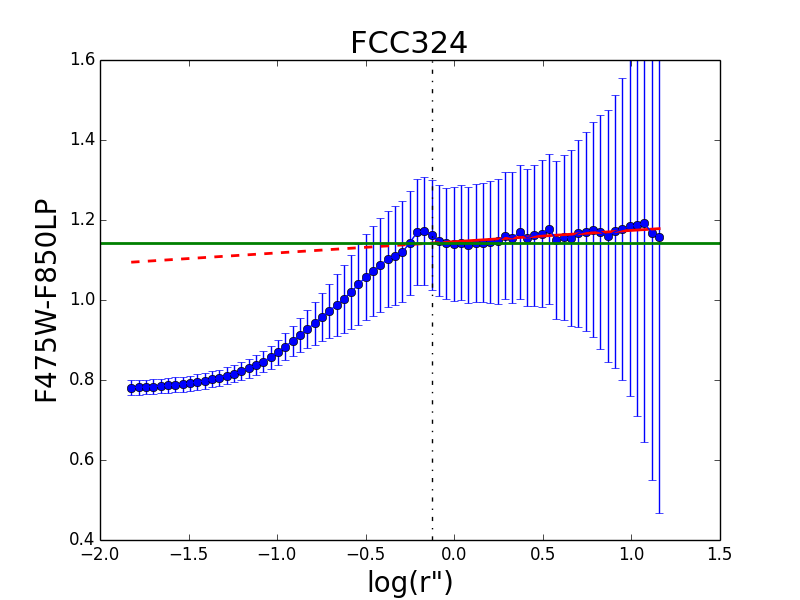}
	\includegraphics[width=4.3cm,height=3.8cm]{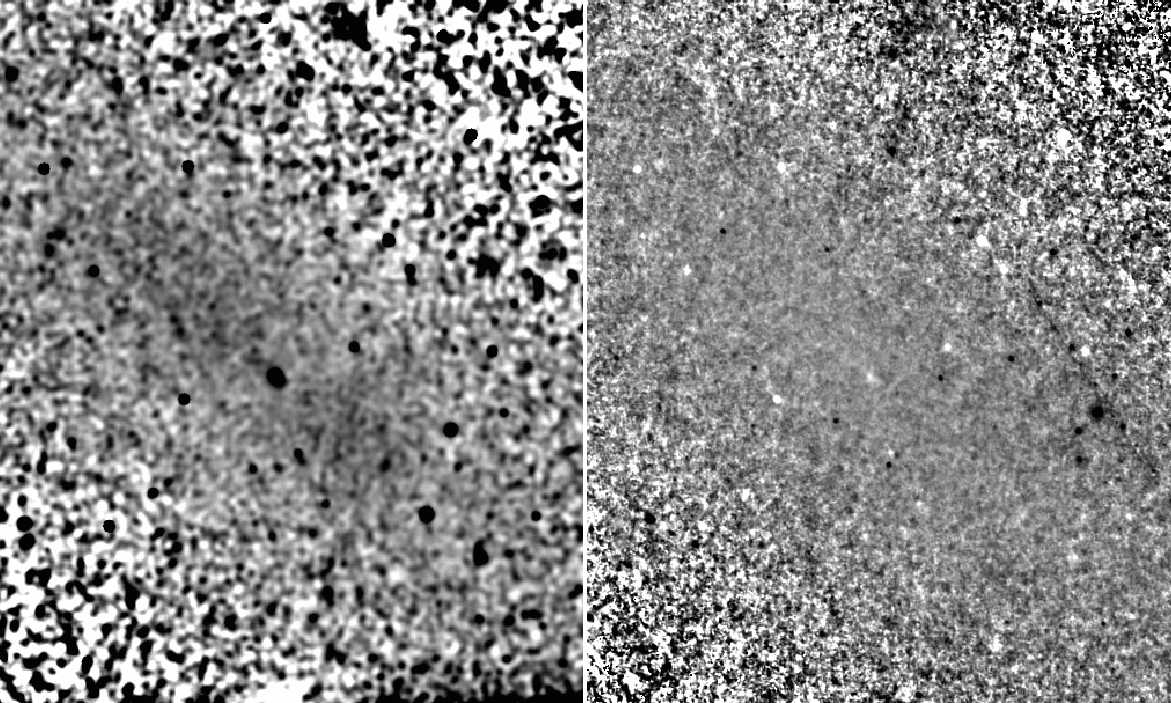}
	\includegraphics[width=4.3cm,height=3.8cm]{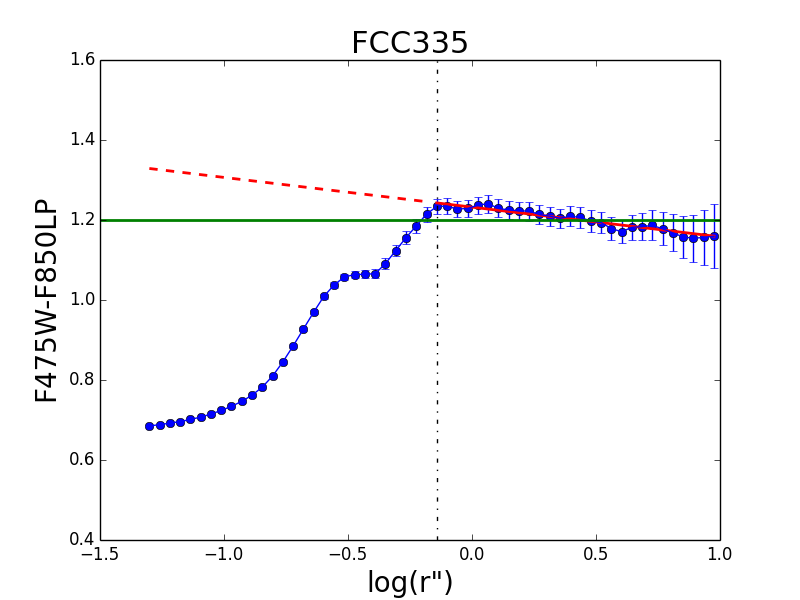}
	\includegraphics[width=4.3cm,height=3.8cm]{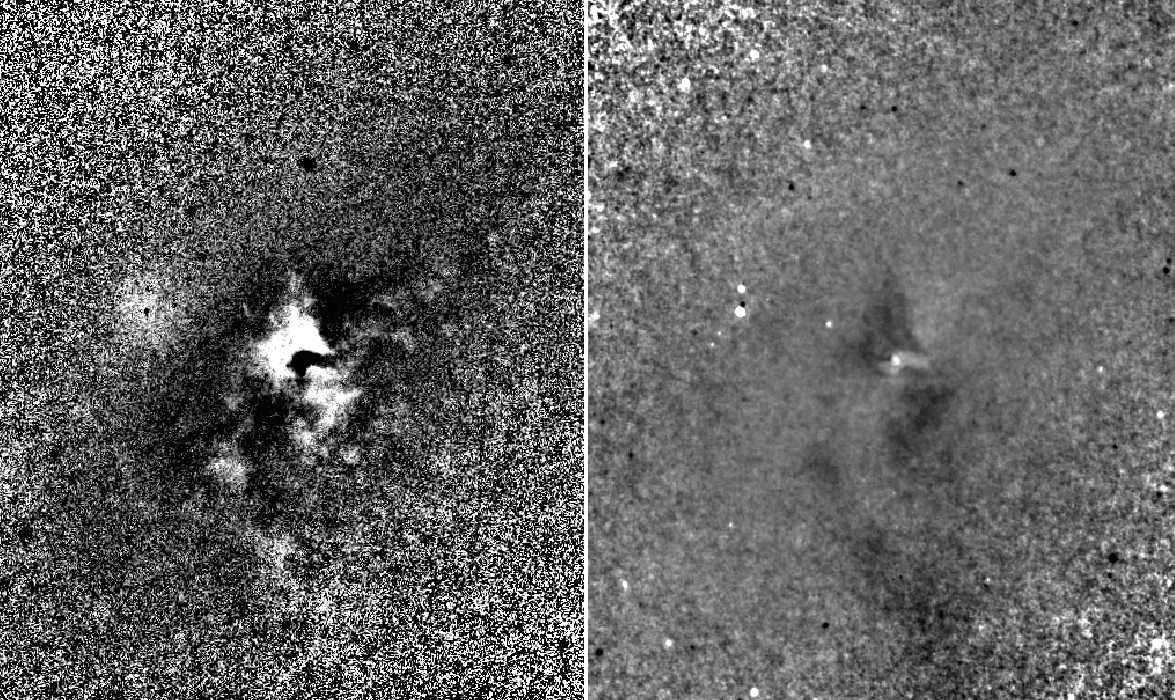}
	\caption{Fig. 8 continued}
	\label{Fig:color profile of Fornax_continue2}
\end{figure*}

\subsection{Radial color profiles}
\label{subsection:Radial color profiles}
\par The color profiles of 26 early-type dwarf galaxies in the Fornax cluster are shown in Fig.~\ref{Fig:color profile of Fornax}. In these profiles, the color is $(F475W-F850LP)$ in AB-mag. The radius (r) is the circularized distance of each fitted ellipse, $r=\sqrt{ab}$, where {\it a} and {\it b} are semi-major and semi-minor axes of the ellipse. We fitted a color-log(r) line to the color of the outer part of the galaxy, the red line in the color profiles of Figures \ref{Fig:color profile of Fornax}, \ref{Fig:color profile of Fornax_continue 1} and \ref{Fig:color profile of Fornax_continue2}, which fits better to the outer parts of the color profiles similar to the way explained by \citet{denBrok2011}. We used an error weighted least-squares method to fit these lines. For each color profile, the red fitted lines are extrapolated inward. We started our color profiles at 1 pixel equal to $0.05^{\prime\prime}$ and continued in radius as far as $0.8R_{e}$. The effective radii and total magnitudes in $g^{\prime}$ are from Table~\ref{tab:Fornax properties}.

\par For most dEs in the Fornax cluster sample, the color profiles are getting slightly bluer going outward. \citet{denBrok2011} and also \citet{Peletier2012} interpreted the outer bluing as an effect of old stellar populations for which the metallicity is slowly decreasing outward. The green line in Fig.~\ref{Fig:color profile of Fornax} shows the color this galaxy would have if it were exactly on the CMR of Fornax (Table~\ref{Tab:CMR}). In the center of the galaxies, the profile drops down making a blue central region. The black vertical lines show the starting point of the core inside which the color profile is not following the slope of the outer part. These points were chosen by eye. \citet{denBrok2011} defined it as the center which has been excluded in their color gradient calculation to have a better fitted line to the outer part of a galaxy. In some cases, the color profiles do not follow this shape and they have a very extended bluer area, like FCC 26, FCC 90 and FCC 152. For these galaxies, it is clear from their unsharp masked and color images that they contain a large and extended amount of dust and young stellar populations. Since the previous scheme is not appropriate in such cases, we did not fit a line to their profiles. 

\par The unsharp masked and color images of each galaxy are shown in the right side of its color profile in Fig.~\ref{Fig:color profile of Fornax}. The width of their boxes are $26^{\prime\prime}$. Galaxies with extended blue cores have young stellar populations and dusty regions in their unsharp masked and color images, such as FCC26, FCC90, FCC119, FCC152 and FCC335. According to the color images, some of the galaxies have a very well-defined blue centers; FCC148, FCC204. 

\par To quantize these color profiles better, we determined various parameters. The color gradient, the slope of the fitted red line, is one of the calculated parameters. A positive (negative) color gradient indicates that the galaxy has a redder (bluer) color as it goes outward. Another parameter is the excess light of the core from the fitted line to the outer region. In the color profiles, it is the average of the difference between the color of the center, separated by the black vertical line, and extrapolated dashed red line:

\begin{ceqn}
\begin{equation} 
\begin{aligned}
X_{outer~fit}=&<Color_{galaxy}-Color_{red~line}> \\ 
=&-2.5 \log_{10} \frac{\int10^{-0.4(color_{galaxy}-color_{red~line})}~I_{F475W}~dA}{\int I_{F475W}~dA} \label{Eq:formula of excess from fitted line}
\end{aligned}
\end{equation}
\end{ceqn}
where  $I_{F475W}$ is the intensity in F475W band. This is calculated between the start point of the radius of color profile and the dashed black line, e.g the central part of the galaxy. Positive (negative) amounts generally show that the center of the galaxy is redder (bluer) than the fitted line to the outer part. A similar formula is used to find the excess light from the CMR line, the green line in color profiles, for the whole profile, $X_{CMR}=<Color_{galaxy}-Color_{CMR}>$. In this parameter which we call excess light from the CMR line, a positive (negative) sign implies the galaxy in total is redder (bluer) than the CMR.

\begin{table*}
\caption{Table of Fornax dEs color profile parameters}
\centering
\begin{tabular}{P{.7in}P{.5in}P{.5in}P{.5in}P{.7in}P{.7in}P{.9in}}
\centering
FCC & Blue & Blue & $r_{core}$(pc)& $X_{outer~fit}$ & $X_{CMR}$ & $\bigtriangledown(g^{\prime}-z^{\prime})$  \\
number & galaxy & core& & &  & \\
(1) & (2) & (3) & (4) & (5) & (6) & (7) \\
\hline
		FCC 19  & Y & Y & 44 & -0.09 & -0.07 & 0.063 $\pm$ 0.005  \\
		FCC 26  & Y& Y & 707 & -0.47 & -0.47 & n  \\
		FCC 55  & Y & Y & 34 & -0.04 & 0.00 & -0.003 $\pm$ 0.002   \\
		FCC 90  & Y & Y & 746 & -0.40 & -0.40 & n    \\
		FCC 95  &  N & Y & 36 & -0.05 & 0.09 & -0.029 $\pm$ 0.003  \\
		FCC 100  & Y & Y & 36 & -0.11 & 0.00 & 0.063 $\pm$ 0.007  \\
		FCC 106  & N & Y & 33 & -0.05 & 0.06 & -0.060 $\pm$ 0.004   \\
		FCC 119  & Y & Y & 85 & -0.24 & -0.00 & 0.095 $\pm$ 0.010  \\
		FCC 136 & N &  Y & 36 & -0.14 & 0.09 & -0.057 $\pm$ 0.004  \\
		FCC 143  & N & Y & 36 & -0.03 & 0.12 & -0.092 $\pm$ 0.004  \\
		FCC 148 & Y & Y & 322 & -0.11 & -0.08 & -0.008 $\pm$ 0.004 \\
		FCC 152  & Y & Y & 389 & -0.13 & -0.13 & n  \\
		FCC 182  & N & Y & 85 & -0.06 & 0.17 & -0.104 $\pm$ 0.004  \\
		FCC 190  & N & Y & 44 & -0.10 & 0.11 & -0.001 $\pm$ 0.005   \\
		FCC 202  & N & Y & 44 & -0.19 & 0.11 & -0.005 $\pm$ 0.009  \\
		FCC 203  & N & Y & 44 & -0.13 & 0.05 & -0.001 $\pm$ 0.006  \\
		FCC 204  & N & Y & 37 & -0.08 & 0.07 & 0.024 $\pm$ 0.005  \\
		FCC 249  & N & Y & 36 & -0.04 & 0.07 & -0.085 $\pm$ 0.005  \\
		FCC 255 & Y & Y & 36 & -0.07 & -0.03 & -0.002 $\pm$ 0.002 \\
		FCC 277  & N & Y &   36 & -0.03 & 0.07 & -0.074 $\pm$ 0.002   \\
		FCC 288  & Y & Y & 41 & -0.12 & -0.04 & -0.017 $\pm$ 0.004  \\
		FCC 301  & N & Y & 30 & -0.10 & 0.02 & -0.033 $\pm$ 0.002   \\
		FCC 303  & Y & Y & 36 & -0.15 & -0.01 & -0.007 $\pm$ 0.004   \\
		FCC 310 & N & Y & 36 & -0.01 & 0.04 & -0.097 $\pm$ 0.004  \\
		FCC 324 & Y & Y & 73 & -0.07 & 0.03 & 0.028 $\pm$ 0.005  \\
		FCC 335 & Y & Y & 70 & -0.34 & -0.03 & -0.074 $\pm$ 0.005 
	\end{tabular}
\tablefoot{Parameters of the color profiles of the dEs in Fornax:\\
		(1) Name of the galaxy from the Fornax Cluster Catalog\\
		(2) Indicates whether the galaxy is blue (Y) or not (N). A galaxy is blue if its total color is bluer than CMR.\\
		(3) If the center of the galaxy is blue (or not), it is shown by Y (N). A galaxy has a blue core when the color of its central part is bluer than the fitted line to the outer color of the galaxy (shown by $X_{outer~fit}>0$ for red and $X_{outer~fit}<0$ for blue centers).\\
		(4) Size of the blue or red center of each galaxy, separated by the black vertical line with an estimated error of $\sim20\%$.\\
		(5) $<Color_{data}-Color_{red~line}>$: the average excess light in the center of the galaxy from the red fitted line in the color profiles (Formula~\ref{Eq:formula of excess from fitted line})\\
		(6) $<Color_{data}-Color_{CMR}>$: the average excess light of the galaxy from the CMR line in the color profiles (the green line)\\
		(7) Color gradient of each galaxy excluding the red or blue centers\\
		If a galaxy is completely blue without any fitted line, or does not have a blue or red center, this is indicated with 'n'. }
	\label{Tab: Fornax parameters}
\end{table*}

\par Table~\ref{Tab: Fornax parameters} lists the measured parameters for our sample of the Fornax cluster. For the three galaxies without the fitted red line, we considered the green line to be their reference color of older population where we can calculate the excess bluer light of the center from there. In this case, $X_{outer~fit}=X_{CMR}$. Their core sizes are equal to the end point of the color profile. The color gradient of the fitted line for these extended cases are set to $n$ in the table.

\par For the Virgo cluster, we applied the same procedure. The color profiles of the dEs in Virgo with the same color-log(r) line fitted to the data are shown in Figure~\ref{Fig:color profile of Virgo} in the appendix. The resulting parameters are listed in Table~\ref{Tab: Virgo parameters} similar to Table~\ref{Tab: Fornax parameters}. Notice that in Table~\ref{Tab: Virgo parameters}, when there is no fitted line to the color profile, as is the case for the galaxies with extended blue centers, the core size, the excess light from the fitted line and the color gradient are as explained in previous paragraph. For the galaxies without a core, the core size and the excess light from the fitted line are displayed by $n$ in the table.
\par The position of the galaxies on the CMR of the Virgo galaxies is given in Table~\ref{Tab:CMR} and is shown by the green horizontal line in the color profiles. For some cases like VCC21, \cite{Cote2006} noted that the color of its nucleus is quite blue (F475W-F850LP)= 0.3 which is consistent with its color profile in Fig~\ref{Fig:color profile of Virgo} of this paper. They interpreted this galaxy as dIrr/dE transition type with a blue nucleus younger than 1 Gyr for any selected metallicity.
\par The output of the Galphot task in the two different bands for each galaxy from \citet{denBrok2011} is used to re-plot the color profiles to determine the same parameters for our Coma sample. A color-log(r) line is fitted to the color profile of the galaxies in Coma to calculate the excess light in the same way as in Fornax and Virgo. A similar color-magnitude relation is used for the Coma cluster, Table~\ref{Tab:CMR}. Since the two filters which have been used in \cite{denBrok2011} for Coma cluster are F475W and F814W, we used formula~\ref{eq:transfor color} to transform$(F475W-F814W)$ to $(F475W-F850LP)$ to be compatible to the colors in our Fornax and Virgo samples. The core size, excess light from the fitted line and from the CMR and the color gradient are calculated in the same way as in the Fornax and Virgo clusters to make a similar table for Coma; see Table~\ref{Tab:Coma parameters} in the appendix.
\begin{figure}
	\includegraphics[width=\columnwidth,height=5.9cm]{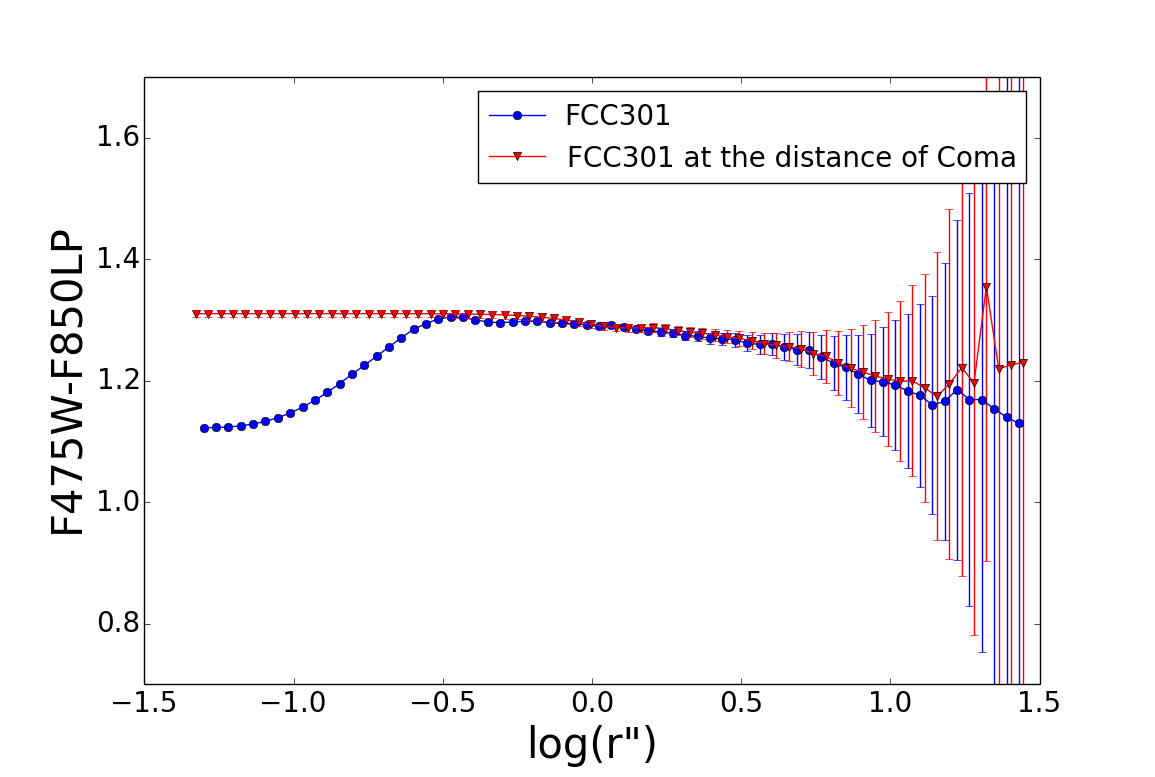}
	\caption{Simulation of the color profile of FCC301 at the distance of Coma in red, together with its original color profile in blue.}
	\label{Fig: FCC301 at the distance of Coma}
\end{figure}
\begin{table*}
\caption{Table of Virgo dEs color profile parameters}
\centering
\begin{tabular}{P{.7in}P{.5in}P{.5in}P{.5in}P{.7in}P{.7in}P{.9in}}
		VCC & Blue & Blue & $r_{core}(pc)$ & $X_{outer~fit}$ & $X_{CMR}$ & $\bigtriangledown(g^{\prime}-z^{\prime})$  \\
		number & galaxy & core & & & &  \\
		(1) & (2) & (3) & (4) & (5) & (6) & (7)  \\	
		\hline
		VCC 9     & Y & Y & 21 & -0.11 & -0.34 & -0.024 $\pm$ 0.005 \\
		VCC 21    & Y & Y & 34 & -0.06 & -0.32 & 0.187 $\pm$ 0.010 \\
		VCC 33    & Y & Y & 25 & -0.07 & -0.44 & 0.022 $\pm$ 0.005 \\
		VCC 140   & N & Y & 21 & -0.05 & -0.06 & -0.008 $\pm$ 0.002 \\
		VCC 200   & N & Y & 34 & -0.04 & 0.10 & -0.035 $\pm$ 0.003 \\
		VCC 437   & N & Y & 34 & -0.12 & -0.03 & 0.053 $\pm$ 0.006 \\
		VCC 543   & N & Y & 34 & -0.03 & -0.04 & -0.019 $\pm$ 0.005 \\
		VCC 571   & Y & Y & 41 & -0.20 & -0.07 & -0.045 $\pm$ 0.026  \\
		VCC 698   & N & Y & 80 & -0.04 & -0.02 & -0.037 $\pm$ 0.003  \\
		VCC 751   & N & Y & 34 & -0.03 & 0.12 & -0.005 $\pm$ 0.004 \\
		VCC 778   & Y &  Y &  11 & -0.03 & -0.53 & 0.032 $\pm$ 0.002 \\
		VCC 828   & Y & N  &  21 & 0.10 & -0.51  & 0.047 $\pm$ 0.003 \\
		VCC 856   & N & Y & 31 & -0.07 & -0.03 & 0.006 $\pm$ 0.002  \\
		VCC 1025  & N & Y & 21 & -0.02 & 0.00 & -0.123 $\pm$ 0.002 \\
		VCC 1049  & Y & Y & n  & n & -0.09 & n  \\
		VCC 1075  & N & Y & 28 & -0.18 & 0.07 & 0.006 $\pm$ 0.006 \\
		VCC 1087  & N & Y & 34 & -0.04 & -0.01 & -0.043 $\pm$ 0.003  \\
		VCC 1125  & N & Y & 106 & -0.02 & -0.02 & 0.002 $\pm$ 0.005  \\
		VCC 1146  & Y & Y  &  13 & -0.09 & -0.39 & -0.002 $\pm$ 0.011 \\
		VCC 1178  & N & Y & 28 & -0.07 & 0.08 & -0.030 $\pm$ 0.003  \\
		VCC 1192  & N & Y & 49 & -0.07 & 0.28 & -0.005 $\pm$ 0.006  \\
		VCC 1261  & Y & Y & 21 & -0.08 & -0.16 & -0.010 $\pm$ 0.002 \\
		VCC 1283  & N & Y & 21 & -0.06 & 0.04 & -0.056 $\pm$ 0.003  \\
		VCC 1297  & N & Y & 34 & -0.04 & 0.28 & -0.056 $\pm$ 0.006 \\
		VCC 1303  & N & Y & 28 & -0.09 & 0.01 & -0.102 $\pm$ 0.005  \\
		VCC 1327  & N & N &  n & n & -0.15 & n  \\
		VCC 1355  & N & Y &  41 & -0.17 & 0.01 & -0.044 $\pm$ 0.006  \\
		VCC 1407  & N & Y &  41 & -0.06 & 0.07 & 0.024 $\pm$ 0.005 \\
		VCC 1422  & N & Y & 21 & -0.07 & -0.10 & -0.032 $\pm$ 0.003  \\
		VCC 1431  & N & Y & 128 & -0.07 & 0.13 & -0.058 $\pm$ 0.007 \\
		VCC 1440  & N & Y &  28 & -0.04 & 0.09 & -0.109 $\pm$ 0.003 \\
		VCC 1475  & N & Y & 25 & -0.03 & -0.04 & -0.046 $\pm$ 0.003 \\
		VCC 1488  & Y & Y & n & n & -0.26 & n\\
		VCC 1499  & Y & Y &  n & n & -0.60 & n \\
		VCC 1528  & N & Y & 34 & -0.05 & 0.08 & -0.033 $\pm$ 0.003  \\
		VCC 1537  & Y & Y &  21 & -0.04 & -0.53 & 0.035 $\pm$ 0.004 \\
		VCC 1539  & N & n & n & n & 0.05 & 0.113 $\pm$ 0.005 \\
		VCC 1545  & N & Y & 34 & -0.04 & 0.16 & -0.121 $\pm$ 0.005 \\
		VCC 1627  & N & Y &  28 & -0.04 & 0.29 & -0.068 $\pm$ 0.002\\
		VCC 1630  & Y  & Y  &  21 & -0.15 & -0.57 & 0.079 $\pm$ 0.008 \\
		VCC 1661  & N & Y &  54 & -0.12 & 0.06 & 0.062 $\pm$ 0.006 \\
		VCC 1695  & Y & N & 19 & 0.05 & -0.13 & 0.030 $\pm$ 0.003 \\
		VCC 1779  & Y & Y & n & n & -0.22 & n \\
		VCC 1828  & N & Y &  49 & -0.08 & 0.10 & -0.041 $\pm$ 0.006 \\
		VCC 1833  & Y & n & n & n & -0.02 & 0.047 $\pm$ 0.001  \\
		VCC 1857  & Y & n & n & n & -0.20 & -0.001 $\pm$ 0.003 \\
		VCC 1861  & N & Y &  66 & -0.10 & 0.06 & -0.005 $\pm$ 0.006 \\
		VCC 1871  & N & Y & 28 & -0.09 & 0.19 & -0.040 $\pm$ 0.002  \\
		VCC 1895  & Y & Y & 21 & -0.03 & -0.02 & 0.015 $\pm$ 0.003 \\
		VCC 1910  & N & Y & 34 & -0.13 & 0.14 & 0.017 $\pm$ 0.004 \\
		VCC 1913  & N & Y & 34 & -0.04 & 0.02 & -0.104 $\pm$ 0.003 \\
		VCC 2019  & N & Y & 34 & -0.07 & -0.02 & -0.026 $\pm$ 0.004 \\
		VCC 2048  & Y & Y & 23 & -0.03 & -0.09 & 0.002 $\pm$ 0.003 \\
		VCC 2050  & N & n & n  & n & 0.10 & -0.035 $\pm$ 0.004
	\end{tabular}
\tablefoot{Parameters of the color profiles of the dEs in Virgo. For the definition of the columns, see Table \ref{Tab: Fornax parameters}}
\label{Tab: Virgo parameters}	
\end{table*}

\begin{figure*}
	\includegraphics[width=17.9cm,height=8.9cm]{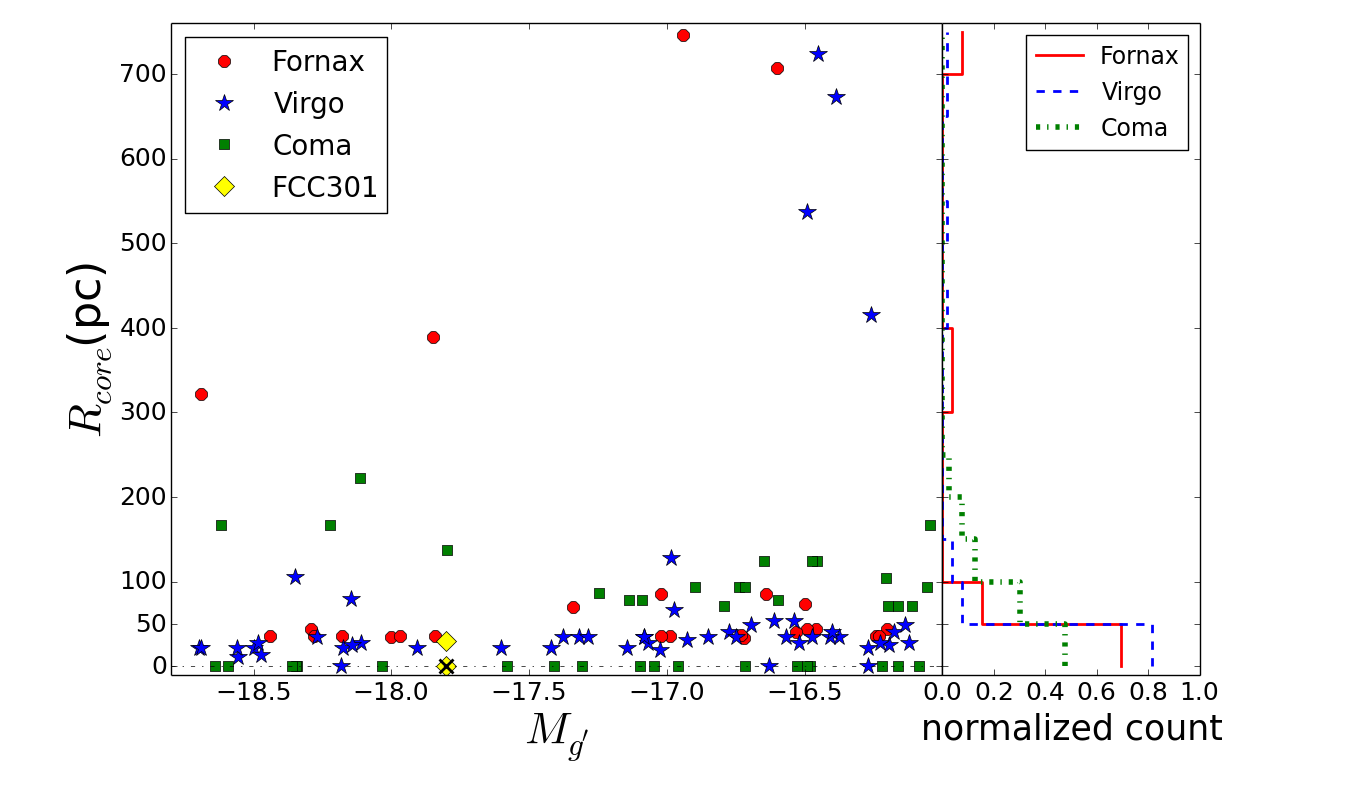}
	\caption{Core size versus magnitude in the Fornax, Virgo and Coma clusters. The yellow markers show the change of the core size for FCC301 before and after (with 'x' sign on it) putting it at the distance of Coma. The zero core size could be a non-detectable core or a galaxy without a core.}
	\label{Fig: Center size}
\end{figure*}

\subsection{Presence of blue centers in the galaxies}
\label{Blue center of early-type dwarf galaxies}
\par In this paper, we used both the \textit{blue core} and \textit{blue galaxy} definitions as follows: if the core of a galaxy, as separated by the black dashed vertical line, is below the red fitted line in its color profile, it has a \textit{blue core} and when the whole color profile of a galaxy is below the green color-magnitude relation line, it is a \textit{blue galaxy}. The definition of a blue center in \citet{Pak} and \citet{Lisker2007} could be a blue galaxy or just a blue center in our study since they do not separate the center as we do. In Figure~\ref{Fig:color profile of Fornax} all galaxies with a fitted line have a blue core, sometimes a very small one, however they are not necessarily blue galaxies. These blue centers are confirmed by aperture photometry and for some of them it is clear in their color and unsharp masked images.

\par As it can be found in the color profiles and the table of the galaxies in Virgo, there are two galaxies with a red center: VCC828 and VCC1695. Most likely they have a dusty center or as \cite{Ferrarese2006} mentioned for VCC1695, for which spiral structure and dust is visible in their color images. There are some completely blue galaxies for which we did not fit a line, just like in the Fornax sample: VCC1049, VCC1488, VCC1499, VCC1779 and some galaxies which do not have any distinct and distinguishable center: VCC1539, VCC1833, VCC1857, VCC2050. Apart from the mentioned galaxies and VCC1327 that we did not fit a line to it, all 43 out of 54 other galaxies have a blue center. Studying color profiles with SDSS data, \citet{Lisker2006} classified VCC21, VCC1779, VCC1488, VCC1499 as early-type dwarf galaxies with blue centers which are also very blue ones in our color profiles of the Virgo galaxies.

\par In our sample of the Coma cluster, 19 galaxies do not have any detectable core $(\sim47\% \pm 3\%)$. It could be that because of the distance to Coma a small core exists but is not detectable, or these galaxies do not have a core at all. There are no galaxies with a red core in our Coma sample. Finally, 29 out of 42 early-type dwarfs have a blue center in our sample for this cluster.
\begin{figure*}
\includegraphics[width=16.9cm,height=9.9cm]{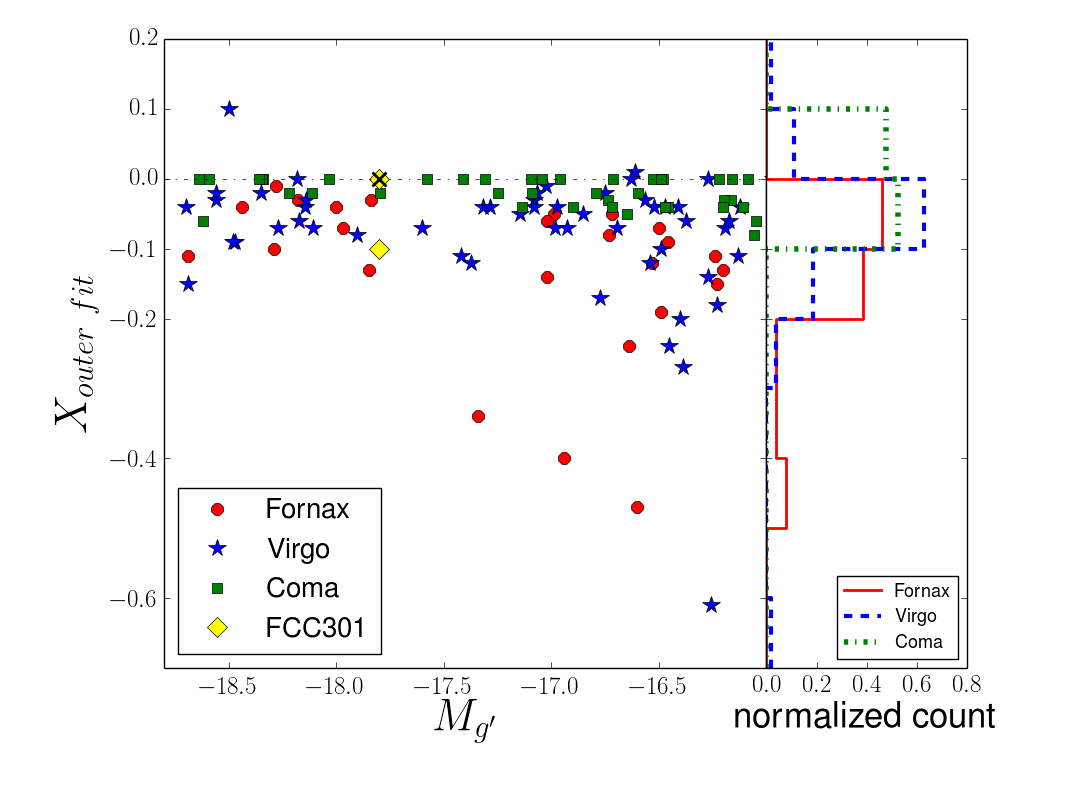}
\caption{Excess from the fitted red line $(X_{outer~fit})$ versus magnitude in the Fornax, Virgo and Coma clusters. As the yellow makers show for FCC301, the excess light drops to zero after putting it at the distance of Coma (the one at the distance of Coma has a 'x' sign on it). A negative (positive) excess is driven by a bluer (redder) center than the galaxy. The very large core sizes are galaxies with extended blue cores.}
\label{Fig: excess from fitted line}
\end{figure*}

\subsection{Comparison between the clusters}
\label{subsection:Comparison of the Clusters}
\par The three clusters that are used in this study, Fornax, Virgo and Coma, have distinct characteristics. Comparing Fornax and Virgo with the Coma cluster, one should keep in mind that Fornax and Virgo with distances of 16.5 and 20.0 Mpc \citep{Blakeslee2009}, are almost at the same distance when compared to the Coma cluster, which is at 100 Mpc  \citep{Carter2008}. To understand how the distance is affecting our parameters in Coma, we simulated the impact of distance on the Fornax and Virgo images, i.e. what would happen if we put a galaxy from Fornax or Virgo at the distance of Coma. As an example we chose one galaxy from the Fornax sample; FCC301. This galaxy is convolved with a Gaussian function to see the effect of distance on our method to determine the blue center. The core size of the galaxy is not detectable in the distance of the Coma cluster, see Figure~\ref{Fig: FCC301 at the distance of Coma}. One should take this into account when comparing the Coma cluster with Fornax and Virgo. We show more results of this convolution in the next figures.
\begin{figure*}
\includegraphics[width=16.9cm,height=8.9cm]{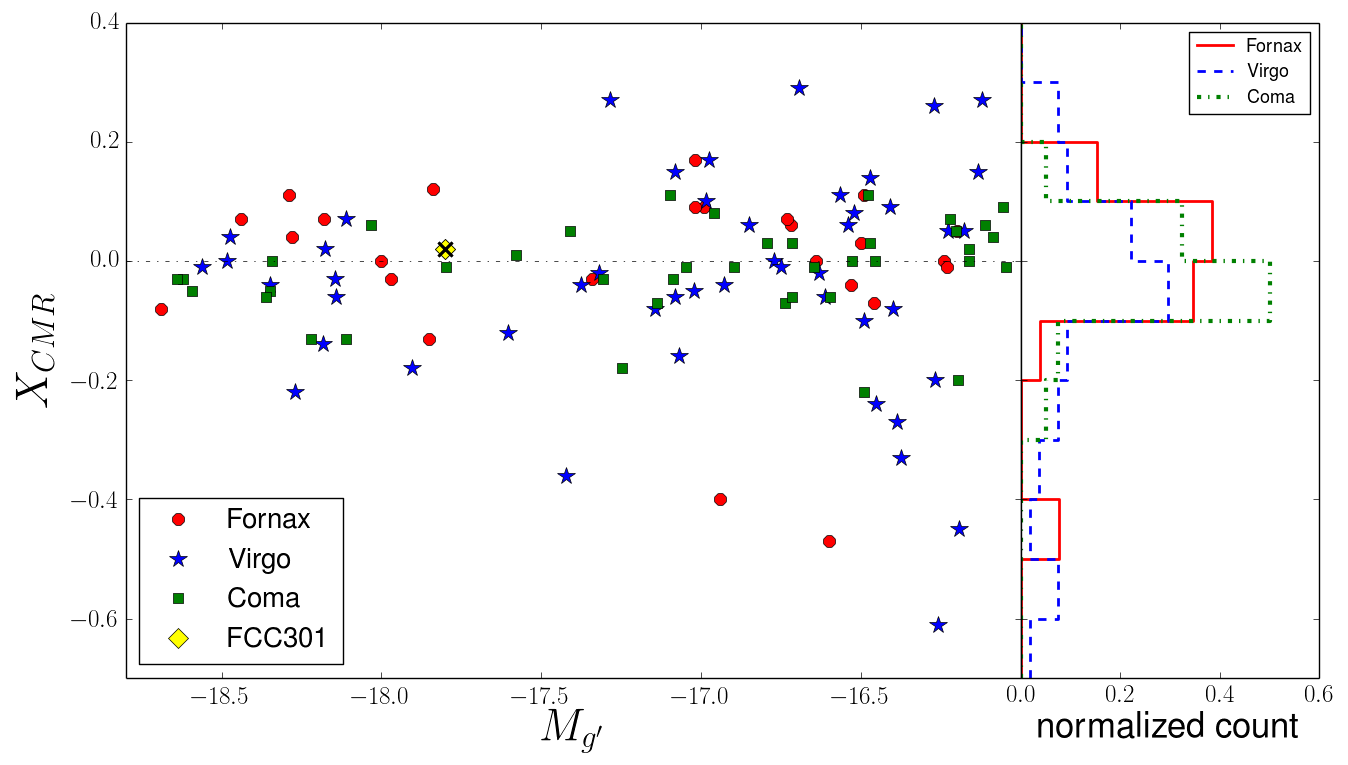}
\caption{Excess light from the CMR line vs magnitude in the Fornax, Virgo and Coma clusters. The yellow point with 'x' sign on it shows FCC301 before and after it was simulated to be at the distance of Coma. $X_{CMR}$ does not change after convolution. The extreme negative points are galaxies with young stellar populations and the most positive ones are dusty galaxies.}
	\label{Fig: excess from CMR}
\end{figure*}
\begin{figure*}
\centering
\includegraphics[width=16.9cm,height=8.9cm]{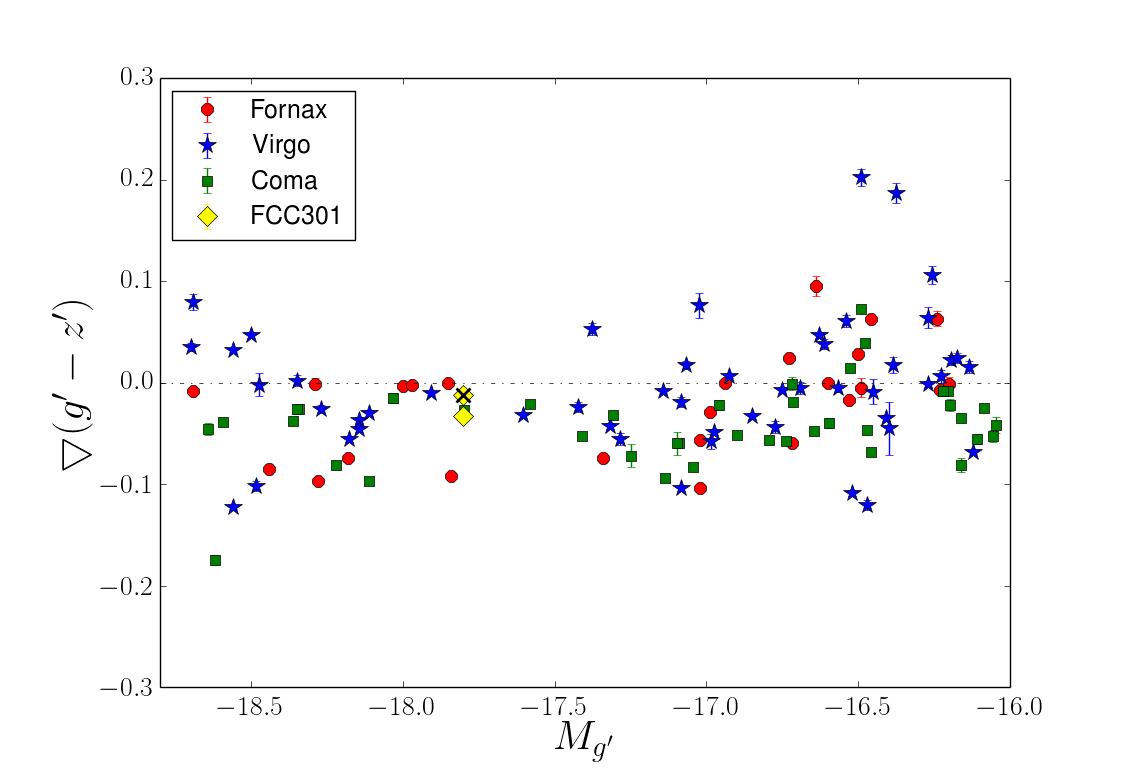}
\caption{Color gradient versus magnitude for the Fornax, Virgo and Coma clusters. The yellow points show FCC301 before and after (with 'x' sign on it) simulated at the Coma distance.}
\label{Fig: color gradient}
\end{figure*}
\par To see the effect of the cluster environment on the calculated parameters, we compared these parameters in the three clusters in Figures~\ref{Fig: Center size}, \ref{Fig: excess from fitted line}, \ref{Fig: excess from CMR} and \ref{Fig: color gradient}. Here is the description of each plot:\\
\textbf{(A)} Figure~\ref{Fig: Center size} shows the size of the blue or red core region in parsec versus the magnitude of its galaxy and its histogram for the three clusters. The estimated errors of the center sizes are about $20\%$. 50 parsecs corresponds to about 11, 12 and 2 pixels at the distance of Fornax, Virgo and Coma respectively. For Fornax and Virgo, most galaxies have a core size between 0 and 50 parsec whereas the core sizes in the Coma cluster peak at values between 50 and 100 parsec. The yellow diamonds are FCC 301 at its true distance and when simulated to be at the distance of the Coma cluster (with a 'x'- sign on it). The core size of the galaxy is smaller at the distance of Coma, and becomes smaller than 50 parsec (see Figure~\ref{Fig: FCC301 at the distance of Coma}).  The larger size of the cores in Coma has been noted before \citep{denBrok2014}, although that study noted that it was not clear if this size difference  was due to the quality of the data, incomplete knowledge of the PSF or actual physical differences. Note that the galaxies located at y=0, the dotted line, are either a galaxy with a small non-detectable core or a galaxy without any specific core. Any special trend is not obvious in this plot. The galaxies which have very large cores are the ones with extended blue regions.\\
\textbf{(B)} Figure~\ref{Fig: excess from fitted line}; The equation used to calculate the excess light, $X_{outer fit}$, is explained in Formula~\ref{Eq:formula of excess from fitted line}. The sign of the excess light from the fitted line to the outer part is an indication of the color of the center; a positive sign refers to a red core and a negative one shows a blue core. Most of the galaxies have a negative excess light, as a result of their blue core. This plot shows that there is no galaxy with a red center in the Fornax and Coma clusters. In Virgo, there are two galaxies with a red core; VCC828 and VCC1695. The peak of excess light from the fitted line in all of these three clusters is in the rage of -0.05 to 0. The excess light of FCC301 becomes zero after moving it to the distance of Coma, as yellow markers show, which is a predictable result due to the Coma's distance. This can explain the smaller excess light of the galaxies in the Coma sample in the same range of magnitude, as the difference between the color of the core and the rest of the galaxy is less distinguishable. In general we see a trend from fainter galaxies with larger and bluer cores to brighter ones with smaller and less blue cores. This result is compatible with \citet{Turner2012} who noted that redder nuclei reside in more luminous hosts in the Fornax cluster and also most of the cores are bluer than their host galaxies. The galaxies with very large excess are always extended, which is clear in their color profiles in Figure~\ref{Fig:color profile of Fornax}.\\
\textbf{(C)} Figure~\ref{Fig: excess from CMR}; this plot is the excess light from the CMR, $X_{CMR}$, of each galaxy versus its magnitude in $g^{\prime}$-band. It is comparable to Figure~\ref{fig:CMD} and \ref{fig:color residual histo} whereas here we can study the distribution of the galaxies regarding to the CMR of each cluster better. The Coma sample is more concentrated around zero and have galaxies in both sides. In comparison, Virgo is more diffuse with both redder and bluer galaxies regarding to the CMR. We emphasis here that the positive and negative signs represent red and blue color of the total profile and not just the core. In this plot, very red galaxies are mostly compact ellipticals and very blue ones are star-forming galaxies as discussed in Section~\ref{Subection:Reasons for the Scatter in the CMR}. Important to note that the excess light value after convolution of FCC 301 is unchanged, so the comparison of the clusters is more reliable in this parameter. It is remarkable that galaxies which are brighter than $M_{g^{\prime}}=-17.5$ are more concentrated around the CMRs. Any trend is not clear in this plot.\\
\textbf{(D)} In Figure~\ref{Fig: color gradient}, the color gradient of Fornax, Virgo and Coma color gradient are plotted. Here the yellow markers show the change from before to after convolution for FCC301. One can notice that the gradients for most galaxies are negative, although for the faintest galaxies many are also positive. Since the young centers have been removed here, this is all about the old stellar populations in the outer part of the galaxies. Virgo and Fornax have more galaxies with positive color gradient, which is a hint that the star-forming regions are extended even excluding the centers. A trend from negative and brighter galaxies to positive and fainter can be distinguish here. The Spearman's rank correlation coefficient for all the galaxies together in this plot is 0.2, which shows a weak correlation. This coefficient is the highest for Fornax 0.49 and lower for Virgo and Coma, 0.06 and 0.20.

\section{Discussion}
\label{section:Discussion}

\par In this paper, we calculated and tabulated the magnitudes, colors and effective radii of the bright early-type dwarfs in the Fornax cluster from high quality HST observation. We then compared their CMD with similar galaxies in Virgo and Coma. Since the observational errors in the HST data are small, we are able to study the scatter in the CMR in detail and to compare the three clusters. We found that the scatter in the CMR of Virgo is considerably larger than in Fornax and Coma. The scatter is due to the presence of the young stars, causing galaxies to lie on the blue side of the CMR, and compact early-type galaxies, which lie on the red side of it. We also found that the CMR of Fornax and Virgo are somewhat bluer than Coma.

\par In addition to colors, we presented the color profiles of dEs in the Fornax and Virgo clusters and we used \cite{denBrok2011}'s color profiles for the Coma cluster. These clusters were observed using ACS on HST, giving us an opportunity to study the color of the innermost parts of the dEs, which is almost impossible from the ground. We parameterized these radial color profiles to study their differences in the three clusters and to find out how the cluster environment can possibly affect them.

\subsection{Outliers and scatter about the CMR}
\par The CMR of the early-type galaxies in nearby clusters and its scatter have been studied for several decades \citep[e.g.][]{Sandage1972,Sandage1978}. For massive galaxies \cite{Schweizer1992} showed that the blueward deviation from the CMR correlates with a so-called fine structure index. It is a morphological indicator of recent mergers, with presence of younger stars, establishing that galaxies evolve towards a tight CMR, which was later called the red sequence \citep{Bell2004}. Our results in Section~\ref{Subection:Reasons for the Scatter in the CMR} confirm the picture of \cite{JanzLisker} and \cite{Roediger2017} that the CMR of Virgo has non-zero scatter. Whereas in this paper we used HST data, which allows us to study the young stellar populations and the extinction in more details.

\par In our sample, all the red outliers are compacts and the projected locations of most of them are very close to a massive companion galaxy. Compact elliptical galaxies (cEs), of which M32 is the 'prototype', have often been thought to be truncated due to their interactions with larger companions \citep[e.g.][]{King1962,Bekki2001,Price2009}. There are not many known cEs. \cite{Chilingarian2009} performed a search using the virtual observatory finding 14 confirmed compact elliptical galaxies, none of which showing any young stellar populations. In an IFU study of compact ellipticals, \cite{Guerou2015} also found that only one of their 8 targets has a stellar population of $\sim3$Gyr, while the rest is older than 6 Gyr. Two of their seven objects in common are clearly redder than the red sequence in our paper. Using a sample of compact stellar system and comparing them with globular clusters, giant and dwarf galaxies, \cite{Janz2016} claimed that cEs are stripped massive galaxies, considering the fact that they have similar metallicities to dwarf nuclei and the center of massive early types. They also reported that the outliers in the mass-metallicity relation are cEs \citep[see][Figure 5]{Janz2016}, similar to what we showed here for red outlier compact dwarfs. Although the above scenario seems to explain the way the cEs form in Virgo, it is not the only formation theory.

\par All the red outliers in Virgo are compact early-types, based on their effective radii and are close companions to massive neighbors. Given the number of the red outlier cEs in Virgo cluster, one could estimate similar amount for the Coma and Fornax clusters. If the probability of finding red outlier cEs is the same in all three clusters, we would expect to have $2\pm1$ in Coma and $1\pm1$ in Fornax. The probability of finding zero red outliers in Coma is $10\%$ and for Fornax, it is $23\%$. As discussed in Section~\ref{Subection:Reasons for the Scatter in the CMR}, we did not find any in Coma and Fornax. Different statistics of compact dwarf galaxies in these clusters can provide a hint to their formation mechanism. One should keep in mind that Ultra Compact Dwarfs (UCDs) are not considered here, since they fall outside our magnitude range \citep{DrinkwaterJones2000}.

\subsection{Blue-cored dEs}
\label{Blue-cored dEs}
\par The most important result of this paper is the large fraction of dEs with young stellar populations in their centers. As discussed in Section~\ref{Blue center of early-type dwarf galaxies}, the number of blue-cored dEs changes from cluster to cluster. In Fornax, all the galaxies in our magnitude range have a bluer center than their host galaxies. In Virgo it is about $85\% \pm 2\%$. In the Coma Cluster, a minimum of $53\% \pm 3\%$ of the sample have a clear blue core. This is a lower limit for Coma cluster due to its distance as explained in Section \ref{subsection:Comparison of the Clusters}. Some of these galaxies contain a considerable amount of gas which has not been blown away such as FCC335 and FCC90 which \cite{Zabel} reported to detect the molecular CO. The galaxies which have a straight color profile, i.e. without a core, and mostly in the Coma sample, could be galaxies with a small non-detectable core or a galaxy without any bluer or redder center, see Figure~\ref{Fig: FCC301 at the distance of Coma} .

\par For all three clusters most galaxies have an excess light from the fitted line, $X_{outer fit}$, in the interval [-0.05,0], as can be seen in Figure~\ref{Fig: excess from fitted line}. There are just a few galaxies with redder centers in the Virgo cluster and none in the other two clusters. We showed that the bluest cores are find in fainter galaxies. This result is compatible with the previous studies on the nuclei of dEs, except that there generally no distinction has been made between the core, i.e. the blue region, and the nucleus, the region with an enhanced surface brightness profile. \cite{Lotz2004}, for the Leo Group, Fornax and Virgo clusters $(-12<M_{B}<-18)$, showed that the nuclei of dEs are bluer than their residing galaxies, $\delta (V - I ) = 0.1-0.15$ mag, and that redder and brighter dEs have redder nuclei. They failed to find any correlation between nuclear properties such as magnitude and color and projected distance of the galaxy from the center. We also did not find any trend between the excess light from the fitted line versus the projected distance of each galaxy to the cluster center.

\par In the literature, there are several studies who report blue-cored early-type galaxies in various environments, associating them with recent star-formation in these areas. Beside their blue cores, the young stellar populations is confirmed using different tools such as $H\alpha$ emission and UV flux. \citet{Pak} studied a sample of 166 galaxies in the Ursa Major cluster located at 17.4 Mpc \citep{Tully2012} (between those of Virgo and Fornax), with ground-based SDSS Data and data from the GALEX satellite. Their sample is chosen in the magnitude range of $-21.5 < M_{r} < -13.5$. In this range 16 galaxies, out of 23 dEs, i.e., $(\sim 70 \%)$ showed a blue core based on their radial NUV-r color profiles. \citet{Pak} claimed that since the UV flux is particularly sensitive to the presence of young stars (<1 Gyr), the blue UV-optical colors of the blue-cored early-type dwarfs indicate that these galaxies have experienced recent or ongoing star formation activities in their central regions. To confirm the recent star-formation, they looked at the $H\alpha$ emission of these galaxies  which is a good indicator of O- and B-type stars with an age less than a few million years old \citep{Kennicutt1998}. They reported that about $75\%$ of them have $H\alpha$ emission with equivalent width of more than $2\AA$. 

\par In the Fornax cluster \citet{DeRijcke}, identified an offset blue core in FCC46, with $H_{\alpha}$ and H\,{\sc i} emission indicating ongoing star formation. In the Virgo cluster, in \cite{Lisker2006} sample the galaxies that overlap with this paper, we also found blue cores (Section~\ref{Blue center of early-type dwarf galaxies}). The difference between the fractions of galaxies with blue cores is purely caused by a difference in resolution, and the fact that that the PSF of HST is so well known. Similar studies have confirmed ongoing star formation by $H_{\alpha}$ imaging \citep{Boselli2008}.

\subsubsection{Size of the blue cores and comparison to nuclear clusters}
\label{subsubsection:Nucleus or blue/red core}
\par Our method to find the blue or red cores in this paper is different from what has been used to detect the nuclei of early-type galaxies in HST studies of Fornax, Virgo and Coma \citep{Turner2012,Cote2006,Lotz2004,denBrok2014}. Their detected nuclei are the excess light in the center of the galaxy when fitting a S\'ersic function to the surface brightness profile. Here, on the other hand, we used the color contrast to define the inner blue regions. The effective radii of the nuclei, determined from the surface brightness profiles, are normally much smaller than our blue cores, indicating that they are not measuring the same thing.
\par
\begin{figure}
\centering
\includegraphics[width=0.5\textwidth,height=5.9cm]{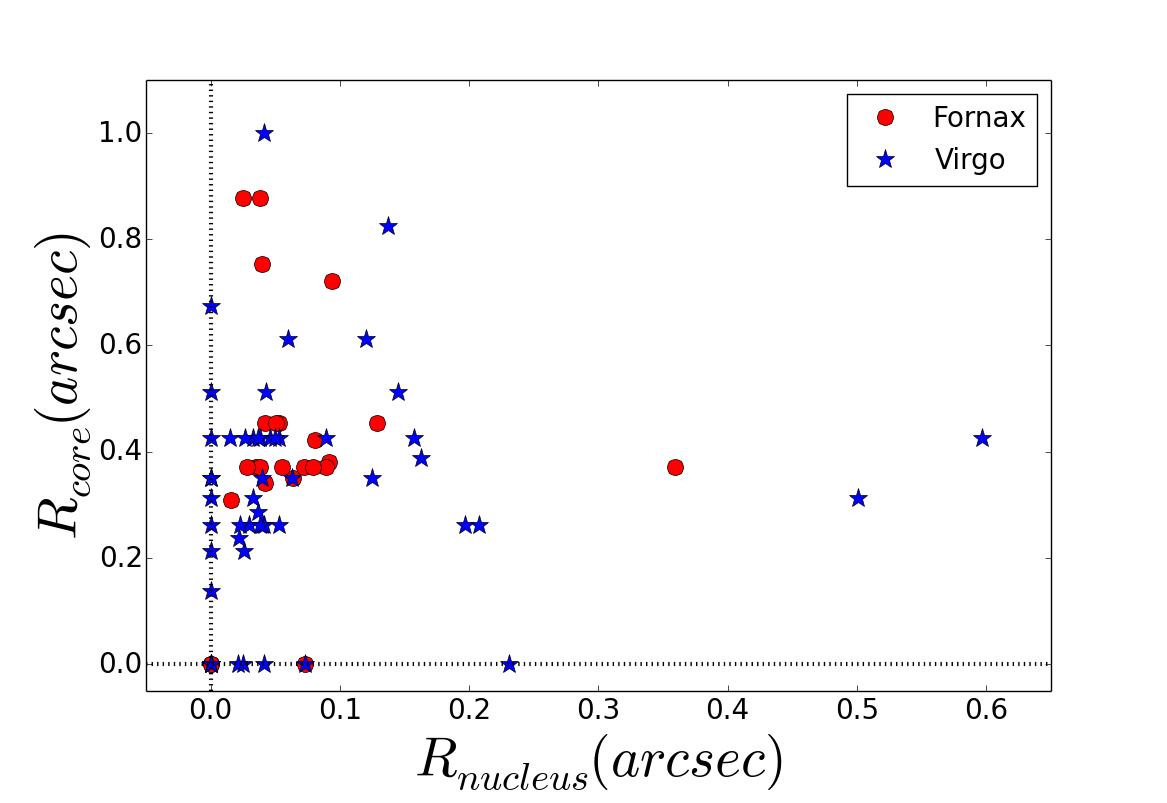}
\caption{Effective radii of the nuclei in the g-band versus the radii of the cores of the dEs in Fornax and Virgo. The effective radii of the nuclei in Fornax have been taken from \protect\cite{Turner2012} and the ones in Virgo were taken from \protect\cite{Cote2006}. The core sizes are tabulated in Table~\ref{Tab: Fornax parameters} and \ref{Tab: Virgo parameters}. }
\label{Fig: nuclei vs core}
\end{figure}

 Figure~\ref{Fig: nuclei vs core} shows the measured core size of the early-types in this paper versus the effective radii of the nuclei of the same galaxies from \cite{Cote2006} (Virgo) and \cite{Turner2012} (Fornax). As it is seen in this image, our blue cores are generally larger than the nuclei. The two exceptions in Virgo, which have much larger nuclei than blue core radii are VCC1630 and VCC1913. Galaxies without a detectable core or nucleus are shown as upper limits on the x=0 or y=0 lines. Despite the fact that the nuclear clusters do not correspond one-to-one with the blue core regions, the literature papers have shown that the nuclear clusters mostly are bluer than their host galaxies. They also noted a (weak) correlation between the color of the nucleus and the luminosity of its galaxy in a way that bluer nuclei reside in fainter galaxies. This is comparable to the trend that we discussed in Fig~\ref{Fig: excess from fitted line}; fainter dEs on average show a larger blue excess.

\par We found that the size of the blue/red core of the galaxy does not correlate with the luminosity of the host galaxy, Figure~\ref{Fig: Center size}. Apart from the fact that the fraction of galaxies with blue or red cores differs from cluster to cluster, the properties of the cores vary so much from cluster to cluster that no general trends are seen. The only clear conclusion here is that no large core radii ($>$ 300 pc) are seen in the Coma cluster.

\subsection{Color gradients}

\par The excellent spatial resolution also made it possible to determine accurate radial color gradients. We excluded the centers while calculating it, similar to \cite{denBrok2011}, since the color of the center of dE is different from the rest of the galaxy. The statement that brighter galaxies mostly have negative gradient observed by \cite{denBrok2011}, after removing the blue cores, is seen in Figure~\ref{Fig: color gradient}. The cluster environment can affect the color gradient by changing the galaxy's star formation rate with various interactions that can happen. There are more galaxies with positive color gradients in Fornax and Virgo, whereas the color gradients of the galaxies in Coma are mostly negative. These positive gradients, common in fainter galaxies, can mean that our simple recipe, which assumes that the region containing young stellar populations is centrally concentrated, and that further out one sees a metallicity gradient in the old stellar populations, is not valid here any more. Note, however, that such galaxies are seen much more in Virgo and Fornax than in Coma. In addition, \cite{denBrok2011} did not find any relation of the gradient with cluster centric radius. COMAi13005.684p275535.20, is one of the galaxies which has positive gradient in their sample and they explained it as a transition dwarf early-type with a spiral structure and central disc. This galaxy is shown as the only blue outlier of the Coma sample in Figure~\ref{fig:CMD}.

\subsection{Comparison between the clusters and possible formation scenarios}
\label{subsection:Comparison between the clusters and Possible formation scenarios}

\par In this paper, we studied three clusters with distinct characteristics. Comparing the cluster's environment, Coma is a relaxed, dense and rich cluster at a distance of 100 Mpc \citep{Carter2008}. It is about a factor 10 more massive than the Virgo cluster. Fornax at a distance of 20 Mpc is six times less massive, smaller and more regular in shape with a shallower potential well compared to Virgo \citep{JordanF2007} (see Table~\ref{Tab:Clusters parameters}). We see several differences in the galaxy properties. Can we understand these purely from the differences in cluster mass?

\par To study the effects of the cluster environment on the formation of galaxies, we used dwarf ellipticals as test objects, since their shallow gravitational potential causes them to interact more effectively with their environment. The most important external processes that can work here are ram pressure stripping (RPS) \citep{Gunn} and harassment \citep{Moore1996}. They can remove the ISM of the galaxy and quench the star formation, in the meantime even re-ignite the central star formation activity \citep{Kronberger2008}. 

\par For the external factors that can affect dEs, \cite{Urich2017} discussed harassment and ram pressure stripping in the context of the formation of blue-cored dEs. They suggested that galaxy harassment is less effective in clusters like Virgo, given their large velocity dispersion, and more efficient in group environments \citep{Yozin2015}. They claimed that the ram pressure stripping is therefore more likely to explain their blue core dEs in the Virgo cluster. According to the model of \cite{Vollmer2009}, ram pressure stripping has a hard time in removing the gas from the center of the galaxies and it can even raise the star formation rate in the cores \citep{Kronberger2008}.

\par We found that the fraction of blue galaxies, measured as outliers from the red sequence on the blue side, is smaller in Coma than in Virgo and Fornax. We saw a similar effect when studying the fraction of galaxies with blue cores, and their sizes. Interesting enough, Fornax does not seem to be very different here from Virgo, except that its CMR is slightly bluer than that of Virgo. If Coma has fewer blue outliers than Virgo, and fewer and smaller blue core regions, then Virgo should have fewer blue outliers etc. than Fornax. Since this is not the case, we probably have to look at the differences in the internal structures of Virgo and Fornax. The galaxy number density in the center of Fornax is twice as high as in the Virgo cluster \citep{Jordan2007}, while the velocity dispersion in the Virgo cluster is two times higher ($\sim 760 km~s^{-1}$) as in Fornax, making the harassment time scales in the Virgo cluster four times longer than in the Fornax cluster \citep{Venhola2018}. On the other hand, the density of the X-ray gas in the center of Virgo \citep{Simionescu2017} is roughly five times higher than in the Fornax cluster \citep{Paolillo2002}, which combined with the two times higher velocities of the Virgo dwarfs makes the ram-pressure in the Virgo cluster $\sim$ 20 times higher than in the Fornax cluster. To this we have to add that in Fornax the ratio of ellipticals versus spirals is larger than in Virgo \citep{Ferguson1988}. Virgo is an active and irregular cluster with a young dynamical nature \citep{Boselli2014,Mei2007,Binggeli1993,Drinkwater2001,Conselice2001}. \cite{Sybilska2017} showed that Virgo has a non symmetric density distribution by studying the number density map of the cluster. 

\par Hence, with harassment being more effective in the central regions of Fornax and ram pressure stripping more effective in Virgo, the effects seem to more or less cancel each other out. However, one thing is very different in these clusters: the fraction of red outliers of the CMR. Since the objects causing this, compact ellipticals, are often caused by galaxy harassment or interactions, our observational result could indeed mean that harassment is much stronger in Virgo, so that bright nuclear dwarfs have a much higher probability to lose their outer parts. It would also mean that we expect more compact ellipticals in Coma, but maybe this is not the case since the cluster is much more relaxed now, and such objects are indistinguishable from stars with HST at this moment. Detailed simulations will have to show whether this hypothesis is true, and observations covering a larger magnitude range will have to make stronger predictions.

\par We have to note that our sample of dEs in Coma used in this study is mostly from the central region of the cluster as the ACS camera failed during the later stage of the survey. Studies showed that the color of a galaxy correlates with the density of its environment; redder galaxies reside in denser environments \citep{Gavazzi2013,BoselliGavazzi2014}. Since the galaxies in the center of the cluster have entered the environment earlier and have gone through more environmental influences, they have less star formation \citep{Sybilska2017,Lisker2013}. This could well be the reason behind fewer blue outliers in Coma in Figure~\ref{fig:CMD}.

\par Investigating the formation of dEs with a blue core, transition-type dwarf galaxies (TTDs) can be important. \cite{DeLooze2013} chose their sample of 36 TTDs based on SDSS spectral in the Virgo cluster. From this sample, 13 of them were detected by the Herschel Space Observatory \citep{Pilbratt2010} of which three galaxies are in common with our sample: VCC571, VCC1488, VCC1499. VCC571 has $H_{\alpha}$ absorption which is a sign of young stellar populations and VCC1488 and VCC1499 have strong $H_{\alpha}$ absorption which shows relatively young $(\sim 1 Gyr)$ stellar populations without current star formation, probably indicating a rapid truncation of the star formation activity. TTDs are spiral galaxies which are loosing their gas entering the cluster environment and turning to quiescent dEs \citep{Boselli2008,Lisker2006}. In this process they can have properties of both early and late-type galaxies. They claimed that several TTDs have blue central regions, which is a hint of outside-in gas removal. Regarding the blue-cored TTDs, they suggested that some of the dEs should be formed by infalling of the field's late-types to the cluster and their transformation to early-types. The authors concluded that interaction with the cluster environment, especially through ram pressure striping, is guiding the evolution of these TTD.

\par What do these new results tell us about the formation of nuclear clusters? Similar to our result, studies on the nuclei of the dEs, profiting of the high resolution of HST, analyzed the parameters of the nuclei and their projected distances from the cluster center. They did not find any evidence that their properties depend on galaxy cluster distance \citep{Turner2012,Cote2004}. In addition, the similarities between the nuclei of the early-type galaxies in our three clusters with distinct properties leads the authors to the outcome that the environment of the clusters does not have a crucial effect on their formation and evolution. Considering these results, one can conclude that the formation of nuclei in dEs is mainly determined by internal and local factors. There are two  main possible scenarios for their formation: Infall of globular clusters or infall of gas to the center of the galaxy. In the first theory, globular clusters around the core experience dynamical friction, merge and spiral to the center. This is not a new idea; \cite{Tremaine1975} suggested that the nucleus of M31, formed in the same way. Its consequence is a lack of or less globular cluster around the center \citep{Capuzzo1999}. Calculating dynamical friction time scale in numerical simulations, \cite{Turner2012} concluded that the nuclei of the galaxies in the range of low to intermediate mass can be formed by this method and it is appropriate for forming the nuclei of non-massive galaxies in their Virgo and Fornax samples \citep[see more in e.g.][]{Lotz2001,Milosav2004}. 

\par The second process that can form the nuclei is gas accretion into the center, which causes star formation in the core of the galaxy \citep{Bergh1986}. There are some theories explaining how gas is transported to the inner parts of the galaxy \citep{Milosav2004,Mihos1994,Hopkins2011}. \cite{Bekki2006} and \cite{Bekki2007} with chemodynamic simulation of inner 1 kpc of dwarf galaxies showed that nuclei are younger and more metal rich than their hosts. Since feedback is more effective in low mass galaxies, the time that gas needs to settle in the center increase for fainter galaxies. This leads to the bluer and younger nuclei in low mass galaxies.

\section{Conclusion}
\par We conclude stating the main results of this paper: \\
$\bullet$ The most important result is the large number of the blue-cored dEs in all the three clusters; all galaxies in the Fornax sample, $85\% \pm 2\%$ in Virgo and a lower limit of $53\% \pm 3\%$ in Coma cluster which is discussed in detail in Section~\ref{Blue center of early-type dwarf galaxies}. Considering the very distinct characteristic of these clusters and their distribution in the cluster, their formation and evolution are influenced by different factors and it can not be explained by a single theory. The possible scenarios are presented in Section \ref{subsection:Comparison between the clusters and Possible formation scenarios}.\\
$\bullet$ We found that in the magnitude interval of $-18.7 \leq M_{g^{\prime}} \leq -16.0$, the scatter in the CMR is the highest in Virgo, 0.144 mag. In Fornax and Coma the scatter is 0.060 and 0.065 mag. The large number of outliers on its blue side shows that the Virgo cluster dwarfs are still quite actively forming stars. The same holds for the Fornax cluster. The scatter on the blue side of CMR is due to galaxies with young stellar populations. This is mainly the case in Fornax and Virgo. Their fractions are $11\% \pm 2\%$, $15\%\pm 3\%$ and $2\% \pm 1\%$of total dwarf early-types in Fornax, Virgo and the central regions of Coma respectively. The galaxies in Fornax and Virgo in general are probably somewhat younger. \\
$\bullet$ The situation with the red outliers is totally different in Virgo, as compared to the other clusters. All the red outliers of the color-magnitude diagram in Virgo cluster are compact ellipticals. Their redder color probably can be explained by tidal interactions with large nearby massive galaxies. There are 3 red outlier compacts in Virgo and zero in the Fornax and Coma clusters. The fact that there are more red outliers in Virgo, a cluster with more interactions going on, and that these are cEs, probably means that the environment is still very active, with many galaxy-galaxy interactions, more than in the other clusters. To fully test this idea, samples with much larger statistics are needed.\\
$\bullet$ A trend that brighter galaxies generally have negative color gradients, which become less negative and even positive for our faintest galaxies can be found in the color gradients of all three clusters. \\
$\bullet$ The bluest cores are found in the faintest galaxies. There is no obvious correlation between the sizes of the blue cores and the luminosity of their host galaxies in any of the clusters.

\section*{Acknowledgements}
\addcontentsline{toc}{section}{Acknowledgements}
RFP acknowledges financial support from the European Union's Horizon 2020 research and innovation programme under Marie Skodowska-Curie grant agreement No 721463 to the SUNDIAL ITN network.
\begin{appendix}
\section{Color profiles of Virgo and Table of Coma}
\label{Appendix}
%
\begin{figure*}
	\centering
	\includegraphics[width=4.3cm,height=3.8cm]{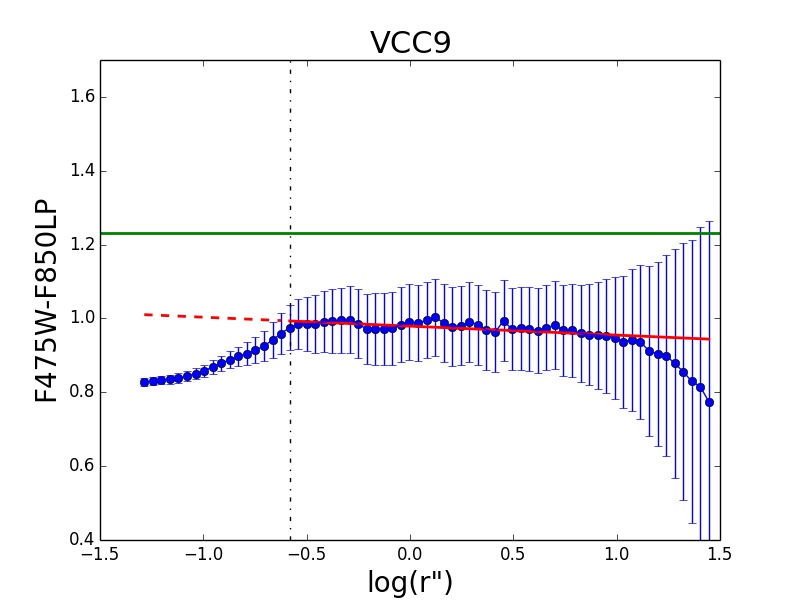}
	\includegraphics[width=4.3cm,height=3.8cm]{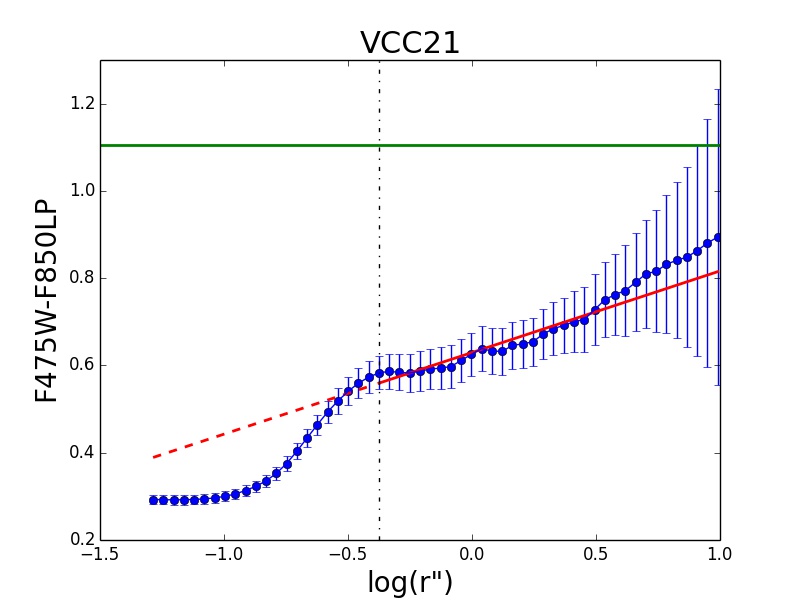}
	\includegraphics[width=4.3cm,height=3.8cm]{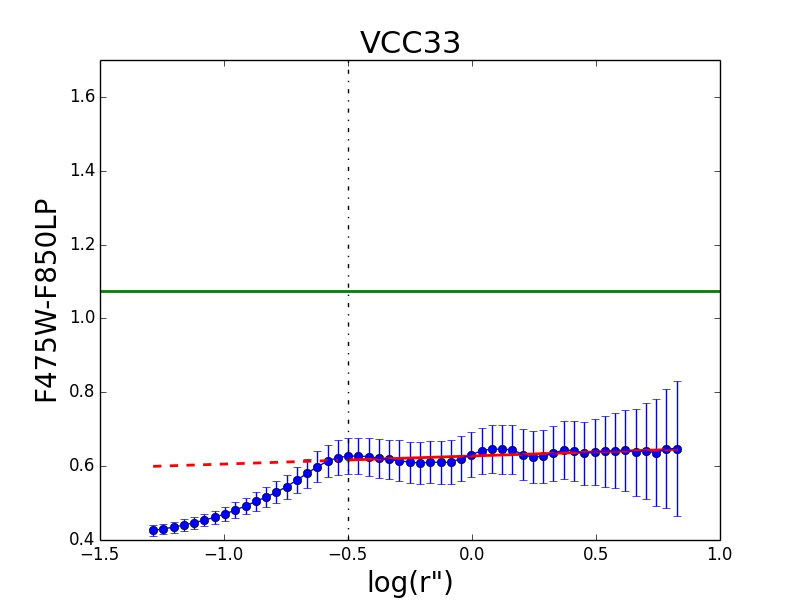}
	\includegraphics[width=4.3cm,height=3.8cm]{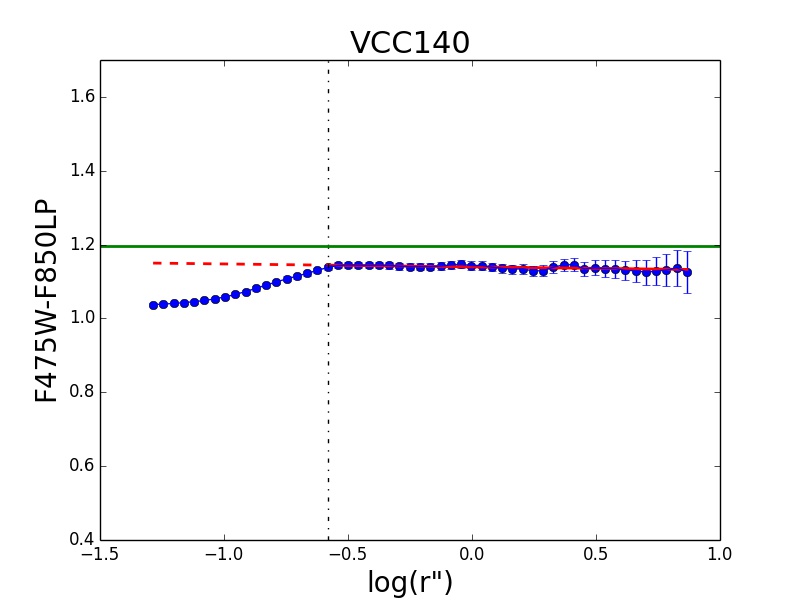}
	\includegraphics[width=4.3cm,height=3.8cm]{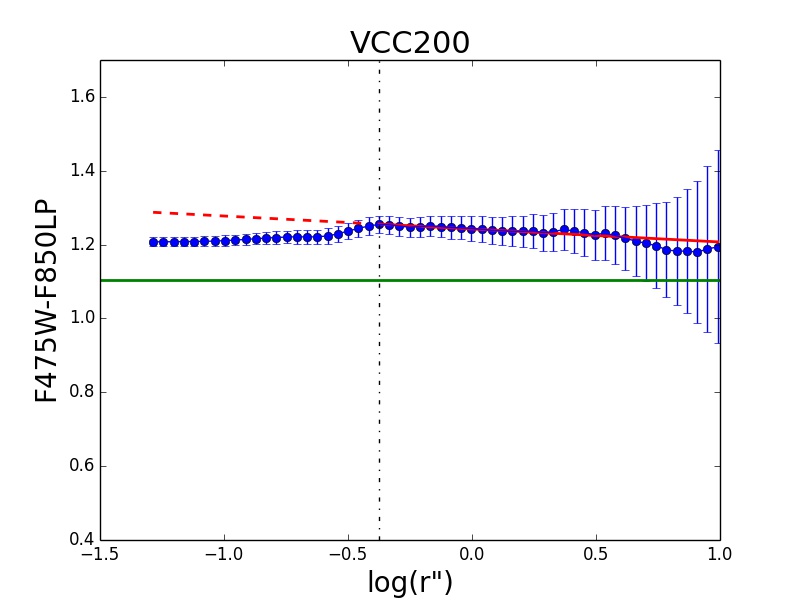}
	\includegraphics[width=4.3cm,height=3.8cm]{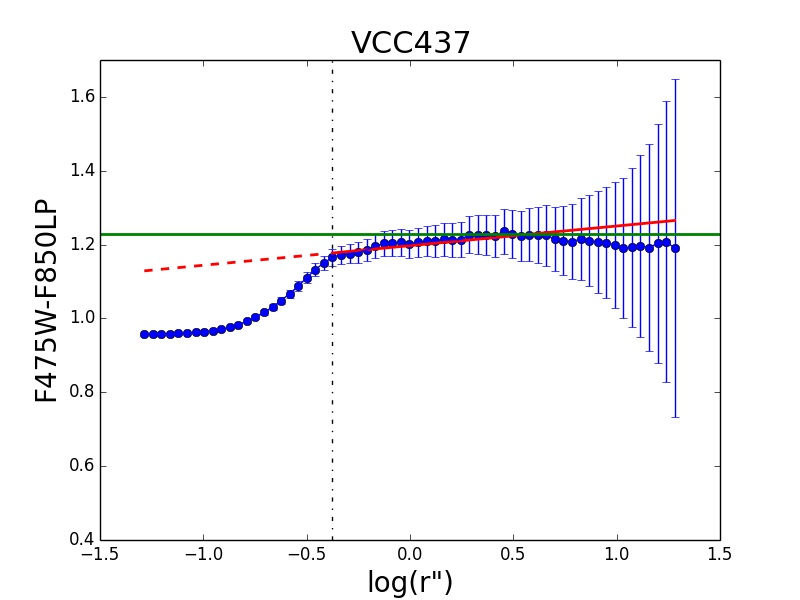}
	\includegraphics[width=4.3cm,height=3.8cm]{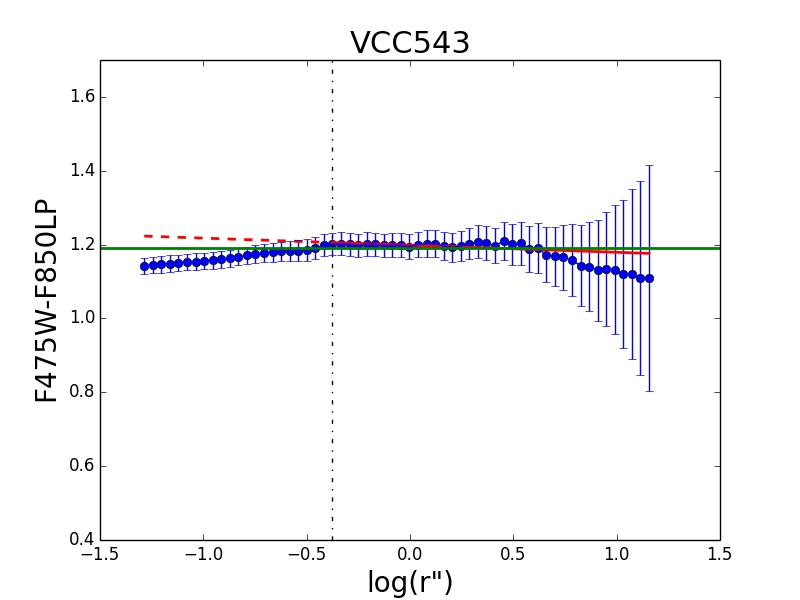}
	\includegraphics[width=4.3cm,height=3.8cm]{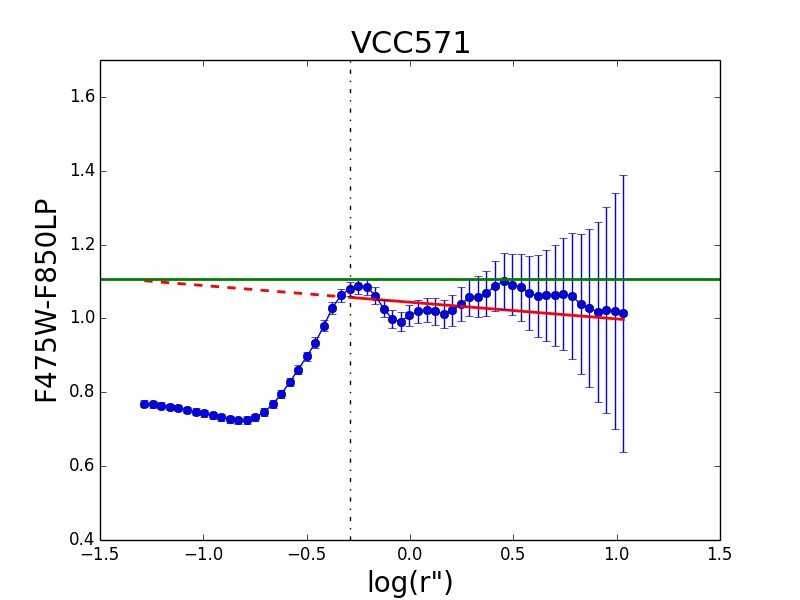}
	\includegraphics[width=4.3cm,height=3.8cm]{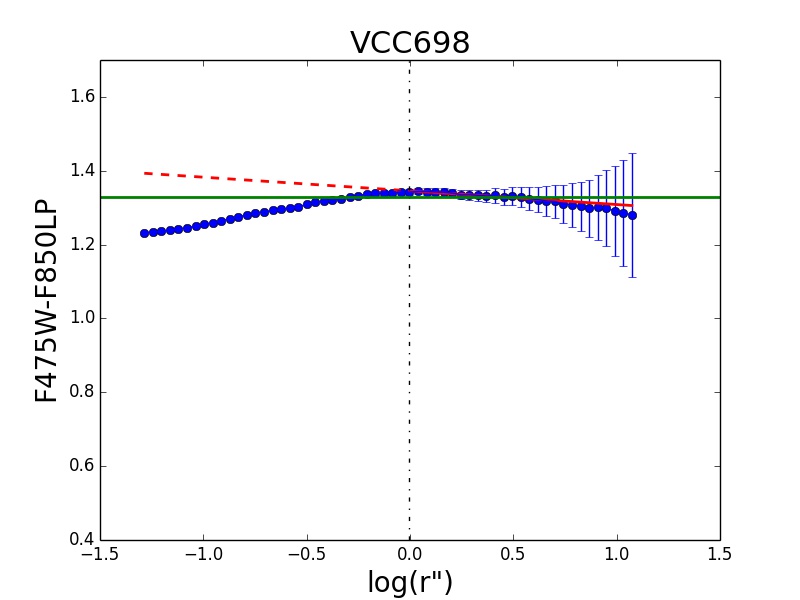}
	\includegraphics[width=4.3cm,height=3.8cm]{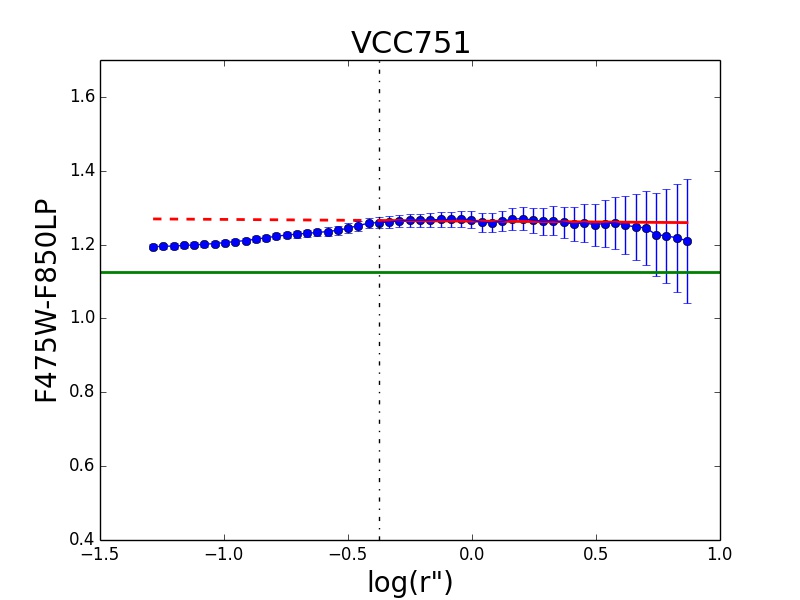}
	\includegraphics[width=4.3cm,height=3.8cm]{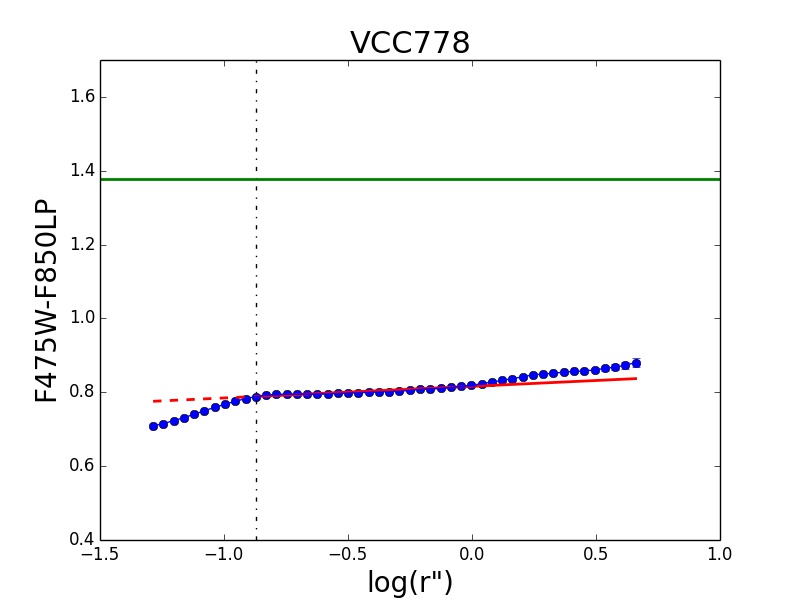}
	\includegraphics[width=4.3cm,height=3.8cm]{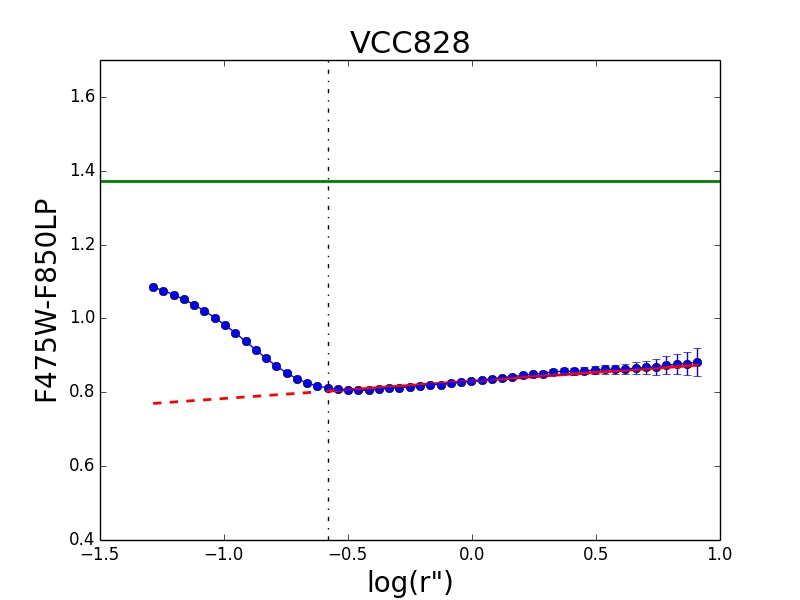}
	\includegraphics[width=4.3cm,height=3.8cm]{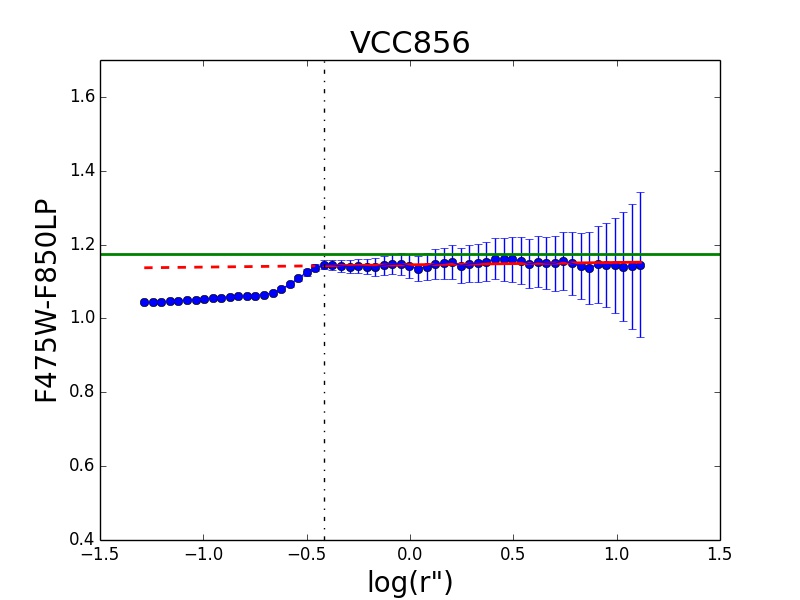}
	\includegraphics[width=4.3cm,height=3.8cm]{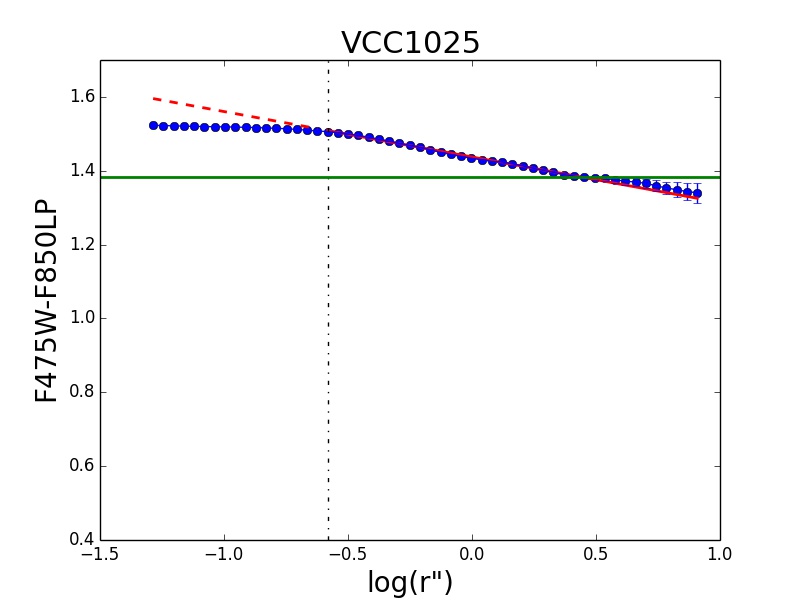}
	\includegraphics[width=4.3cm,height=3.8cm]{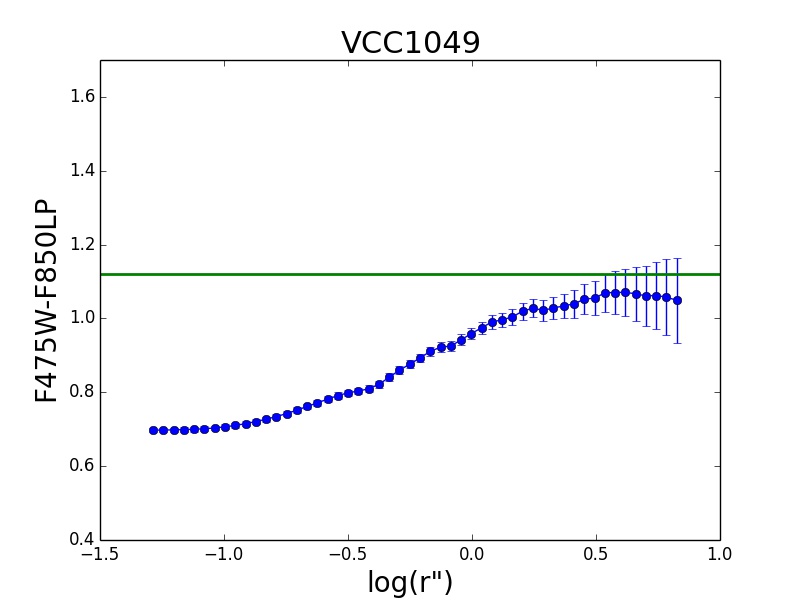}
	\includegraphics[width=4.3cm,height=3.8cm]{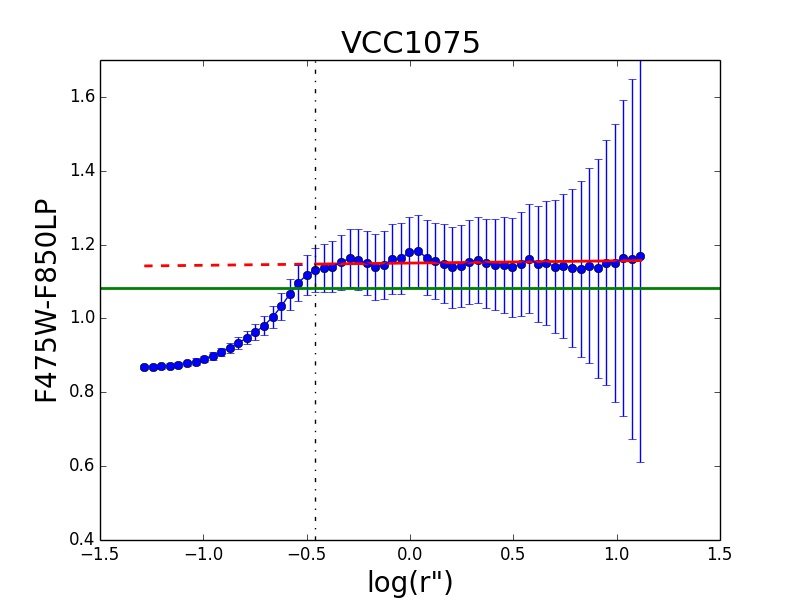}
	\includegraphics[width=4.3cm,height=3.8cm]{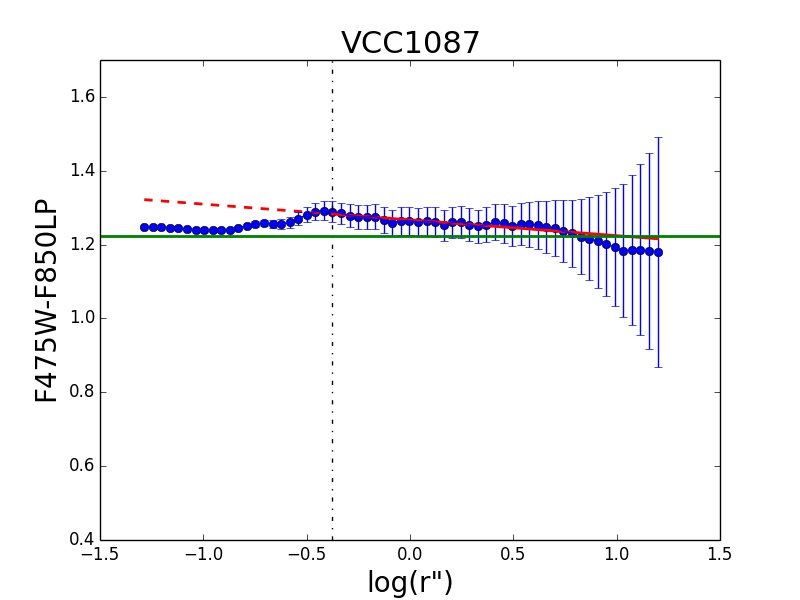}
	\includegraphics[width=4.3cm,height=3.8cm]{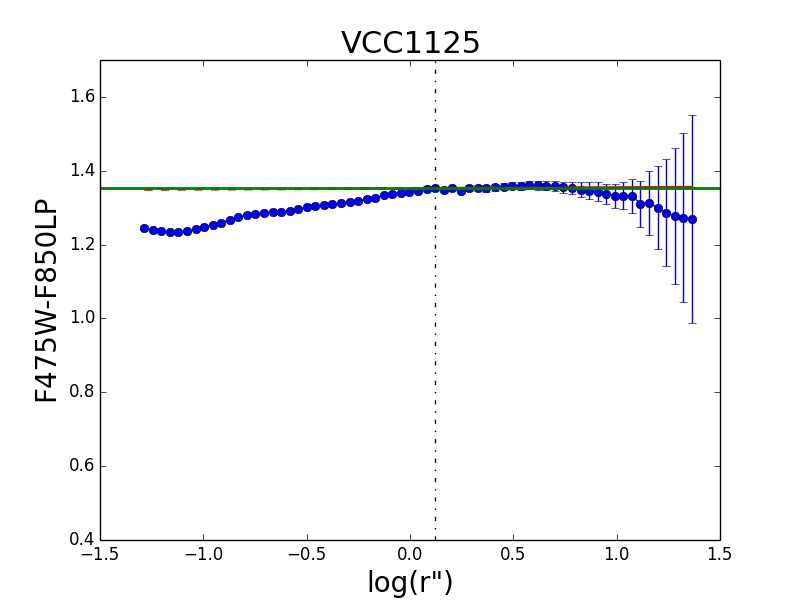}
	\includegraphics[width=4.3cm,height=3.8cm]{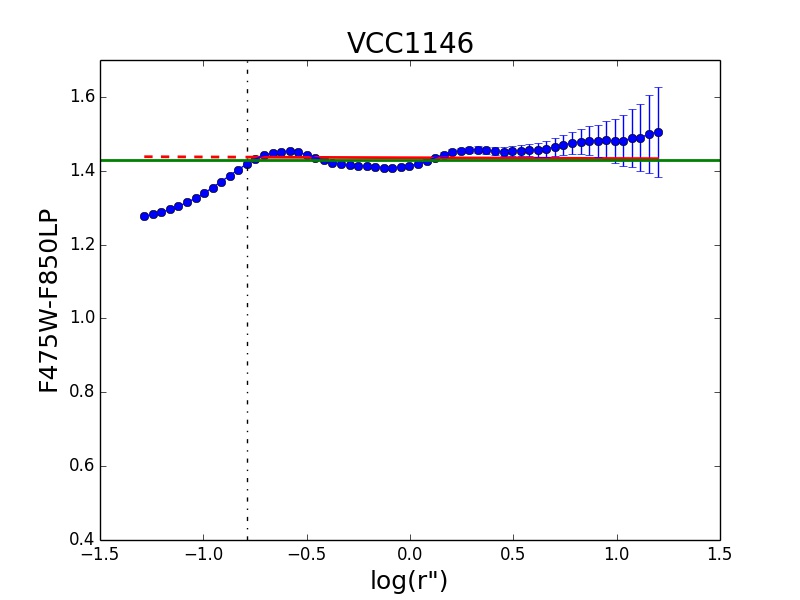}
	\includegraphics[width=4.3cm,height=3.8cm]{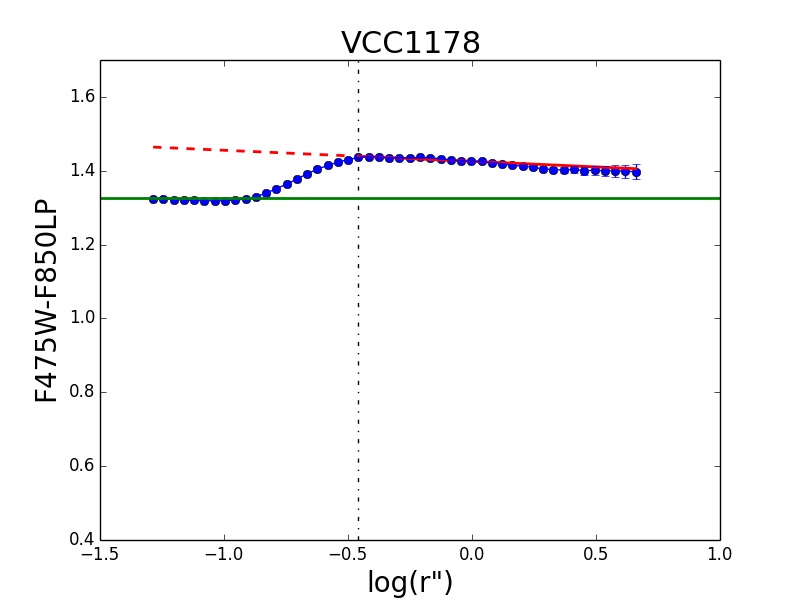}
	\includegraphics[width=4.3cm,height=3.8cm]{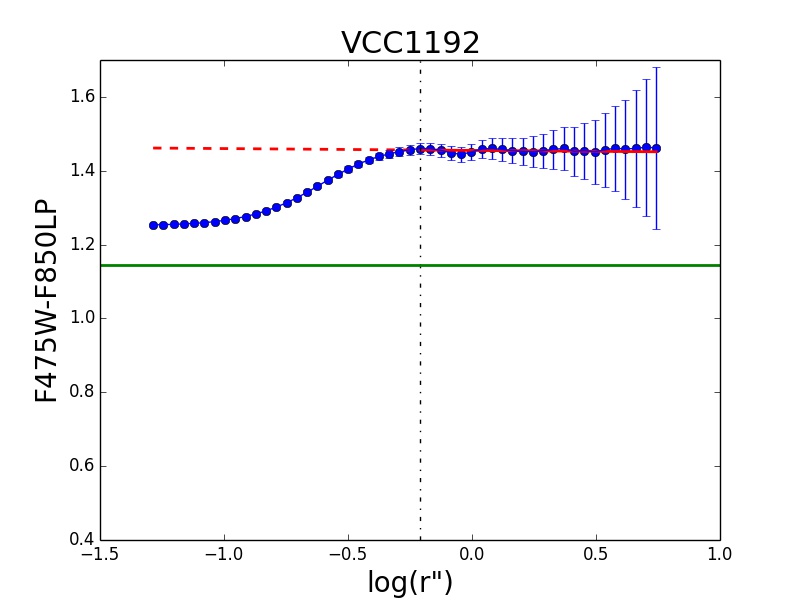}
	\includegraphics[width=4.3cm,height=3.8cm]{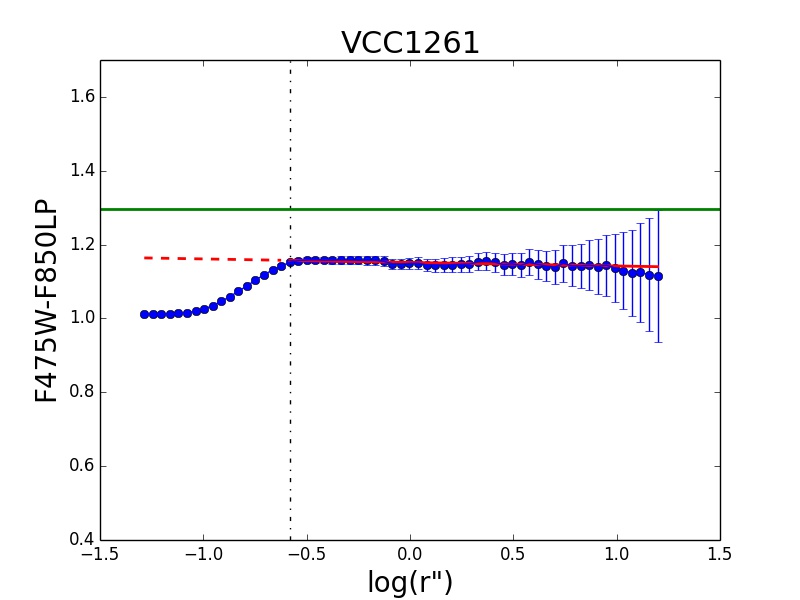}
	\includegraphics[width=4.3cm,height=3.8cm]{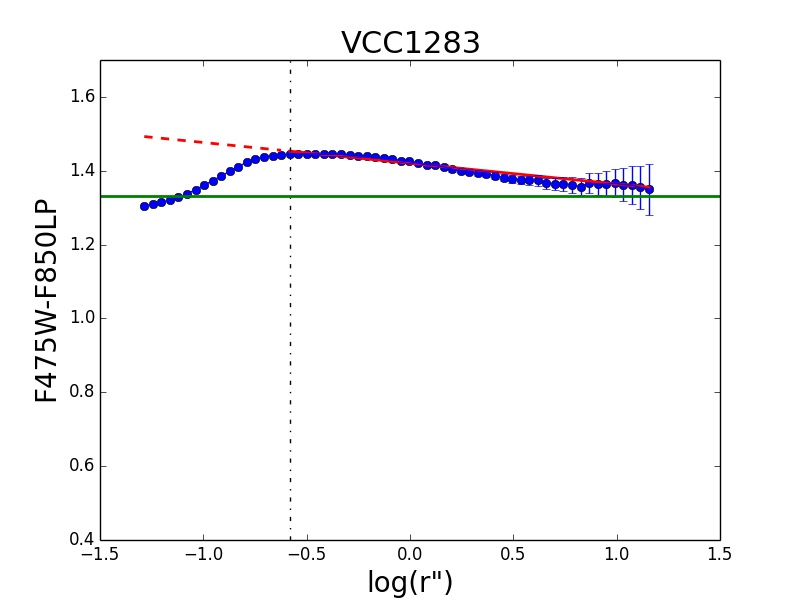}
	\includegraphics[width=4.3cm,height=3.8cm]{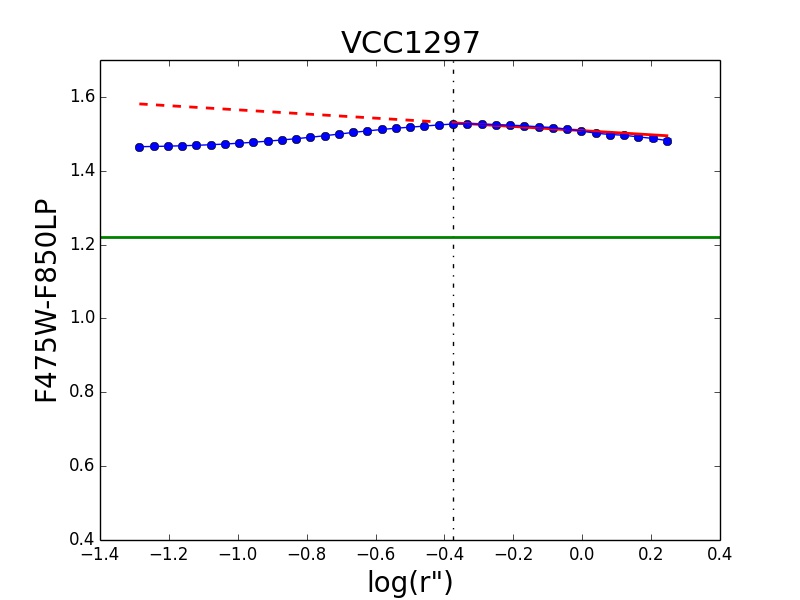}
	\caption{Color profile of Virgo early-type dwarf galaxies}
	\label{Fig:color profile of Virgo}
\end{figure*}
\begin{figure*} 
	\centering
	\includegraphics[width=4.3cm,height=3.8cm]{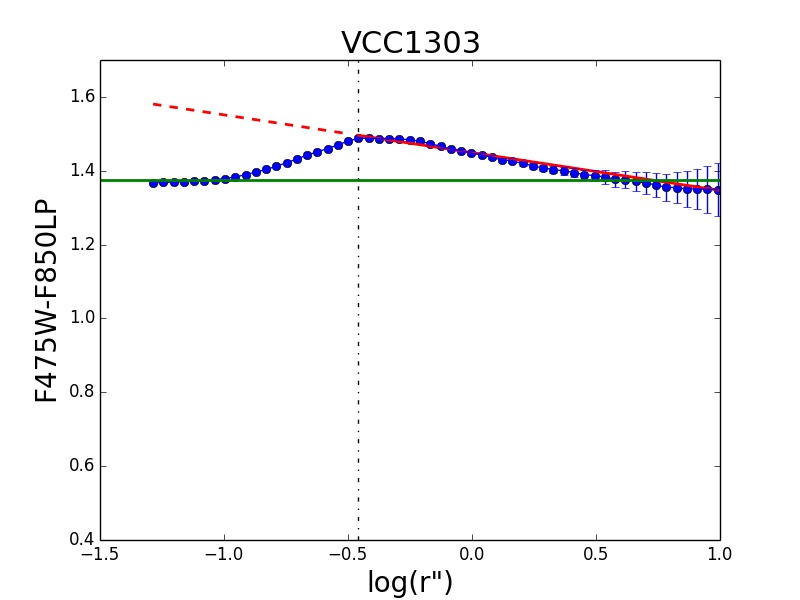}
	\includegraphics[width=4.3cm,height=3.8cm]{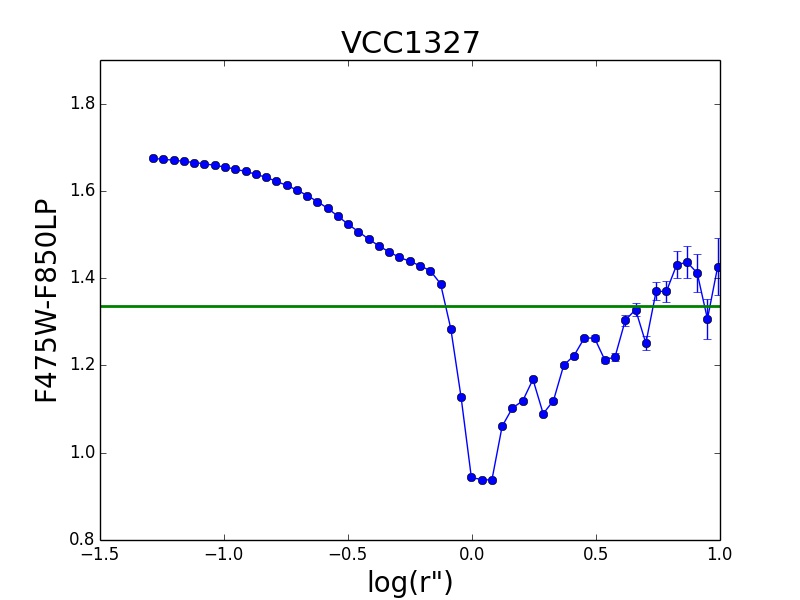}
	\includegraphics[width=4.3cm,height=3.8cm]{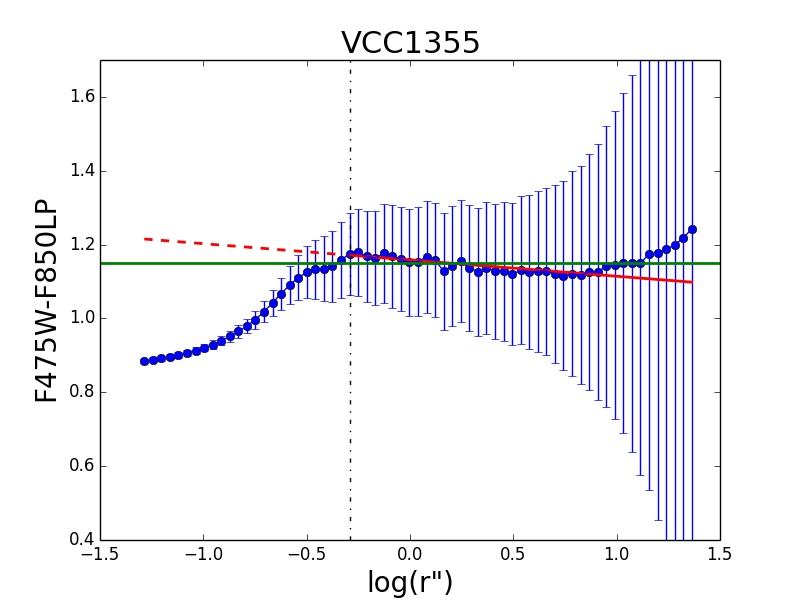}
	\includegraphics[width=4.3cm,height=3.8cm]{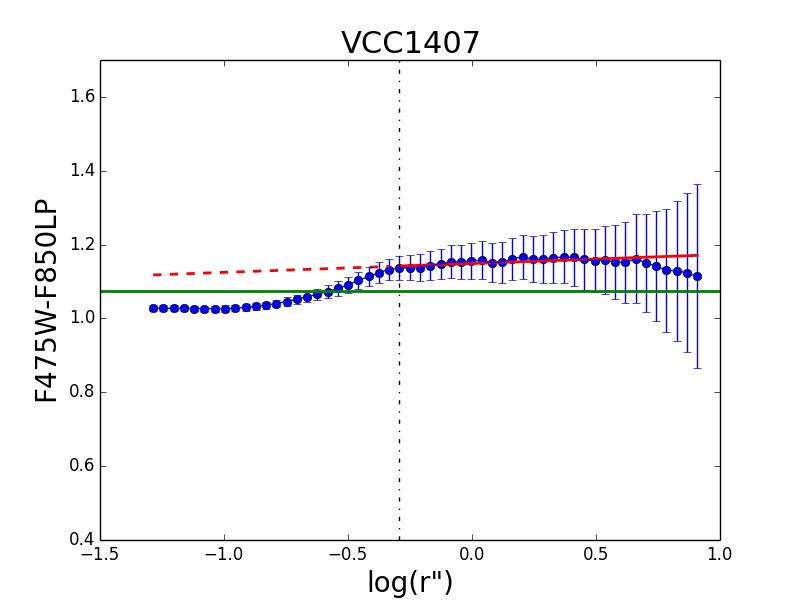}
	\includegraphics[width=4.3cm,height=3.8cm]{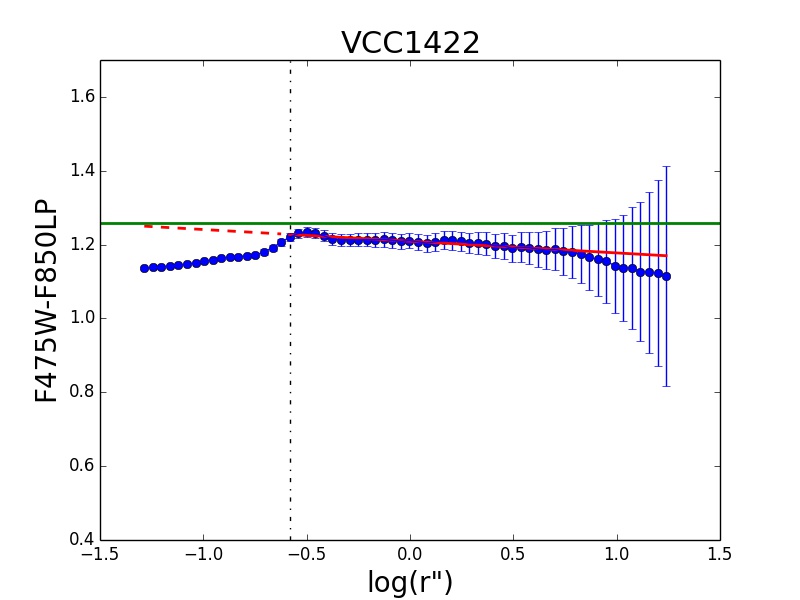}
	\includegraphics[width=4.3cm,height=3.8cm]{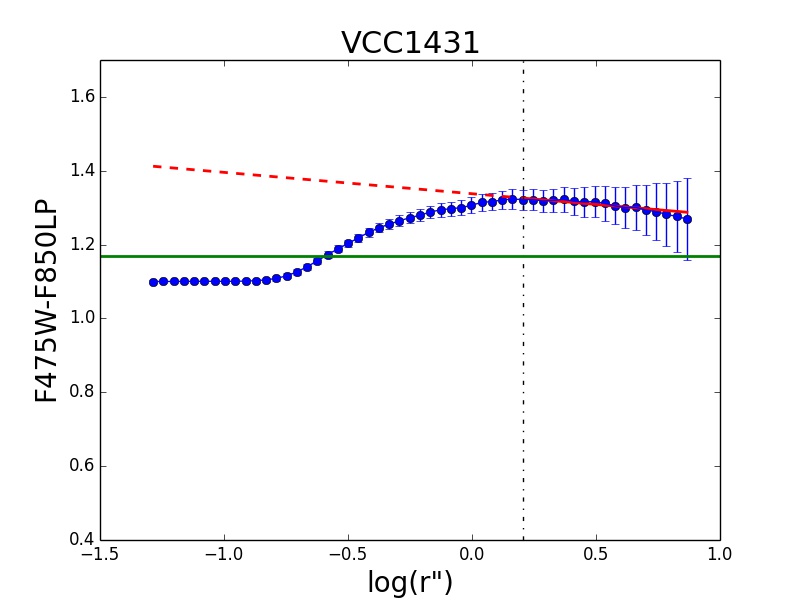}
	\includegraphics[width=4.3cm,height=3.8cm]{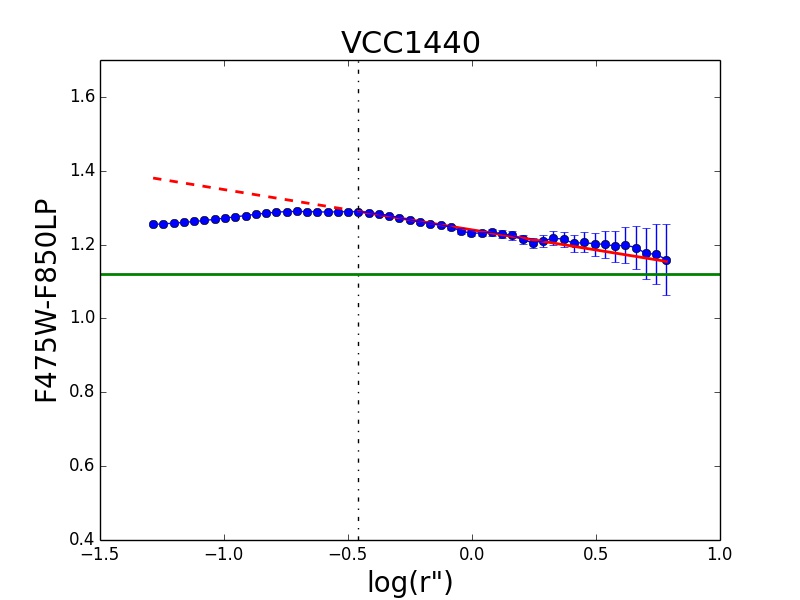}
	\includegraphics[width=4.3cm,height=3.8cm]{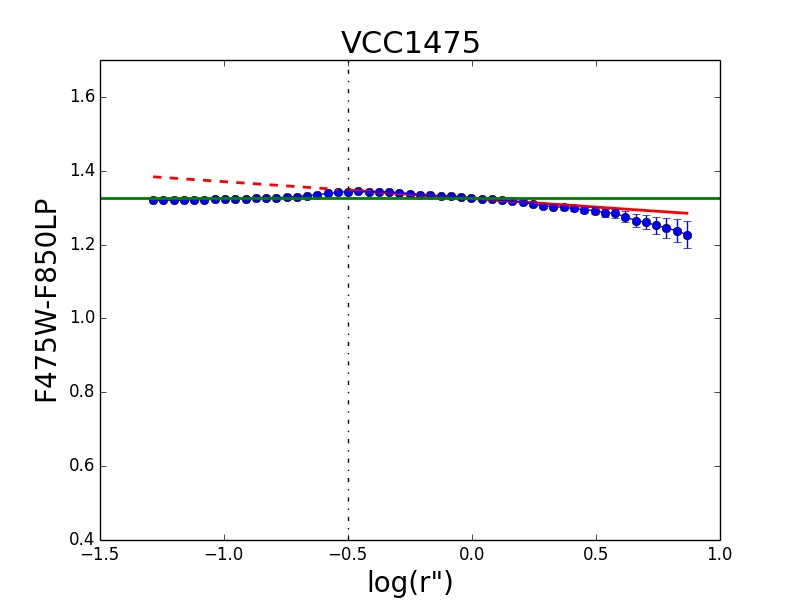}
	\includegraphics[width=4.3cm,height=3.8cm]{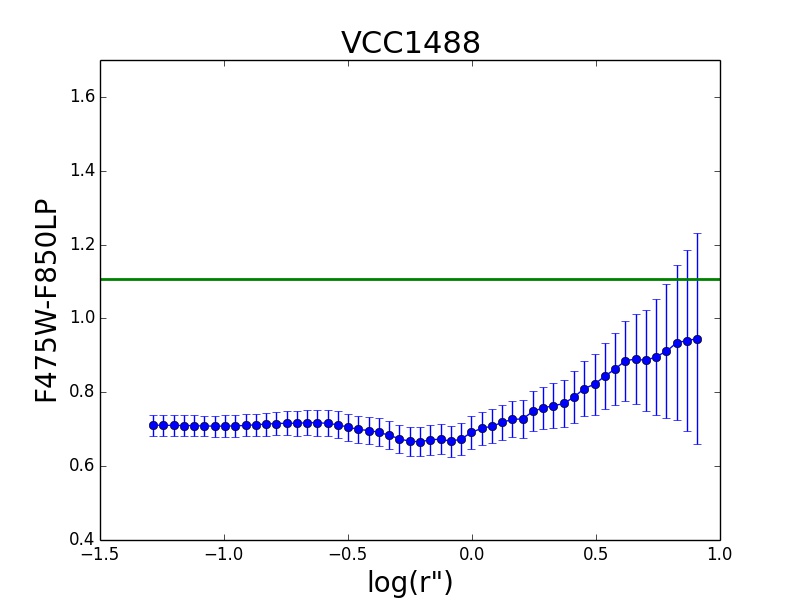}
	\includegraphics[width=4.3cm,height=3.8cm]{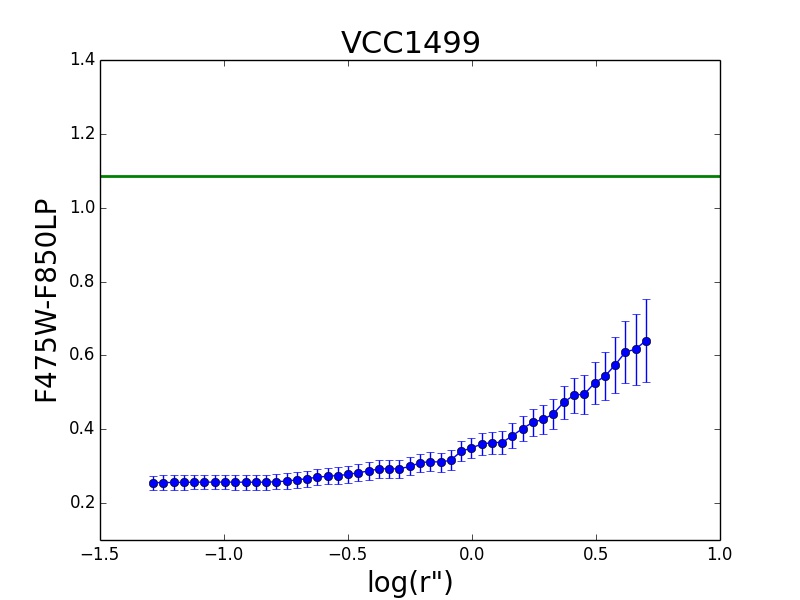}
	\includegraphics[width=4.3cm,height=3.8cm]{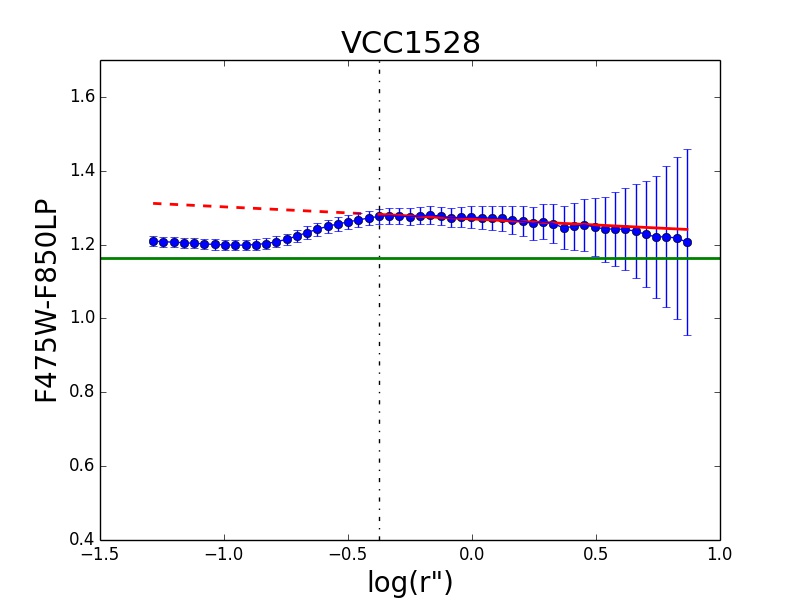}
	\includegraphics[width=4.3cm,height=3.8cm]{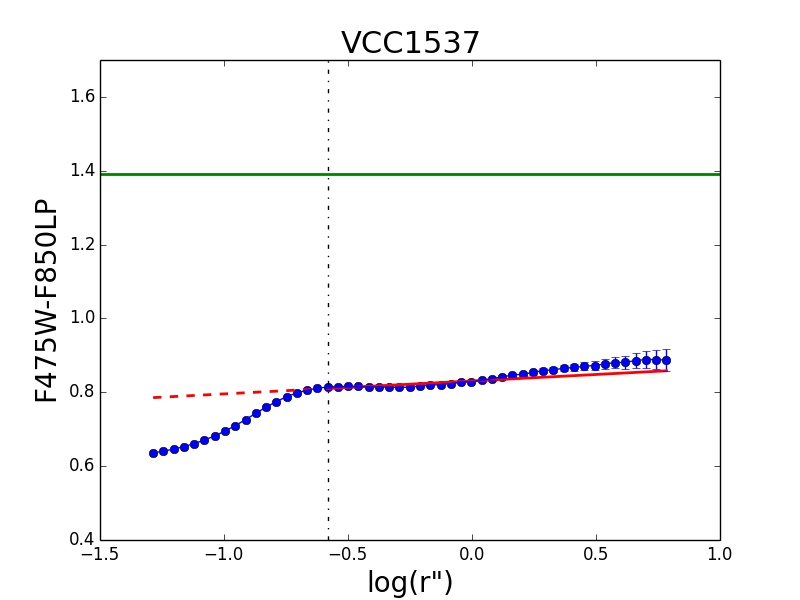}
	\includegraphics[width=4.3cm,height=3.8cm]{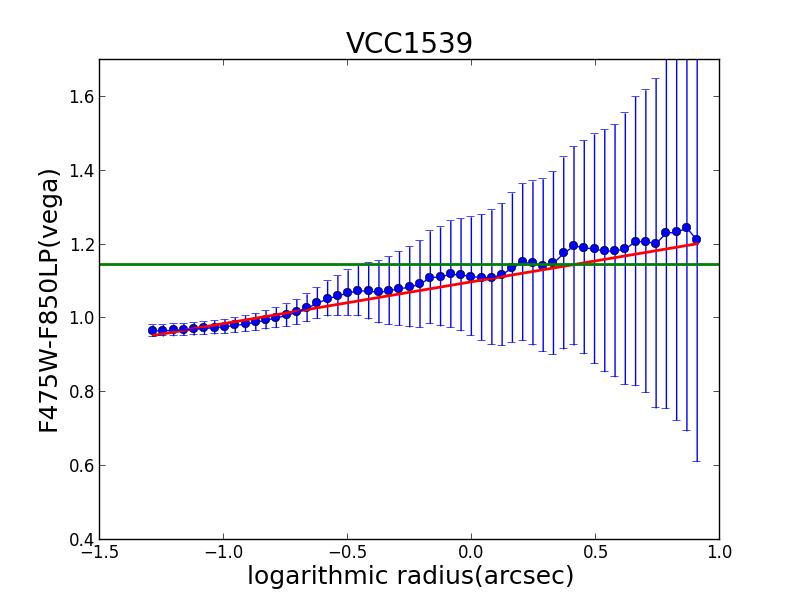}
	\includegraphics[width=4.3cm,height=3.8cm]{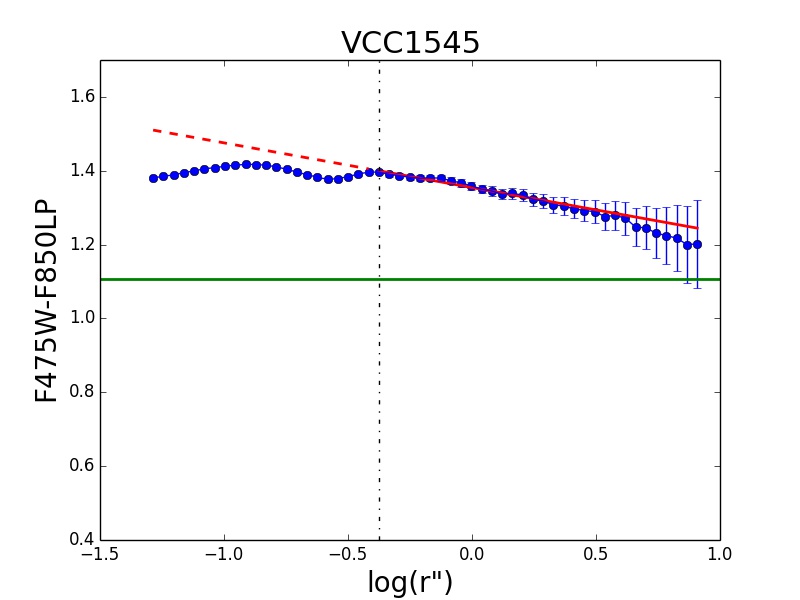}
	\includegraphics[width=4.3cm,height=3.8cm]{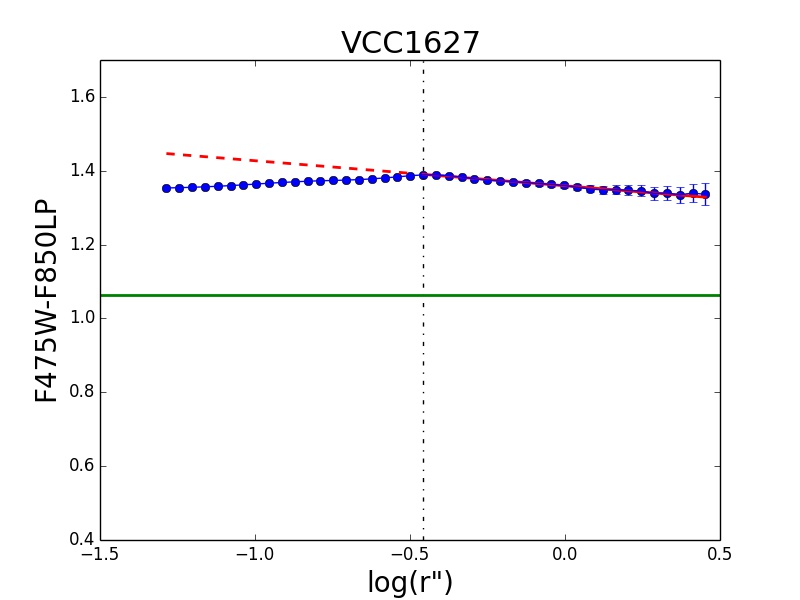}
	\includegraphics[width=4.3cm,height=3.8cm]{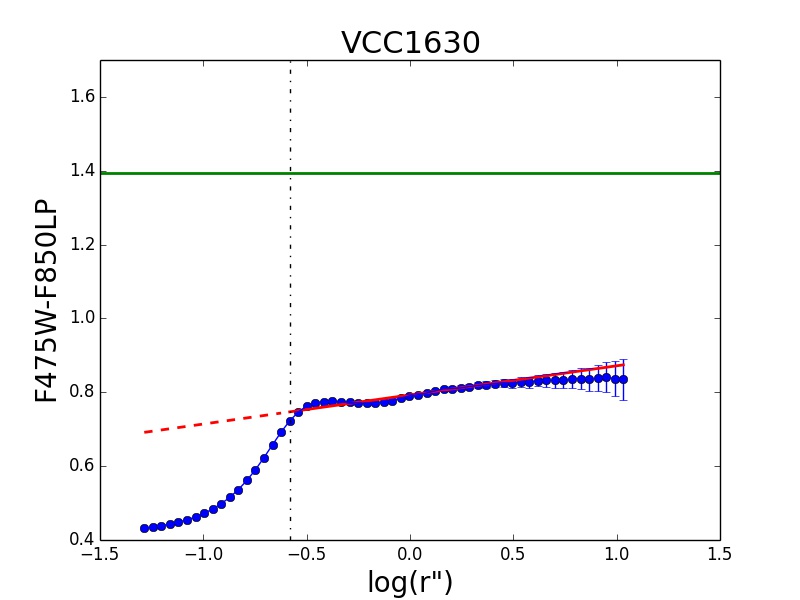}
	\includegraphics[width=4.3cm,height=3.8cm]{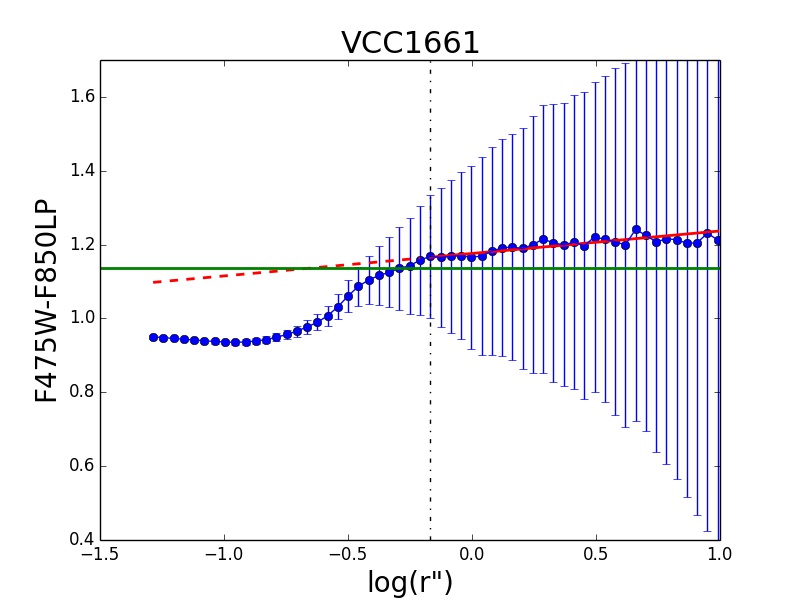}
	\includegraphics[width=4.3cm,height=3.8cm]{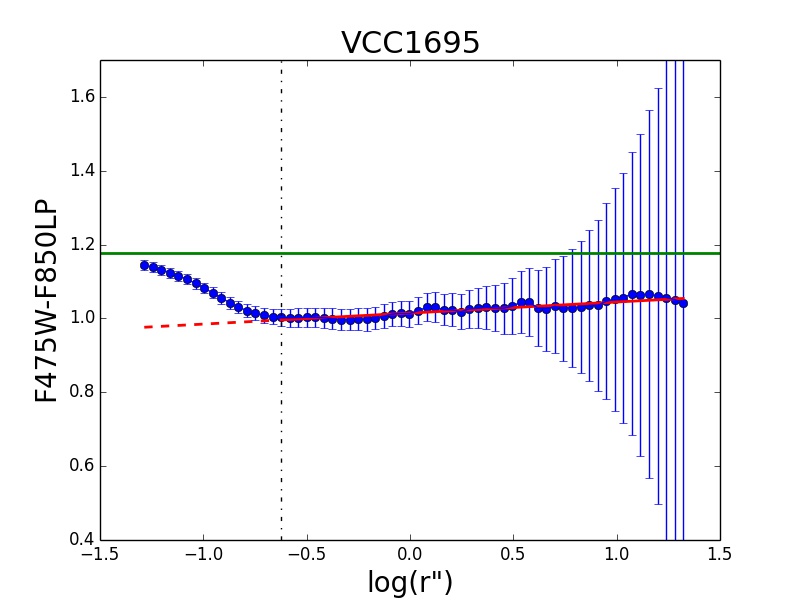}
	\includegraphics[width=4.3cm,height=3.8cm]{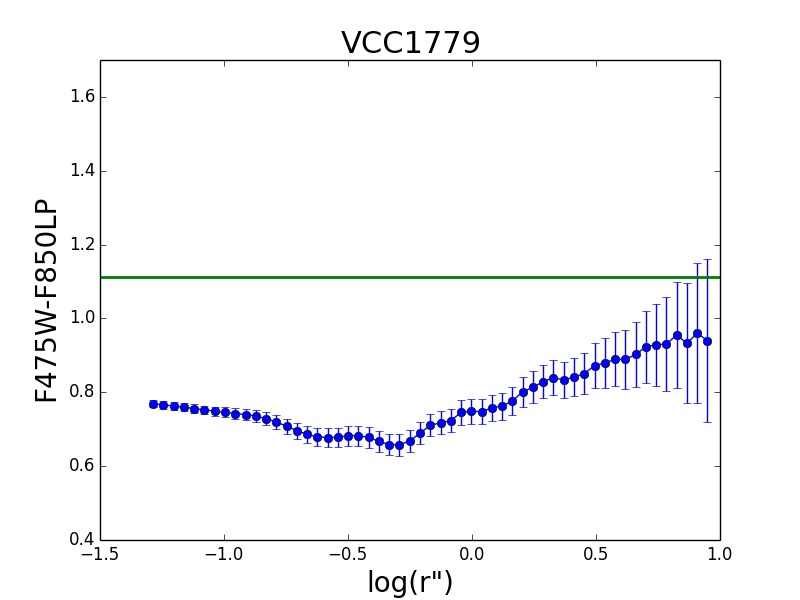}
	\includegraphics[width=4.3cm,height=3.8cm]{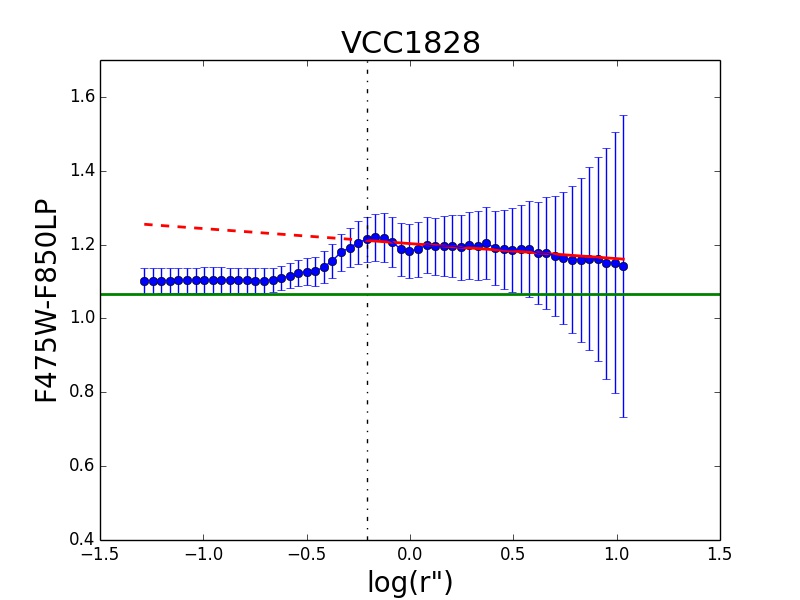}
	\includegraphics[width=4.3cm,height=3.8cm]{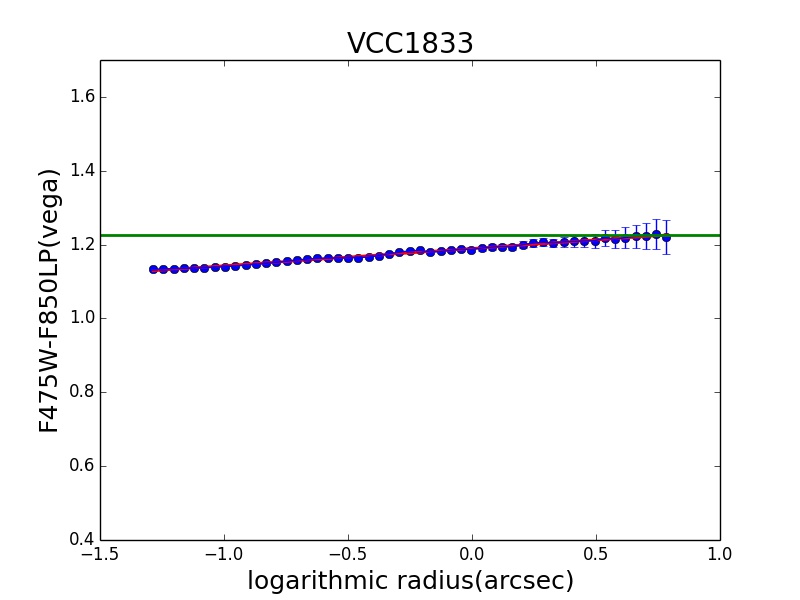}
	\includegraphics[width=4.3cm,height=3.8cm]{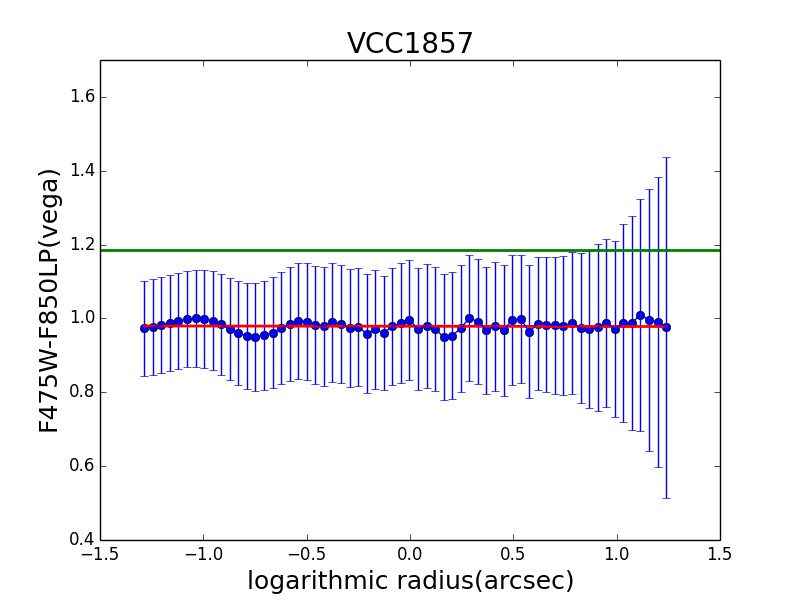}
	\includegraphics[width=4.3cm,height=3.8cm]{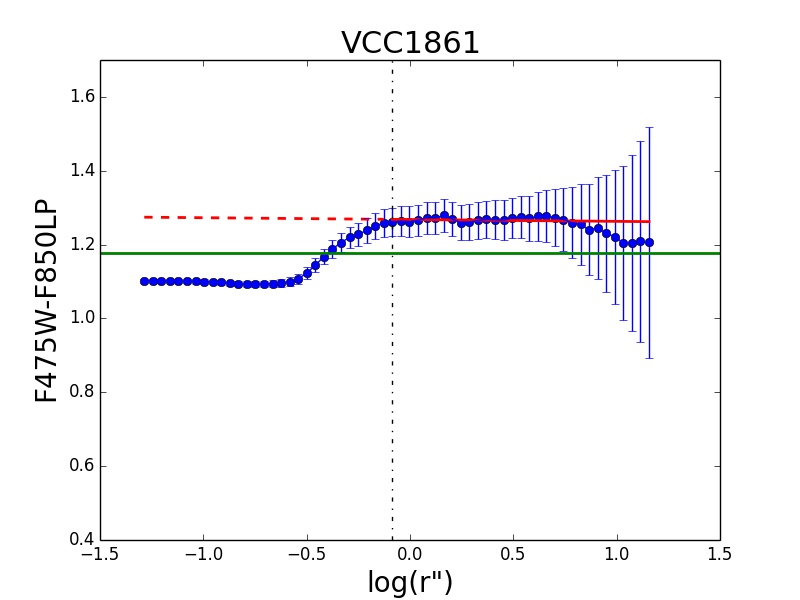}
	\includegraphics[width=4.3cm,height=3.8cm]{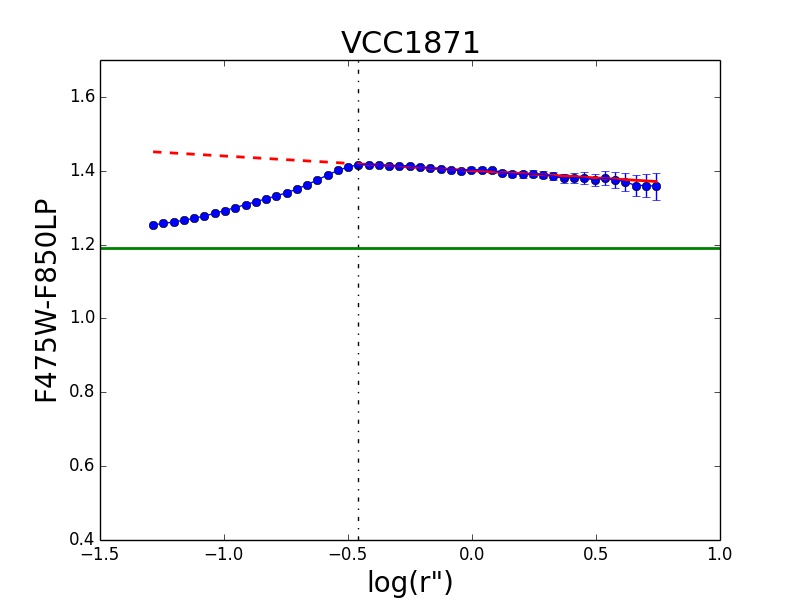}
	\caption{Continue of color profile of Virgo early-type dwarf galaxies1}
\end{figure*}

\begin{figure*} 
	\centering
	\includegraphics[width=4.3cm,height=3.8cm]{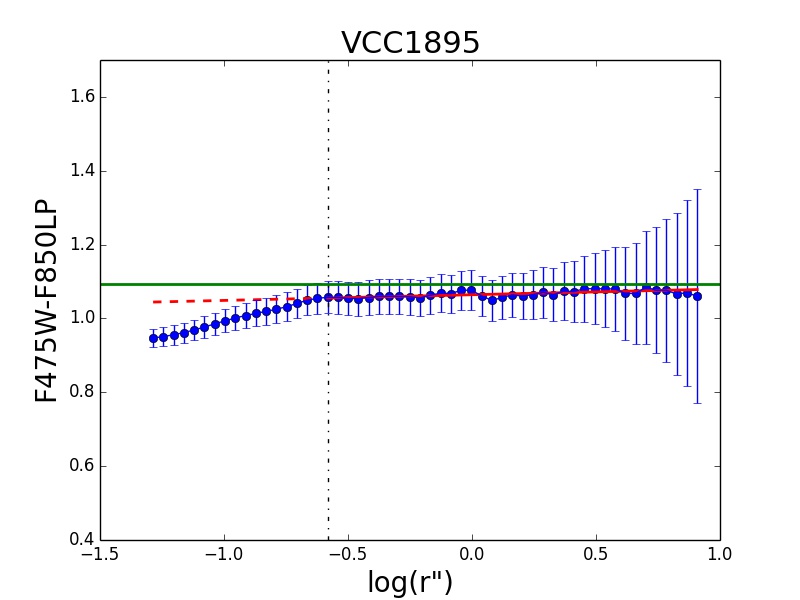}
	\includegraphics[width=4.3cm,height=3.8cm]{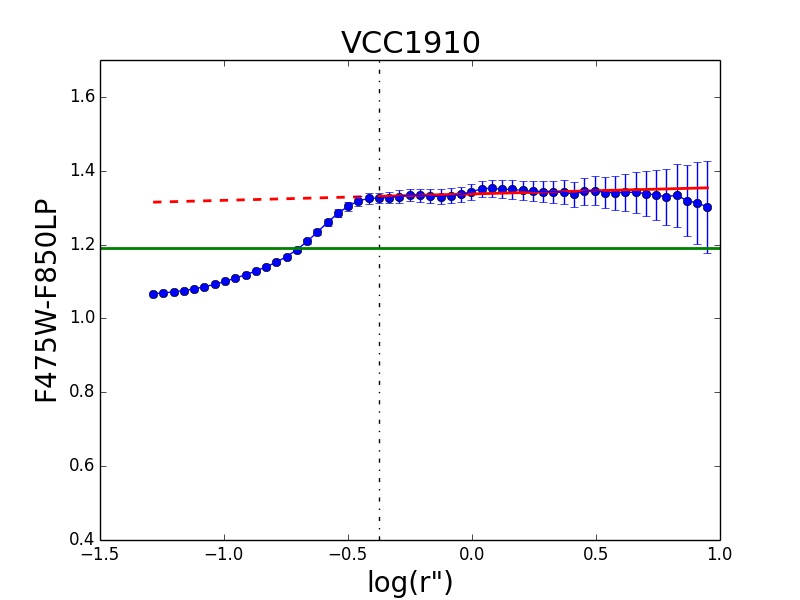} 
	\includegraphics[width=4.3cm,height=3.8cm]{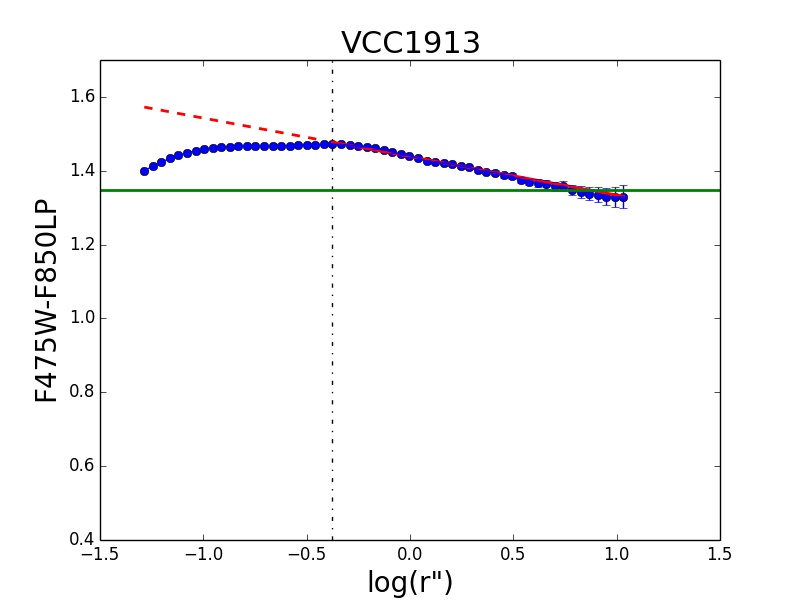}
	\includegraphics[width=4.3cm,height=3.8cm]{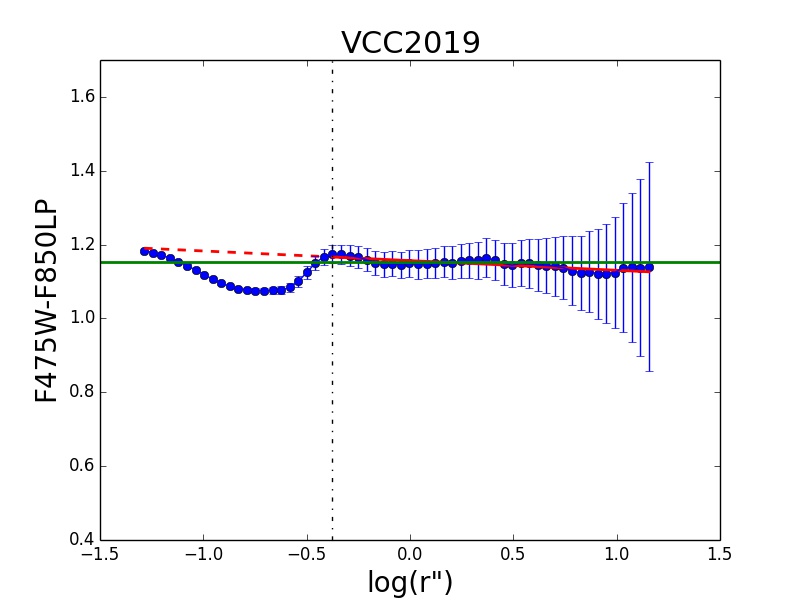}
	\includegraphics[width=4.3cm,height=3.8cm]{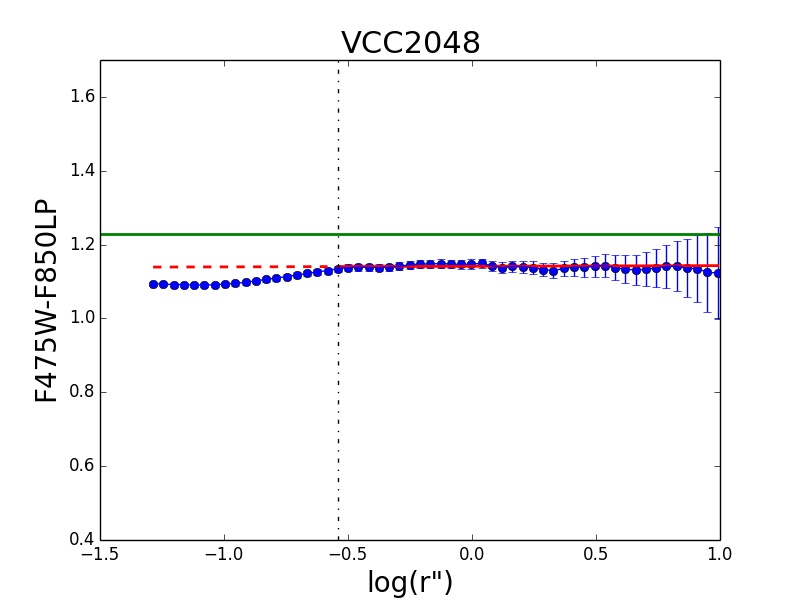}
	\includegraphics[width=4.3cm,height=3.8cm]{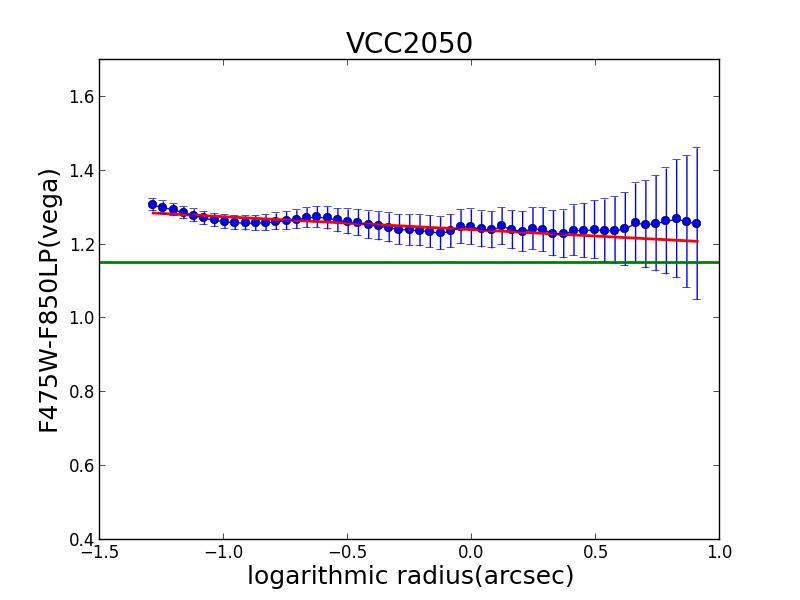}
	\caption{Continue of color profile of Virgo early-type dwarf galaxies2}
\end{figure*}

%
\newpage
\begin{table*}
\caption{Table of Coma dEs color profiles parameters}
\centering
\begin{tabular}{p{1.7in}P{.5in}P{.9in}P{.9in}P{.9in}}
		Coma-ID & $r_{center}(pc)$ & $X_{OuterFit}$ & $X_{CMR}$ & $\bigtriangledown(g^{\prime}-z^{\prime})$ \\
		(1) & (2) & (3) & (4) & (5)  \\	
		\hline
COMAi13030.949p28630.18 & 125 & -0.04 & -0.00 & -0.068 $\pm$ 0.003 \\
COMAi13024.823p275535.89 & 94 & -0.04 & -0.01 & -0.052 $\pm$ 0.003 \\
COMAi13018.782p275613.47 & 222 & -0.02 & -0.13 & -0.097 $\pm$ 0.003 \\
COMAi13021.672p275354.80 & 78 & -0.02 & -0.03 & -0.060 $\pm$ 0.002 \\
COMAi125930.268p28115.17 & 71 & -0.04 & 0.06 & -0.056 $\pm$ 0.005 \\
COMAi125937.988p28003.56 & 78 & -0.04 & -0.07 & -0.094 $\pm$ 0.002 \\
COMAi125820.533p272546.3 & 86 & -0.02 & -0.18 & -0.072 $\pm$ 0.011 \\
COMAi125815.275p272752.96 & 94 & -0.08 & 0.09 & -0.053 $\pm$ 0.006 \\
COMAi125704.336p273133.26 & 138 & -0.02 & -0.01 & -0.027 $\pm$ 0.004 \\
COMAi13035.418p275634.5 & 94 & -0.03 & -0.07 & -0.058 $\pm$ 0.002 \\
COMAi125845.544p274513.68 & 125 & -0.05 & -0.01 & -0.048 $\pm$ 0.004 \\
COMAi13041.192p28242.38 & 71 & -0.02 & 0.03 & -0.057 $\pm$ 0.003 \\
COMAi13018.543p28549.48 & 94 & -0.04 & 0.03 & -0.001 $\pm$ 0.006 \\
COMAi13005.403p28128.24 & 167 & -0.02 & -0.13 & -0.081 $\pm$ 0.003 \\
COMAi125950.181p275445.54 & 125 & -0.04 & 0.03 & -0.047 $\pm$ 0.003 \\
COMAi125942.306p275529.11 & 167 & -0.06 & -0.03 & -0.175 $\pm$ 0.005 \\
COMAi125926.459p275124.76 & 71 & -0.03 & 0.02 & -0.035 $\pm$ 0.003 \\
COMAi125948.590p275858.1 & 71 & -0.03 & -0.20 & -0.022 $\pm$ 0.006 \\
COMAi125953.930p275813.76 & 104 & -0.04 & 0.05 & -0.008 $\pm$ 0.004 \\
COMAi125946.941p275930.84 & 78 & -0.02 & -0.06 & -0.040 $\pm$ 0.005 \\
COMAi13026.152p28032.2 & 167 & -0.06 & -0.01 & -0.042 $\pm$ 0.008 \\
COMAi13044.634p28602.30 & n & n & -0.05 & -0.039 $\pm$ 0.002  \\
COMAi125904.792p28301.21 & n & n & -0.03 & -0.032 $\pm$ 0.002 \\
COMAi125909.465p28227.38 & n & n & 0.05 & -0.053 $\pm$ 0.004  \\
COMAi125911.545p28033.38 & n & n & 0.11 & -0.060 $\pm$ 0.011 \\
COMAi125940.278p275805.73 & n & n & 0.06 & -0.015 $\pm$ 0.005  \\
COMAi125939.657p275713.86 & n & n & 0.00 & -0.026 $\pm$ 0.002 \\
COMAi125931.115p275717.73 & n & n & 0.11 & 0.039 $\pm$ 0.001 \\
COMAi13034.427p275604.97 & n & n & 0.00 & 0.014 $\pm$ 0.002 \\
COMAi125937.10p28106.95 & n & n & 0.07 & -0.008 $\pm$ 0.001  \\
COMAi13011.143p28354.92 & n & n & -0.01 & -0.083 $\pm$ 0.002  \\
COMAi125935.286p275149.16 & n & n & 0.01 & -0.021 $\pm$ 0.002  \\
COMAi125959.476p275626.4 & n & n & -0.06 & -0.019 $\pm$ 0.003  \\
COMAi13005.684p275535.20 & n & n & -0.22 & 0.072 $\pm$ 0.002  \\
COMAi125944.182p275730.39 & n & n & -0.05 & -0.026 $\pm$ 0.003  \\
COMAi13006.399p28015.86 & n & n & -0.03 & -0.046 $\pm$ 0.006  \\
COMAi125944.217p275730.29 & n & n & -0.06 & -0.038 $\pm$ 0.003  \\
COMAi13017.643p275915.26 & n & n & 0.04 & -0.025 $\pm$ 0.003  \\
COMAi13018.873p28033.38 & n & n & 0.08 & -0.022 $\pm$ 0.005 \\
COMAi13007.123p275551.49 & n & n & -0.00 & -0.081 $\pm$ 0.007

\end{tabular}
\tablefoot{Coma dEs color profiles determined parameters:\\
	(1) The name of the galaxy from the Coma Cluster Survey.\\
	(2) The size of the blue or red center of each galaxy, separated by the black vertical line with an estimated error of $\sim20\%$\\
	(3) $<Color_{data}-Color_{red~line}>$: the average excess light in the center of the galaxy from the red fitted line in the color profiles\\
	(4) $<Color_{data}-Color_{CMR}>$: the average excess light of the galaxy from the CMR line in the color profiles (the green line)\\
	(5) Color gradient of each galaxy excluding the red or blue centers.\\
	Whether a galaxy is totally blue without any fitted line or it does not have a blue or red center, the undefinable parameters are shown by $n$.}
\label{Tab:Coma parameters}
\end{table*}
\end{appendix}


\clearpage
\bibliographystyle{aa} 
\bibliography{mbib1}

\end{document}